\documentclass[10pt,conference,letterpaper]{IEEEtran}
\usepackage{times,amsmath,epsfig}
\usepackage{graphicx}
\usepackage{balance}  % for  \balance command ON LAST PAGE  (only there!)

\usepackage{booktabs} % For formal tables

\usepackage{import}
\usepackage{booktabs}
\usepackage{hyperref}
\usepackage{graphicx}
\usepackage{subfig}
\let\labelindent\relax
\usepackage[inline]{enumitem}   %uncomment for vldb
\usepackage{ltablex}
\usepackage{color}	 
\usepackage{amsmath,scalerel}
\usepackage{amsfonts}
\usepackage{url}
\usepackage{balance}
\usepackage{bm}
\usepackage{bbm} %for indicator function
\usepackage{multirow}

\usepackage{colortbl}
\newtoggle{fond}
\providecommand{\chfd}%
    {\iftoggle{fond}%
        {\rowcolor[rgb]{1,.1,1}}%
        {}%
    }%

\togglefalse{fond}

\usepackage{todonotes}
\usepackage{tabularx}
\usepackage{multirow}
\usepackage[ruled, linesnumbered, titlenumbered]{algorithm2e}

\usepackage{amsthm}
\newcommand{\zzcomment}[1]{}

\DeclareMathOperator*{\argmin}{argmin}
\DeclareMathOperator*{\argmax}{argmax}
\newcommand{\dataset}{\mathcal{D}}
\newcommand{\trainset}{\mathcal{R}}
\newcommand{\testset}{\mathcal{T}}

\newcommand{\LT}{\mathcal{L}}

\newcommand{\usersofIteminTrainset}{\mathcal{U}_{i}^{\trainset}}
\newcommand{\usersofIteminTestset}{\mathcal{U}_{i}^{\testset}}
\newcommand{\itemsofUserinTrainset}{\mathcal{I}_{u}^{\trainset}}
\newcommand{\itemsofUserinTestset}{\mathcal{I}_{u}^{\testset}}
\newcommand{\itemsinTrainset}{\mathcal{I}^\trainset}
\newcommand{\itemsinTestset}{\mathcal{I}^\testset}
\newcommand{\relevantItemsofUserinTestSet}{\mathcal{I}_{u}^{\testset +}}

\newcommand{\simpleRisk}{\theta_u^{A}} %A was S
\newcommand{\LTRisk}{\theta_u^{N}}
\newcommand{\heuristicRisk}{\theta_u^{H}}
\newcommand{\tfidfRisk}{\theta_u^{T}}
\newcommand{\size}{N}

\newcommand{\recallName}{\mathrm{Recall@\size}}

\newcommand{\precisionName}{\mathrm{Precision@\size}}

\newcommand{\LTAccuracyName}{\mathrm{LTAccuracy@\size} }

\newcommand{\coverageName}{\mathrm{Coverage@\size}} 

\newcommand{\giniName}{\mathrm{Gini@\size}}

\newcommand{\FmeasureName}{\mathrm{F}\textnormal{-}\mathrm{measure@\size}}

\newcommand{\StratRecallName}{\mathrm{StratRecall@\size}}

\newcommand{\HMName}{\mathrm{HM}(\mathbf{x}, \mathbf{w})}

\newcommand{\recallFormula}{\recallName= \frac{1}{\vert \mathcal{U} \vert} \sum_{u \in \mathcal{U} } \frac{\vert \relevantItemsofUserinTestSet  \cap \mathcal{P}_{u} \vert}{\vert \relevantItemsofUserinTestSet \vert}
}

\newcommand{\precisionFormula}{\precisionName= \frac{1}{\size \vert \mathcal{U} \vert} \sum_{u \in \mathcal{U} } \vert \relevantItemsofUserinTestSet  \cap \mathcal{P}_{u} \vert}

\newcommand{\LTAccuracyFormula}{\LTAccuracyName = \frac{1}{\size \vert \mathcal{U} \vert}\sum_{u \in \mathcal{U}} \vert \LT \cap \mathcal{P}_{u} \vert}

\newcommand{\coverageFormula}{ \coverageName = \frac{ \vert \cup_{u \in \mathcal{U}} \mathcal{P}_{u} \vert }{|\mathcal{I}|}}

\newcommand{\giniFormula}{\giniName = \frac{1}{|\mathcal{I}|}(|\mathcal{I}|+1-2 \frac{\sum_{j=1}^{|\mathcal{I}|} (|\mathcal{I}|+1-j)f[j]}{\sum_{j=1}^{|\mathcal{I}|} f[j]})}

\newcommand{\FmeasureFormula}{\FmeasureName=  \frac{\precisionName . \recallName}{ \precisionName  + \recallName} }

\newcommand{\StratRecallFormula}{\StratRecallName= \frac{\sum_{u \in \mathcal{U} } \sum_{i \in \relevantItemsofUserinTestSet \cap \mathcal{P}_u} \big(\frac{1}{f_i^{\mathcal{R}}} \big)^{\beta} }{{\sum_{u \in \mathcal{U} } \sum_{i \in \relevantItemsofUserinTestSet  } \big(\frac{1}{f_i^{\mathcal{R}}} \big)^{\beta}  }} }

\newcommand{\WeightedHMFormula}{\HMName = \frac{\sum_{i=1}^{n} w_i}{\sum_{i=1}^{n} \frac{w_i}{x_i}}}

\numberwithin{equation}{section}

\newif\ifproposal
\proposalfalse

\newif\iffullpaper
\fullpapertrue

\newif\ifextraparts % should always be false. check with grep -nr 'extraparts' to see where it is used
\extrapartsfalse
\newif\ifbootstrapexp
\bootstrapexptrue
\newcommand{\eat}[1]{} % the eat command just deletes text
\newenvironment{changemargin} [2]{\begin{list}{}{
         \setlength{\topsep}{0pt}\setlength{\leftmargin}{0pt}
         \setlength{\rightmargin}{0pt}
         \setlength{\listparindent}{\parindent}
         \setlength{\itemindent}{\parindent}
         \setlength{\parsep}{0pt plus 1pt}
         \addtolength{\leftmargin}{#1}\addtolength{\rightmargin}{#2}
         }\item }{\end{list}}

  {
    \begin{changemargin}{-8pt}{-0cm}
    \vspace{-13pt}
    \hspace{-8pt}
    \begin{itemize}
    \setlength{\itemsep}{-1pt}
  }
  {
    \end{itemize}
    \end{changemargin}
  }

  {
    \begin{changemargin}{-8pt}{-0cm}
    \vspace{-13pt}
    \hspace{-8pt}
    \begin{enumerate}
    \setlength{\itemsep}{-1pt}
  }
  {
    \end{enumerate}
    \end{changemargin}
  }

\newlength\mystoreparindent

\newtheorem{defin}{Definition}
\newtheorem{prblm}{Problem}
\newtheorem{chal}{Challenge}
\newtheorem{ex}{Example}
\newtheorem{thm}{Theorem}[section]
\newtheorem{lem}{Lemma}[section]
\newtheorem{corol}{Corollary}[section]
\newtheorem{clam}{Claim}[section]

\newenvironment{theorem}{\begin{thm} \nopagebreak}{\end{thm}}

\newenvironment{lemma}{\begin{lem} \nopagebreak}{\end{lem}}

\numberwithin{equation}{section} 
\DeclareMathAlphabet{\mathcal}{OMS}{cmsy}{m}{n}

\title{A Generic  Top-N  Recommendation Framework For Trading-off  Accuracy, Novelty, and Coverage}%

\author{%
{Zainab Zolaktaf{\small $~^{\#1}$}, Reza Babanezhad{\small $~^{\#2}$}, Rachel Pottinger{\small $~^{\#3}$} }%
\vspace{1.6mm}\\
\fontsize{10}{10}\selectfont\itshape
$^{\#}$\,Department of Computer Science, University of British Columbia, Vancouver, B.C, Canada\\
\fontsize{9}{9}\selectfont\ttfamily\upshape
$^{1}$\,zolaktaf@cs.ubc.ca,
$^{2}$\,rezababa@cs.ubc.ca,
$^{3}$\,rap@cs.ubc.ca%
\vspace{1.2mm}\\
\fontsize{10}{10}\selectfont\rmfamily\itshape
}
\begin{document}
\maketitle
\begin{abstract}
\iffalse
In online commerce, recommender systems help consumers find products
they may purchase and help producers increase revenue. We consider
top-$\size$ recommendation in these contexts, where the goal is to
recommend the most appealing set of $\size$ items. Traditional
recommender systems are inherently biased toward advocating popular
items. This is inadequate from both the consumer's and producer's
perspectives. We investigate how individual consumer characteristics
can be exploited to spread demand more evenly between the popular and
niche items, and to ensure both consumer and producer satisfaction.
We propose a generic recommendation framework that targets relevance
of individual top-$\size$ sets and long-tail item promotion with
regard to consumer traits. We then exploit the structure and
properties of our formulated objective function to design efficient
greedy heuristics that obtain near-optimal solutions. Empirical
results also show our proposed framework \begin{enumerate*} \item
succeeds in spreading the demand between the popular and niche items,
and \item significantly increases the coverage of the system while
maintaining consumer satisfaction.\end{enumerate*}

\else
Standard collaborative filtering approaches for top-N
recommendation are biased toward popular items. As a result, they
recommend items that users are likely aware of and
under-represent \emph{long-tail} items.  This is inadequate,
both for consumers who prefer novel
items and because concentrating on popular items poorly covers 
the item space, whereas high item space coverage increases providers'
revenue.

We present an approach that relies on historical rating data to learn
user long-tail novelty preferences.  We integrate these preferences
into a \emph{generic} re-ranking framework that customizes balance
between accuracy and coverage. We empirically validate that 
our proposed
framework increases the novelty of recommendations. Furthermore, by
promoting long-tail items to the right group of users, we
significantly increase the system's coverage while scalably
maintaining accuracy. Our framework also enables personalization of
existing non-personalized algorithms, making them competitive with
existing personalized algorithms in key
performance metrics, including accuracy and coverage.

\end{abstract}

\section{Introduction}
\label{sec:Introduction}

The goal in top-$\size$ recommendation is to recommend to each
consumer a small set of $\size$ items from a large collection of
items~\cite{cremonesi2010performance}.  For example, Netflix may want
to recommend $\size$ appealing movies to each consumer.  Collaborative
Filtering (CF)~\cite{herlocker2002empirical,lee2012comparative} is a
common top-$\size$ recommendation method.  CF infers user interests by
analyzing partially observed user-item interaction data, such as user
ratings on movies or historical purchase
logs~\cite{kanagal2012supercharging}. The main assumption in CF is that
users with similar interaction patterns have similar interests.

Standard CF methods for top-$\size$ recommendation focus on making  suggestions  that accurately reflect the user's preference history. However, as  observed in previous work,  CF recommendations are generally biased toward  popular items, leading to a rich get richer effect~\cite{vargas2014improving,steck2011item}.  The major reasons for this are \textit{popularity bias} and \textit{sparsity} of CF interaction data (detailed in Section~\ref{sec:related-work}). In a nutshell, to maintain  accuracy, recommendations are generated from the dense regions of the data,  where the popular items lie.  

However,  accurately suggesting popular items, may not be satisfactory for the consumers. For example, in Netflix, an accuracy-focused movie recommender may recommend ``Star Wars: The Force Awakens'' to users who have seen ``Star Wars: Rogue One''.  But, those users are probably already aware of ``The Force Awakens''. Considering additional factors, such as novelty of recommendations,  can lead to more effective suggestions~\cite{cremonesi2010performance,Castells2015,zhang2008avoiding,ziegler2005improving,zhang2012auralist}. 
%Second, accuracy-focused models typically achieve a   overall item-space coverage across their recommendations,  whereas high item-space coverage helps providers of the items increase revenue
%, users satisfaction since they are  likely already aware of or can find these items on their own.  

Focusing on popular items also adversely affects the satisfaction of  the providers of the items. This is because  accuracy-focused models typically achieve a  low overall item space coverage across their recommendations, whereas   high item space coverage helps providers of the items increase their revenue~\cite{vargas2014improving,Castells2015,adomavicius2011maximizing,anderson2006thelongtail, yin2012challenging,adomavicius2012improving}.
%accuracy-focused models typically achieve a

In contrast to the relatively small number of popular items, there are copious  {\it long-tail\/} items that have fewer observations (e.g., ratings) available. More precisely,  using the Pareto  principle (i.e.,~the $80/20$ rule),  long-tail items can be defined as items that generate the lower $20\%$ of observations~\cite{yin2012challenging}. Experimentally we found that these items correspond to almost $85\%$ of the items in several datasets (Sections~\ref{sec:Notation} and \ref{sec:Experiments}). %Table~\ref{tab:DatasetStatsticsSmall})

As previously shown, one way to improve the novelty of top-$\size$ sets is to recommend interesting long-tail items~\cite{cremonesi2010performance,ge2010beyond}.  The intuition  is that since they have fewer observations available,  they are more likely to be unseen~\cite{Kaminskas:2016:DSN:3028254.2926720}.  
 %For example, in online commerce,  newly added items are long-tail items that are yet to be discovered.  
Moreover, long-tail item promotion also results in higher overall coverage of the item space%, which increases profits for providers of the items
~\cite{vargas2014improving,Castells2015,zhang2008avoiding,zhang2012auralist,adomavicius2011maximizing,anderson2006thelongtail,yin2012challenging,jambor2010optimizing}. Because long-tail promotion reduces accuracy~\cite{steck2011item}, there are trade-offs to be explored.

%original submitted to ICDE
%This work studies three aspects of top-$\size$ recommendation: accuracy, novelty, and item-space coverage, and examines their trade-offs. In most previous work, predictions of a base recommendation system are re-ranked to handle their trade-offs~\cite{adomavicius2012improving,jambor2010optimizing,zhang2013personalize,wang2009portfolio}. Due to performance considerations, however, these techniques are not customized per user. For example,  parameters that balance the trade-off between novelty and accuracy are cross-validated at a global level.  This can be detrimental since users have varying preferences for  objectives such as long-tail novelty. We explore how to  automatically infer  user  preference for long-tail novelty, and how to leverage  it to correct  the popularity bias in standard recommender models. Our work does not rely on any additional contextual data, although such data, if available, can help promote newly-added long-tail items~\cite{agarwal2009regression,Saveski:2014:ICR:2645710.2645751}.

This work studies three aspects of top-$\size$ recommendation: accuracy, novelty, and item space coverage, and examines their trade-offs. In most previous work, predictions of a base recommendation algorithm are \textit{re-ranked} to handle these trade-offs~\cite{adomavicius2012improving,jambor2010optimizing,zhang2013personalize,wang2009portfolio}. The re-ranking models are computationally efficient but suffer from two drawbacks. First, due to performance considerations,  parameters that balance the trade-off between novelty and accuracy  are not customized per user. Instead they are cross-validated at a global level.  This can be detrimental since users have varying preferences for  objectives such as long-tail novelty. Second,  the re-ranking methods are often limited to a specific base recommender  that may be sensitive to dataset density. 
As a result, the datasets are pruned and the problem is studied in dense settings~\cite{adomavicius2012improving,ho2014likes}; but real world  scenarios are often sparse~\cite{kanagal2012supercharging,liu2017experimental}.   

We address the first limitation by directly inferring  user  preference for long-tail novelty  from interaction data.   Estimating these  preferences  using only item popularity statistics, e.g., the average popularity of rated items as in~\cite{jugovac2017efficient}, disregards additional information, like whether the user found the item interesting or the long-tail preferences of other users  of the items. We propose an approach that  incorporates  this information and  learns the users' long-tail novelty preferences from interaction data.

This approach allows us to customize the re-ranking  per user, and  design a \textit{generic} re-ranking framework, which resolves the second limitation of prior work. In particular, since the long-tail novelty preferences are estimated independently of any base recommender, we can  plug-in an appropriate one w.r.t. different factors, such as the dataset sparsity.

Our top-$\size$ recommendation framework, \textbf{GANC}, is \textbf{G}eneric, and provides customized balance between \textbf{A}ccuracy, \textbf{N}ovelty, and \textbf{C}overage. % Moreover, based on the learned long-tail novelty preferences, we also design a novel algorithm, {\it Ordered Sampling-based Locally Greedy (OSLG)\/}, that relies on the learned long-tail novelty preferences  to scalably correct for popularity bias. 
Our work does not rely on any additional contextual data, although such data, if available, can help promote newly-added long-tail items~\cite{agarwal2009regression,Saveski:2014:ICR:2645710.2645751}. In summary:

%\input{RunningExample}

%We propose a novel approach that allows us to  promote long-tail items in a targeted manner, thereby improving the novelty of top-$\size$ sets, the overall item-space coverage across recommendations, while maintaining reasonable levels of accuracy.

%Next, we integrate these learned preferences  in a generic  top-$\size$ recommendation framework to provide customized balance between accuracy and coverage.

%sequentially make recommendations, while adjusting its parameters with regard to the set of top-$\size$ recommendations made so far. However, since  sequential parameter updates  cause  scalability issues, we propose a sampling based algorithm. This variant of our framework, called {\it Ordered Sampling-based Locally Greedy (OSLG)\/},  allows us to  correct for the popularity bias in recommendations with regard to individual user long-tail preferences. 

%ICDE submission
%Our framework differs with  prior work in the following aspects:  unlike~\cite{adomavicius2011maximizing,adomavicius2012improving,zhang2013personalize,ho2014likes},  the long-tail preference personalization in our framework is learned rather than optimized using cross-validation or parameter tuning. In other words, our personalization method is independent of the underlying base  recommendation models.  Moreover, our framework is  generic. This enables us to  plug-in several base recommenders, and evaluate their  effectiveness without requiring  extensive tuning for the accuracy and coverage trade-off. 

%\vspace{-2.8pt}
\begin{itemize}

\item  We examine various measures for estimating user long-tail novelty preference in Section~\ref{sec:lt-pref} and formulate an optimization problem  to directly learn users' preferences for long-tail  items from interaction data in Section~\ref{sec:learning-lt-pref}. %In addition, we introduce several heuristics for measuring the user preference for less common items from historical rating data.% 

\item  We integrate the user preference estimates into GANC %, a generic re-ranking framework that provides customized balance between accuracy, novelty, and coverage 
(Section~\ref{sec:RiskbasedReranking}), and  introduce {\it Ordered Sampling-based Locally Greedy (OSLG)\/}, a scalable algorithm that relies  on user long-tail preferences to correct the popularity bias (Section~\ref{sec:optimizationAlgorithm}).
%We introduce OSLG, a scalable algorithm that relies  on user long-tail preferences to  maximize item space coverage \textcolor{red}{while maintaining acceptable levels of accuracy} (Section~\ref{sec:optimizationAlgorithm}).

\item   We conduct an extensive empirical study and evaluate performance from  accuracy, novelty, and coverage perspectives (Section~\ref{sec:Experiments}).  We use five  datasets with varying density and difficulty levels. %:  Netflix, MovieTweetings, and MovieLens (100K, 1M, 10M). 
  In contrast to most related work,  our evaluation considers realistic settings that include a large number of infrequent  items and users. %This enables us to study the impact of  data density on the performance trade-offs of several  state of the art top-$\size$ recommendation algorithms. %   %,  and use the all-items ranking protocol~\cite{steck2013evaluation,vargas2014improving}, where performance is measured using all items with train data. to evaluate the performance of several  state of the art top-$\size$ recommendation algorithms 
 
\item Our empirical results confirm that the performance of re-ranking models is impacted by the underlying   base recommender and the dataset density. Our generic approach enables us to easily incorporate a suitable base recommender to devise an effective solution for both dense and sparse settings. In dense settings, we use the same base recommender as existing re-ranking approaches, and we outperform them in accuracy and coverage metrics. For sparse settings, we plug-in a more suitable base recommender, and devise an effective solution that is competitive with existing top-$\size$ recommendation methods in accuracy and novelty. 

%Directly estimating the long-tail novelty preferences allows us to customize re-ranking per user, and  devise a generic framework.   
 
\end{itemize}

Section~\ref{sec:related-work} describes related work. Section~\ref{sec:conclusion} concludes.

\section{Long-tail novelty preference}
\label{sec:lt-pref}
We begin this section by  introducing our notation. We then describe various models  for measuring user long-tail novelty preference (Section~\ref{sec:simple-lt-pref}).% and propose a new  optimization framework to learn these preferences (Section~\ref{sec:learning-lt-pref}). %user long-tail novelty
\iffullpaper
\begin{figure*}[t]
\centering
        \subfloat[ML-100K]
        {					 
		\includegraphics[scale=0.27]{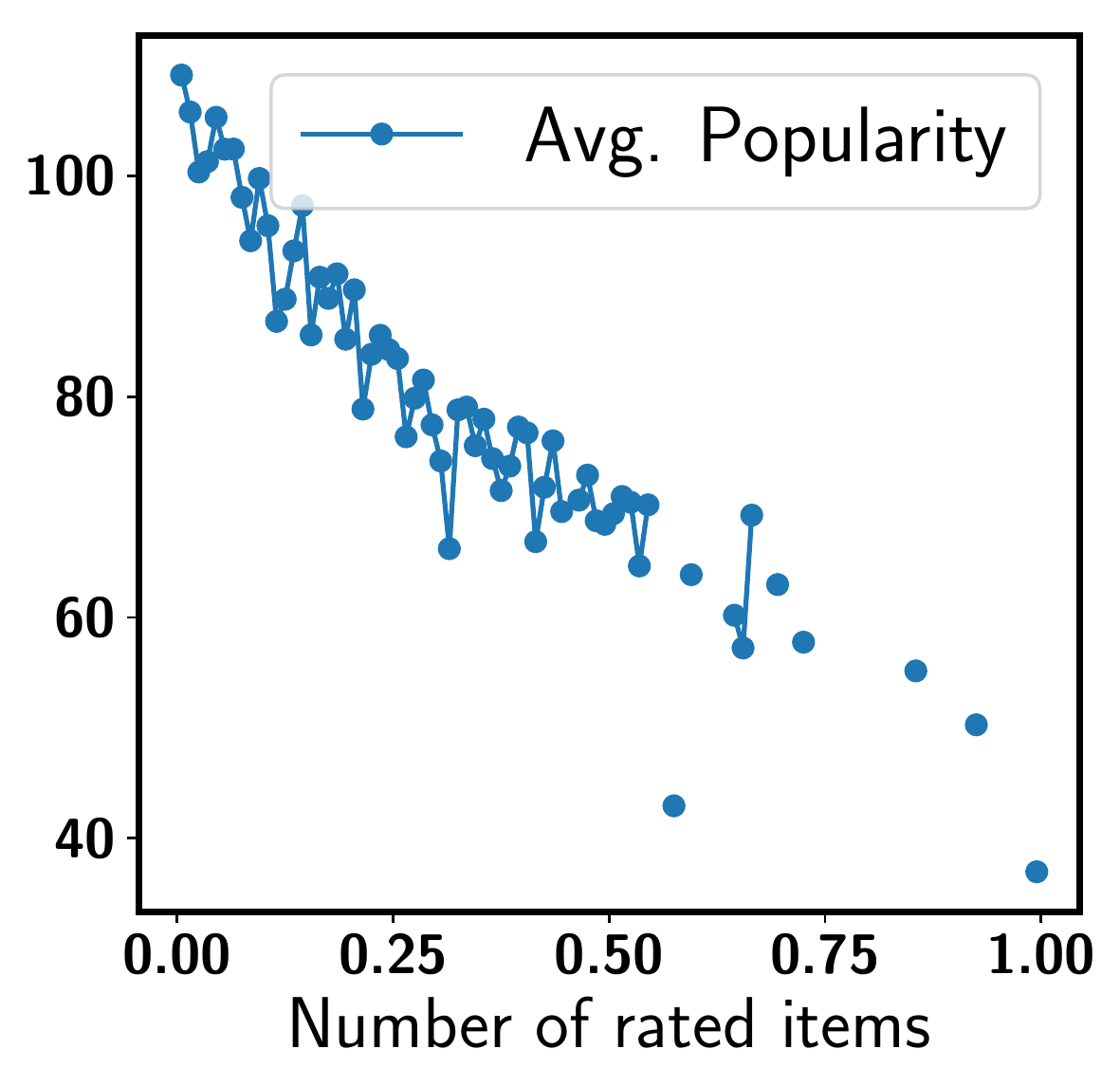}
   		\label{fig:ML10m-pop-vs-number-ratings}
        }
        \subfloat[ML-1M]
        {
		\includegraphics[scale=0.27]{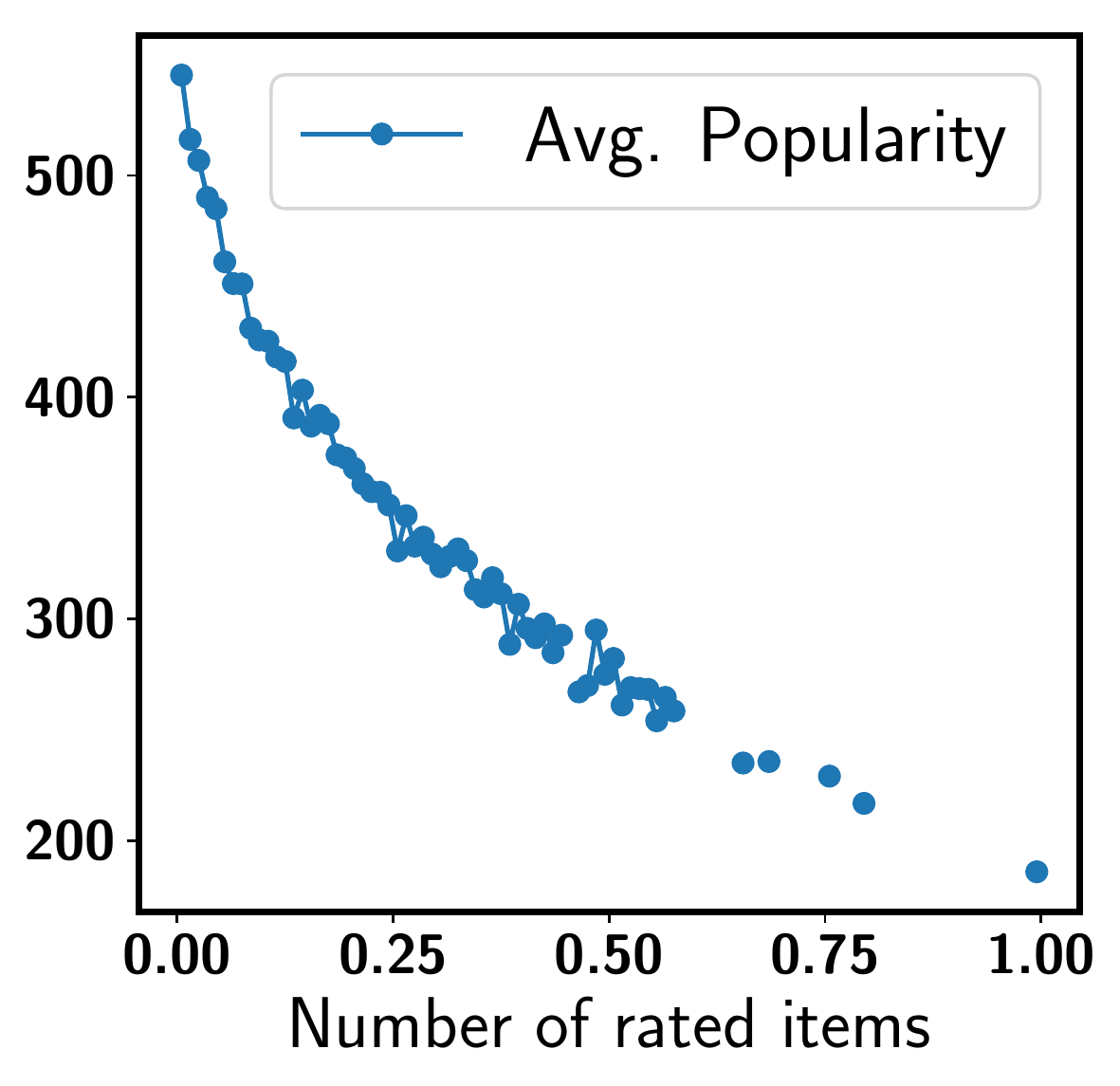}
    		\label{fig:ML1m-pop-vs-number-ratings}
        }
        \subfloat[ML-10M]
        {					 
		\includegraphics[scale=0.27]{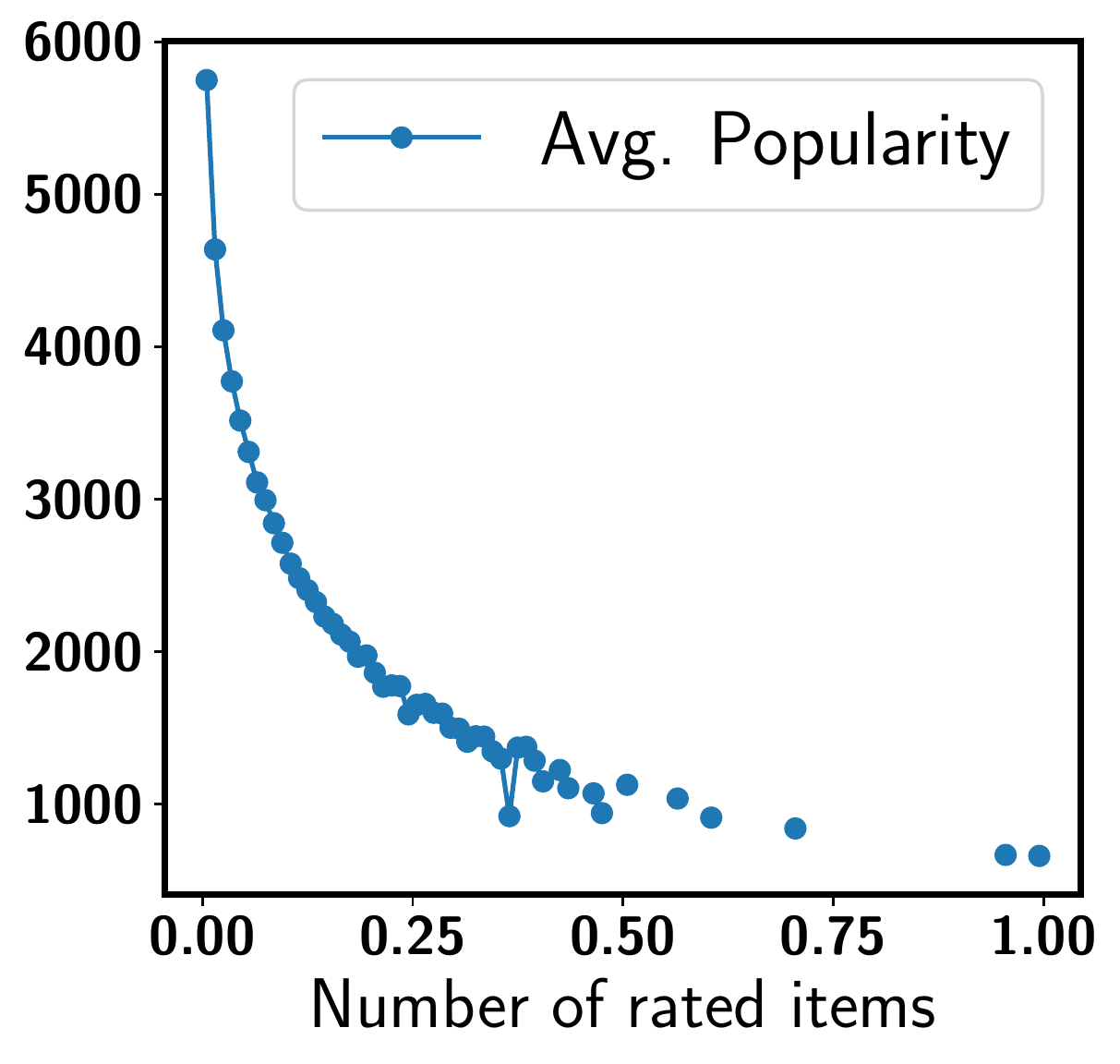}
   		\label{fig:ML10m-pop-vs-number-ratings}
        }
         \subfloat[MT-200K]
        {					 
		\includegraphics[scale=0.27]{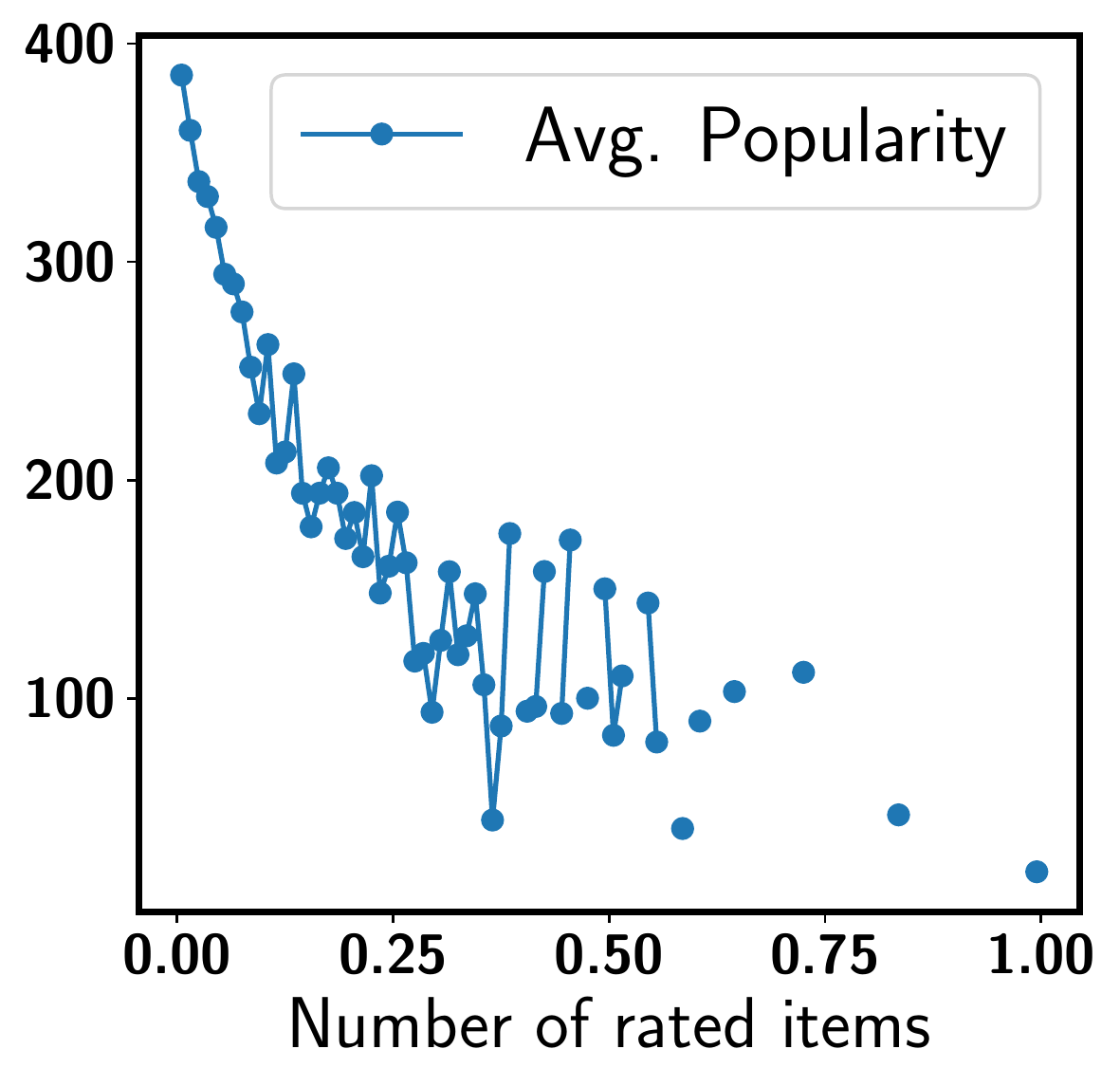}
   		\label{fig:MT-pop-vs-number-ratings}
        }
        \subfloat[Netflix]
        {					 
		\includegraphics[scale=0.27]{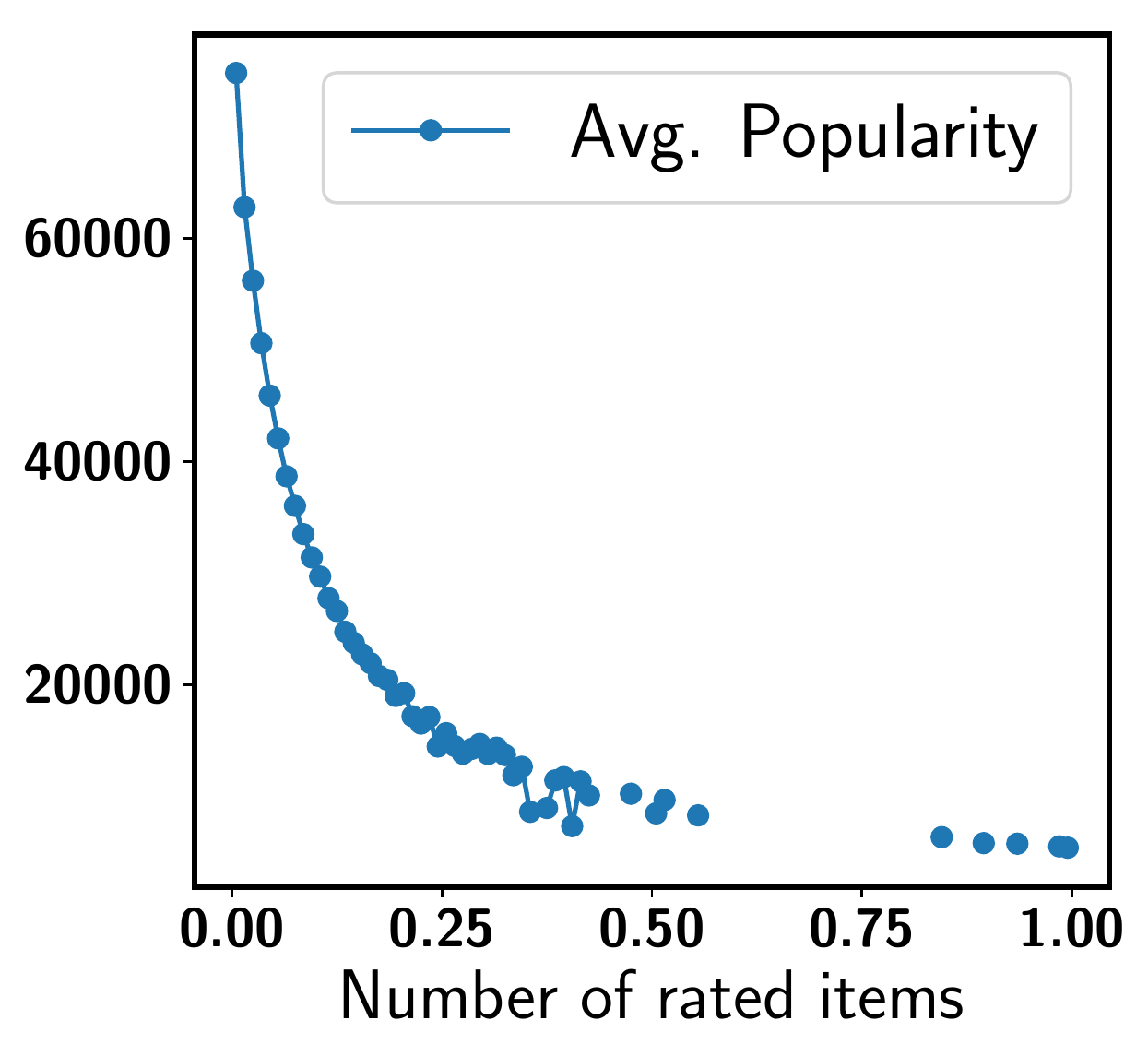}
   		\label{fig:MT-pop-vs-number-ratings}
        }
 		        
\caption{For each user $u$, we consider the set of items rated by $u$ in train $\itemsofUserinTrainset$, and compute its average popularity $ \bar{a} =\frac{1 }{|\itemsofUserinTrainset|} \sum_{i \in \itemsofUserinTrainset} f_i^{\trainset}$. The x-axis shows the binned normalized $|\itemsofUserinTrainset|$, while the y-axis plots the mean of the corresponding $\bar{a}$ values. The average popularity of items rated decreases as the number of items rated increases. }
\label{fig:pop-vs-no-items}
\end{figure*}

\else % if you just want to figures in the paper, use this section.
\begin{figure}[t]
\centering
         \subfloat[ML-1M]
        {
		\includegraphics[scale=0.33]{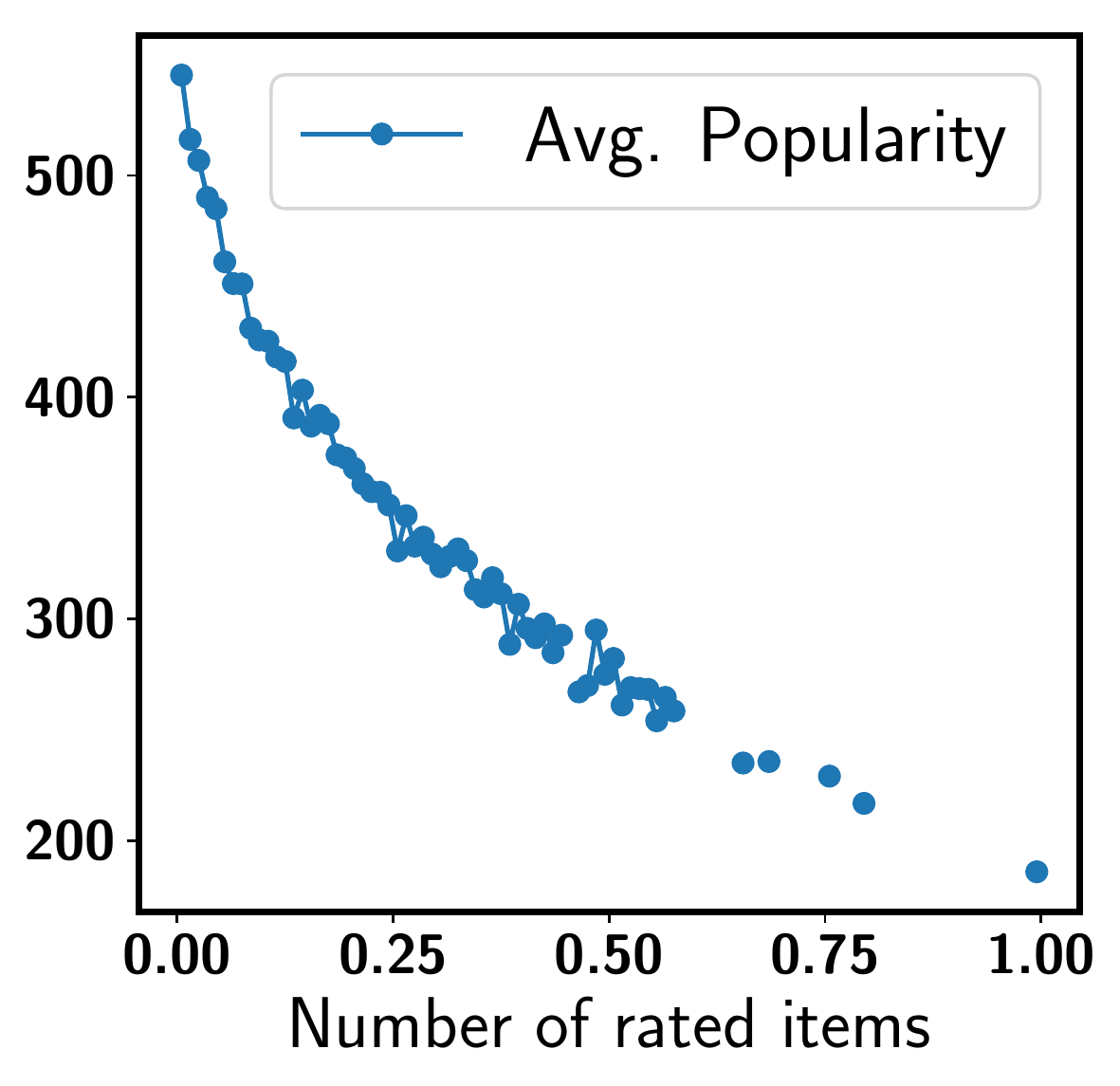}
    		\label{fig:ML1m-pop-vs-number-ratings}
        } 
        \subfloat[Netflix]
        {					 
		\includegraphics[scale=0.33]{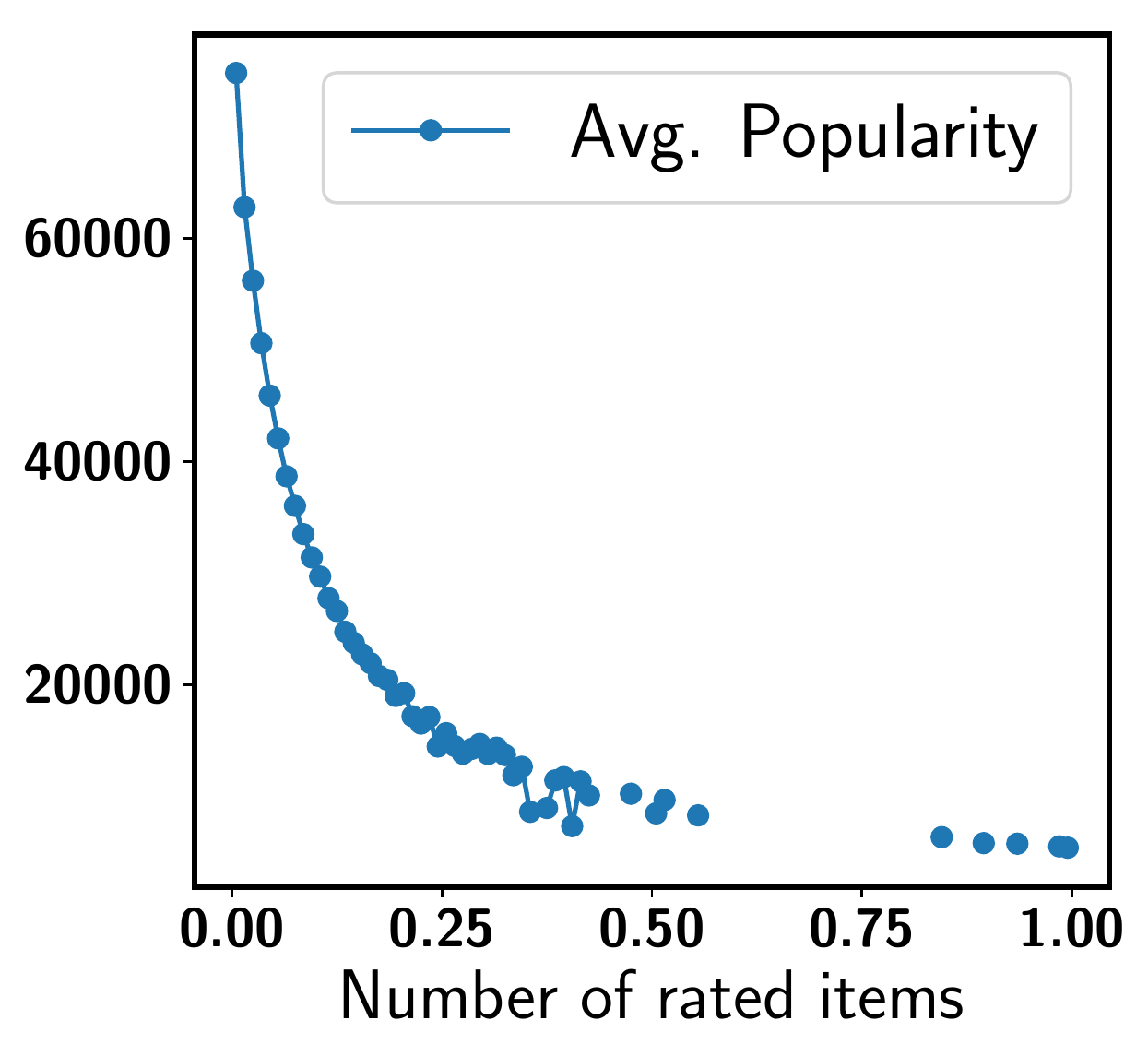}
   		\label{fig:MT-pop-vs-number-ratings}
   		}
\caption{For each user $u$, we consider  $\itemsofUserinTrainset$, the set of items rated by $u$ in train, and compute its average popularity $ \bar{a} =\frac{1 }{|\itemsofUserinTrainset|} \sum_{i \in \itemsofUserinTrainset} f_i^{\trainset}$. The x-axis shows the binned normalized $|\itemsofUserinTrainset|$, while the y-axis plots the mean of the corresponding $\bar{a}$ values. The average popularity of items rated decreases as the number of items rated increases. }
\label{fig:pop-vs-no-items}
\end{figure}
\fi
%\noindent \textbf{Data Model. }
%$\mathcal{U}=\{1, 2, \dots , U\}$ $\mathcal{I} = \{1,2,\dots,I\}$
%\footnote{RAP: Especially since this now comes right after the introduction, there realyl should be some brief outline/explanation as to what you're going to be doing in this section.}

\subsection{Notation and data model}
\label{sec:Notation}

Let $\mathcal{U}$ denote the set of all consumers or users and $\mathcal{I}$ the set of all items.  We reserve $u$, $s$  for indexing users, and $i$, $j$ for indexing items. Our dataset $\dataset$, is a set of ratings of  users on various items, i.e.,  $\dataset = \{r_{ui}: u \in \mathcal{U}, i \in \mathcal{I} \}$. Since every user rates only a small subset of items, $\dataset$ is a small subset of a complete rating matrix $\mathbf{R}$, i.e., $\dataset \subset \mathbf{R} \in \mathbb{R}^{\vert \mathcal{U} \vert \times \vert \mathcal{I} \vert}$. 
  \iffalse
 We split $\dataset$ into a train set $\trainset$ and test set $\testset$, with $\itemsinTrainset$ denoting items in train, and $\itemsinTestset$ denoting items in test. 
   Let  $\itemsofUserinTrainset =\{ i: r_{ui} \in \trainset \}$ denote the items  rated by  user $u$ in the train set,  and $\usersofIteminTrainset =\{ u: r_{ui} \in \trainset \}$  denote users that rated item $i$ in the train set, with similar definitions for $\itemsofUserinTestset$ and $\usersofIteminTestset$.
   For each user, we  generate a top-$\size$ sets by  ranking all items that do not appear in the train set of that user, i.e., $\itemsinTrainset \setminus \itemsofUserinTrainset$).  
   \fi
   
%We split $\dataset$ into a train set $\trainset$ and test set $\testset$. Let  $\itemsinTrainset$ denote items in train,  with $\itemsofUserinTrainset$  the items rated by a single  user $u$ in the train set. Similarly, $\itemsinTestset$ denotes the test  items and  $\itemsofUserinTestset$  denotes the test items of a single user $u$.  Let  $\usersofIteminTrainset$ denote users that rated item $i$ in the train set, and $\usersofIteminTestset$ users that rated the item in test.  For each user, we  generate a top-$\size$ set by  ranking all unseen train items ($\itemsinTrainset \setminus \itemsofUserinTrainset$).  

%alternative shorter
 We split $\dataset$ into a train set $\trainset$ and test set $\testset$. Let  $\itemsinTrainset$ ($\itemsinTestset$) denote items in the train (test) set, with $\itemsofUserinTrainset$ ($\itemsofUserinTestset$)  denoting the items rated by a single  user $u$ in the train (test) set.   Let  $\usersofIteminTrainset$  ($\usersofIteminTestset$) denote users that rated item $i$ in the train (test) set. For each user, we  generate a top-$\size$ set by  ranking all unseen train items, i.e.,~$\itemsinTrainset \setminus \itemsofUserinTrainset$.  

We denote the frequency of item $i$ in a given set $\mathcal{A}$ with $f_i^{\mathcal{A}}$. Following~\cite{adomavicius2012improving}, the popularity of an item $i$  is its frequency in the train set, i.e.,~$f_i^{\trainset} = |\usersofIteminTrainset|$. Based on the Pareto  principle~\cite{yin2012challenging}, or the $80/20$ rule,  we  determine long-tail items, $\LT$,  as those that generate the lower $20\%$ of the total ratings in the train set,  $ \LT \subset \itemsinTrainset$ (i.e,~items are sorted in decreasing popularity).
In our work, we use $x_i = \frac{ x_i - \min(\mathbf{x}) }{ \max(\mathbf{x}) - \min(\mathbf{x}) }
$ for normalizing a generic vector $\mathbf{x}$. 
% Long-tail items are denoted $\LT$, where $ \LT \subset \trainset$. 

\iffullpaper
Table~\ref{tab:notation} summarizes our notation. 
We  typeset the sets (e.g., $\mathcal{A}$), use upper case bold letters for matrices (e.g., $\mathbf{A}$),  lower-case bold letters for vectors (e.g., $\mathbf{a}$), and lower case letters for scalar variables (e.g., $a$). %Subscripts index specific elements of a matrix or vector or set.  
%\noindent \textbf{Preprocessing. }  

\begin{table}[t]
\centering
\small
\begin{tabular}{ll}
  \toprule

  {\bf Parameter} & {\bf Symbol}  \\ 
  \midrule
  Dataset & $\dataset$\\	
  Train dataset & $\trainset$\\
  Test dataset & $\testset$\\  
 
 % Set of long tail items in $\testset$ & $\testsetLT$ \\ 

  Set of users & $\mathcal{U}$ \\
  Set of items & $\mathcal{I}$ \\  
  Set of long tail items in $\trainset$ & $\LT$ \\ 
  
  Specific user & $u$ \\ 
  Specific item & $i$ \\
  Set of items of $u$ in $\trainset$ & $\itemsofUserinTrainset$ \\
  Set of items of $u$ in $\testset$ & $\itemsofUserinTestset$ \\
  Set of users of $i$ in $\trainset$ & $\usersofIteminTrainset$ \\
  Set of users of $i$ in $\testset$ & $\usersofIteminTestset$ \\ 	  
  
  Rating of user $u$ on item $i$ & $r_{ui}$ \\ 
  %Predicted rating of user $u$ on item $i$ & $\hat{r}_{ui}$ \\ 
  Size of top-$\size$ set & $\size$ \\
  
  Top-$\size$ set of $u$ & $\mathcal{P}_u$ \\
  Collection of top-$\size$ sets for all users & $\mathcal{P}$ \\
 
 Long-tail novelty preference of user $u$ acc. model $m$ & $\theta^{m}_u$ \\
  %Focusing degree of user $u$ & $\rho_u$ \\  
    
  Accuracy function  & $a(.)$ \\
 % Diversity function & $d(.)$ \\
  Coverage function  & $c(.)$ \\
  Value function of user $u$ & $v_{u}(.)$ \\
  %Original rating matrix & $\mathbf{R}$  \\
  %Item recommendation frequency vector & $\mathbf{f}$  \\  
  %Completed rating matrix & $\mathbf{\hat{R}}$ \\ 
   
\bottomrule
\end{tabular}
\caption{Notation.}
\label{tab:notation}
\end{table}

\fi

\iffalse
\begin{equation*}
\begin{aligned}
\small
& \mathcal{P}^* &\in & \underset{ \mathcal{P} = (\mathcal{P}_1 , ..., \mathcal{P}_{U})}{\argmax} &&  v(\mathcal{P})  \\ 
&   && s.t.& & \mathcal{P}_{u} \subseteq \mathcal{I},  |\mathcal{P}_{u}|=\size,    \mathcal{P}_{u} \cap \mathcal{I}_{u} = \emptyset, \forall u \in \mathcal{U} . 
\end{aligned}
\end{equation*}

Here, $v(.)$ quantifies the value of an allocation $\mathcal{P}$. 
\fi

\subsection{Simple long-tail novelty preference models}
\label{sec:simple-lt-pref}
 % Measures the an individual's desire to own exclusive goods.
Users have different preferences for discovering long-tail items.  Given the train set $\trainset$, we need  a measure of the user's willingness to explore less popular items.  Let  $\theta_u^m$ denote user $u$'s preference for long-tail novelty as measured by  model $m$.

Figure~\ref{fig:pop-vs-no-items} plots the average popularity of rated items vs.~the number of rated items (or \emph{user activity})  for 
\iffullpaper
our datasets. 
\else
ML-1M and Netflix. 
\fi
 As user activity increases, the average popularity of rated items decreases. This motivates an \textit{Activity} measure $\simpleRisk = \vert \itemsofUserinTrainset \vert$. But,  most users  only rate  a few items, and $\simpleRisk$ does not indicate whether those items were long-tail  or popular.  %it is possible that two users with equal activity focus on different types of items: one rates popular items, while the other rates long-tail items. Furthermore, most users rate only a few items (small $\simpleRisk$). % and $\simpleRisk$ does not indicate whether the user rated popular items, or long-tail ones.% (confirmed empirically in Figure~\ref{fig:Hist-LT-PreferenceModels}).

%Naturally, we expect consumers who rate more items to be more willing to explore unpopular items. We can define a \textit{Simple} measure using $\simpleRisk = \vert \itemsofUserinTrainset \vert$. However, most users rate only a few items and this measure does not distinguish between type of items they rated. 

Instead, we can define a \textit{Normalized Long-tail} measure
\begin{align}
\small\LTRisk = \frac{  \vert \itemsofUserinTrainset  \cap \LT \vert}{\vert \itemsofUserinTrainset \vert}
\label{eq:nlt-risk}
\end{align} 
 % $\LTRisk = \frac{  \vert \itemsofUserinTrainset  \cap \LT \vert}{\vert \itemsofUserinTrainset \vert}$.   
which is the fraction of long-tail items in the user's  rated items.  The higher this fraction, the higher her  preference for long-tail items. However, $\LTRisk$ does not capture the user's interest (e.g,~rating) in the items, and does not distinguish between the various long-tail items.

%In our case, documents correspond to users and the words in each document correspond to the items. The frequency of a word in a document ($tf$) is the rating given to an item by a user ($r_{u,i}$). The $idf$ term is $\frac{n}{f_{i}}$,

To resolve both problems, we can use similar notions as in TFIDF~\cite{salton1988term}.  The rating an item receives from a particular user reflects its importance for that user. To capture the discriminative power of the item among the set of users and control for the fact that some items are generally more popular, we also incorporate an inverse popularity factor  ($|\usersofIteminTrainset|$) that is logarithmically scaled. We define \textit{TFIDF-based} measure using 
\begin{align}
\small
\tfidfRisk = \frac{ 1}{\vert \itemsofUserinTrainset \vert} \sum_{i \in \itemsofUserinTrainset} r_{ui} \log \left( \frac{|\mathcal{U}|}{ |\usersofIteminTrainset|} \right)
\label{eq:tfidf-risk}
\end{align}
This measure increases proportionally to the user rating $r_{ui}$, but is also counterbalanced by the popularity of the item. A higher  $\tfidfRisk$ shows more preference for less popular items. 
Although $\tfidfRisk$ considers both  the user interest $r_{ui}$, and the  item popularity $|\usersofIteminTrainset|$, it has no indication about the preferences of the users in $\usersofIteminTrainset$. 
%The problem is that, two  items $i_1$ and $i_2$, with equal popularity $|\mathcal{U}_{i_1}^{\trainset}| = |\mathcal{U}_{i_2}^{\trainset}|$, can have user sets with  entirely different long-tail preferences.  
To address this limitation, observe Eq.~\ref{eq:tfidf-risk} can be re-written  as
\begin{align}
\small
\tfidfRisk = \frac{ 1}{\vert \itemsofUserinTrainset \vert} \sum_{i \in \itemsofUserinTrainset} \theta_{ui}  = \frac{ \sum_{i \in \mathcal{I}_u} w_i \theta_{ui}}{ \sum_{i \in \mathcal{I}_u} w_i }
\label{eq:tfidf-risk-rewrite}
\end{align}
where $\theta_{ui} = r_{ui} \log \left( \frac{|\mathcal{U}|}{|\usersofIteminTrainset|} \right)$ is a \textit{per-user-item} long-tail preference value, and  $w_{i} = 1$ for all items.  Basically,  $\tfidfRisk$ gives equal importance to all items and is a crude average of $\theta_{ui}$.

%and .

%\input{RunningExample}

We can generalize the idea in Eq.~\ref{eq:tfidf-risk-rewrite}. Specifically, for each user, we consider a  \textit{generalized long-tail novelty} preference estimate, $\theta_u^G$.  We assume $\theta_u^G$ is a  weighted average of $\theta_{ui}$.  However, rather than imposing equal weights, we define  $w_i$ to  indicate an \textit{item importance} weight. Our second assumption is that an item is important when its users do not regard it as a mediocre  item; when their preference for that item deviates  from their generalized preference.  In other words,  an item $i$ is important when $\sum_{u \in \usersofIteminTrainset} (\theta_{ui} - \theta_u^G)^2$ is large.   Since $w_i$ and $\theta^G_u$ influence each other, below we  describe how to learn these variables in a joint optimization objective.

%Before that, we note two important observations: 1.~Our approach is inspired by truth discovery approaches~\cite{},  where the goal is infer trustworthy information. In that setting, sources provide claims (similar to $\theta_{ui}$), each claim has a true value, and sources are assinged reliability scores. The main principle however, is that sources that provide true information more often will be assigned higher reliabiity degrees, and the information provided by reliable sources is regarded as truths.   idea of allowing weights $w_i$, per-user-item preferences $\theta_{ui}$, and a generalized preference for each user $\theta_u^{*}$ is  

%It does not consider  the long-tail preferences of other of users $i$. i.e.,~$\usersofIteminTrainset$.
%\footnote{ We introduce  the item \textit{long-tail importance weight} $w_i$, to distinguish between items based on the long-tail peference of their user groups. Moreover,  we  need to learn a \textit{generalized} long-tail preference estimate,  $\theta_u^*$, for each user, that is capable of differentiating between users based on the types of items they rate. The estimates  $w_i$ and $\theta^G_u$ are dependent and should be learned in a joint optimization objective. }

%In order to distinguish between items based on the long tail preference of their users, we introduce the notion of  item long-tail importance, $w_i$.

\subsection{Learning generalized long-tail novelty preference}
\label{sec:learning-lt-pref}
\iffalse
\begin{align}
\small
\epsilon = \sum_{i \in  \itemsinTrainset}  \sum_{u \in \usersofIteminTrainset}   w_i \epsilon_{ui}^2  =  \sum_{i \in  \itemsinTrainset} \sum_{u \in \usersofIteminTrainset} w_i \big( \theta_{ui} -  \theta_{u}^G \big)^2
\label{eq:totalSquaredError}
\end{align}
\fi
We define  $\epsilon_{i} = \big[ \sum_{u \in \usersofIteminTrainset}  1- \big( \theta_{ui} -\theta_u^G\big)^2 \big] $ as  the \textit{item mediocrity coefficient}. Assuming $|\theta_{ui} - \theta_u^G| \leq 1$ (explained later), the maximum  of $\epsilon_i$ is obtained  when $\theta_{ui} = \theta_u^G$.  We formulate our objective as:
\begin{align}
\ O(\mathbf{w},\bm{\theta}^G)= \sum_{i \in  \itemsinTrainset } w_i \big[ \sum_{u \in \usersofIteminTrainset}  1- \big( \theta_{ui} -\theta_u^G\big)^2 \big] =  \sum_{i \in  \itemsinTrainset } w_i \epsilon_i \nonumber
\end{align}
which is the total weighted mediocrity. Here, $ \itemsinTrainset$ are the items in train, $\usersofIteminTrainset$  denotes users that rated item $i$ in the train set (Section~\ref{sec:Notation}), and $\theta_{ui}$  is the per-user-item preference value,  computed from rating data.   Our objective  has two unknown variables: $\bm{\theta}^G \in \mathbb{R}^{|\mathcal{U}|}$, $\mathbf{w} \in \mathbb{R}^{|\mathcal{I}|}$.
We use an alternating optimization approach for optimizing   $\mathbf{w}$ and  $\bm{\theta}^G$. When optimizing w.r.t. $\mathbf{w}$,  we must minimize the objective function in accordance with our intuition about $w_i$. In particular, for larger mediocrity coefficient, we need smaller weights. On the other hand, when optimizing w.r.t. $\bm{\theta}^G$, we need to increase the closeness between $\theta_u^G$ and all $\theta_{ui}$'s, which is aligned with our intuition about $\theta_u^G$. So, we have to maximize the objective function w.r.t.  $\bm{\theta}^G$.  Our final objective is a minimax problem:  
\begin{align}
&\min_{\mathbf{w}} \max_{\bm{\theta}^G} &  O(\mathbf{w},\bm{\theta}^G)  - \lambda_1 \sum_{i \in \itemsinTrainset } \log{w_i}   
\label{eq:thetaStarOpt}
\end{align}
\iffalse
\begin{align}
\min_{\mathbf{w}} \max_{\bm{\theta}^G}\ O(\mathbf{w},\bm{\theta}^G)= \sum_{i \in  \itemsinTrainset } w_i \big[ \sum_{u \in \usersofIteminTrainset}  - \big( \theta_{ui} -\theta_u^G\big)^2 \big] 
\label{eq:thetaStarOpt}
\end{align}
\fi
where we have added a regularizor term $\sum_{i \in \itemsinTrainset } \log{ w_i}$ to  prevent $w_i$ from approaching 0~\cite{li2015discovery}. % To solve  Eq.~\ref{eq:thetaStarOpt} we use alternating optimization methods~\cite{koren2009matrix}, which iterate between fixing  one set of values and solving for the other until convergence. 

When $\bm{\theta}^G$ is fixed, we need to solve a minimization problem involving only $\mathbf{w}$. By taking the derivative w.r.t. $w_i$ we have  
\begin{align}
w_i = \frac{\lambda_1}{ \sum_{u \in \usersofIteminTrainset} 1 - \big(\theta_{ui} -\theta_u^G\big)^2  } = \frac{\lambda_1}{\epsilon_i}
\label{eq:weightUpdate}
\end{align}
When $\mathbf{w}$ is fixed, we need to solve a maximization problem involving $\bm{\theta}^G$. Taking the derivative w.r.t. $\theta_u^G$ we derive
\begin{align}
\theta_u^G =\frac{\sum_{i \in \itemsofUserinTrainset} w_i \theta_{ui}}{\sum_{i \in \itemsofUserinTrainset} w_i} 
\label{eq:overallRiskUpdate}
\end{align}

Essentially, Eq.~\ref{eq:weightUpdate} characterizes an item's weight,  based on the item mediocrity. The higher the mediocrity, the lower the weight.  Moreover, for every user $u$, $\theta_u^G$ is a weighted average of all $\theta_{ui}$. Note, in Eq.~\ref{eq:overallRiskUpdate}, $\theta_u^G=\tfidfRisk$ when $w_i =1$ for all items.  Our generalized $\theta_u^G$ incorporates both the user interest and popularity of items (via $\theta_{ui}$), and  the preferences of other users of the item (via $w_i$). Furthermore, since we need $|\theta_{ui} - \theta_u^G| \leq 1$, and prefer  $\theta_u^G \in [0,1]$, we project all $\theta_{ui}$ to the $  [0,1]$ interval before applying the algorithm. We also set $\lambda_1=1$.

\section{GANC: Generic Re-ranking Framework}
\label{sec:RiskbasedReranking} 
We define user satisfaction  based on the accuracy of the top-$\size$ set and its coverage of the item space, so as to introduce novelty and serendipity into recommendation sets.   %on the relevance of items in the top-$\size$ set and the promotion of long-tail items. 
We consider three components for our framework: \begin{enumerate*}
\item an accuracy recommender (ARec) that is responsible for suggesting accurate top-$\size$ sets. 
\item a coverage recommender (CRec)  that is responsible for  suggesting top-$\size$ sets that maximize coverage across the item space, and consequently promote long-tail items. 
\item the user preference for long-tail novelty  $\theta_u \in [0,1]$. 
\end{enumerate*} 
We use the template $\texttt{GANC} (\texttt{ARec}, \bm{\theta}, \texttt{CRec)}$ to specify the used components. 

We define individual user value functions  for a top-$\size $ set $\mathcal{P}_u$  as%The  value of a set of items $\mathcal{P}_u$ for user $u$, is quantified using a personalized value function %take the cooperative mixture of experts view, and
\begin{equation} 
\small
v_u(\mathcal{P}_u) =  (1-\theta_u) a(\mathcal{P}_{u})  + \theta_{u} c(\mathcal{P}_{u})  \label{eq:vurelcov}
\end{equation}
% $v_{u}, a, c :\mathcal{I}^{|\mathcal{A}|} \rightarrow [0,1]$ and $\theta_u \in [0,1]$ is a user-specific trade-off value.
where  $a(.)$  measures the score of a set of items according to the accuracy recommender, and $c(.)$  measures the score of a set according to the coverage recommender. With slight abuse of notation, let $a(i)$ and  $c(i)$ denote the accuracy score and  coverage score  of a single item $i$. We ensure $a(i), c(i) \in [0,1]$ to have the same scale. Furthermore, we define  $a(\mathcal{P}_u) = \sum_{i \in \mathcal{P}_u} a(i) $ and $c(\mathcal{P}_u) = \sum_{i \in \mathcal{P}_u} c(i) $.

The user  value function in Eq.~\ref{eq:vurelcov} positively rewards  sets that increase coverage.  Similar intuitions for encouraging solutions with desirable properties,~e.g., diverse solutions, have been used in related work~\cite{ agrawal2009diversifying}. %,qin2013promoting}.  %lin2011class,gollapudi2009axiomatic,
  However, their trade-off parameters  are typically obtained via parameter tuning or cross validation. By contrast, we impose personalization via the user preference estimate, $\theta_u$, that  is learned based on historical rating data. Next, we list the various base recommender models integrated into GANC.
 
\iffalse
Competitive Mixture of Experts :The accuracy recommender and the coverage recommender can be viewed as two experts, and  any of our risk measures can be used as a probabilistic switching indicator. Specifically,  the top-$\size$ set for a user is determined as follows:
\begin{enumerate}
%\item Generate random sample $b \sim \text{Bern}(\theta_u)$
\item With probability $1-\theta_u$ we choose the set suggested by the accuracy recommender, i.e., $v_u(\mathcal{P}_u) =   a(\mathcal{P}_{u}) $. 
\item With probability $\theta_u$ we suggest the top-$\size$ set suggested by the coverage recommender, i.e., $ v_u(\mathcal{P}_u) =   c(\mathcal{P}_{u})$.  
\end{enumerate} 
This model resolves scale issues between the two objective functions. 
\fi
 
\subsection{Accuracy recommender}
\label{sec:accuracyRecommender}
The accuracy recommender provides an accuracy score, $a(i)$, for each  item $i$.  We experiment with three models (Section~\ref{sec:ExpSetup} provides details and setup configurations): 
\begin{itemize}
\item \textbf{Most popular (Pop)} ~\cite{cremonesi2010performance}  is non-personalized and recommends the most popular unseen items. It makes accurate recommendations, but has low coverage and novelty~\cite{cremonesi2010performance}. Since it does  not score items, we define $a(i)=1$ if item $i$ is in the top-$\size$ set suggested by Pop, otherwise $a(i)=0$.

\item \textbf{Regularized SVD (RSVD)}~\cite{zhuang2013fast} learns latent factors for users and items by analyzing  user-item interaction data (e.g.,~ratings). The  factors  are then used to predict the values of unobserved ratings. We use RSVD to compute a predicted rating matrix $\mathbf{\hat{R}} \in \mathbb{R}^{\vert \mathcal{U} \vert \times \vert \mathcal{I} \vert}$. We normalize the predicted rating vectors  of all users  to ensure  $\hat{r}_{ui} \in [0,1]$, and define $a(i) =\hat{r}_{ui}$. %These scores correspond to relevance scores. %(Unless otherwise stated, $\hat{r}$ denotes normalized values.)  % $a(\mathcal{P}_u ; \mathbf{\hat{R}} ) =\frac{1}{\size} \sum_{i \in \mathcal{P}_u}{\hat{r}_{ui}}$. Note,  the scores obtained from MF correspond to relevance scores, 

\item  \textbf{PureSVD (PSVD)}~\cite{cremonesi2010performance} is also a latent factor model.  We follow the same procedure  as RSVD, using PSVD factors~\cite{cremonesi2010performance}. Note, PSVD scores  correspond to associations between users and items.
\end{itemize}

\subsection{Coverage recommender}
The coverage recommender provides  a coverage score, $c(i)$, for each  item $i$. We use  three coverage recommenders: 
\begin{itemize}

\item \textbf{Random (Rand)}  recommends  $\size$ unseen items randomly. It has high coverage, but  low accuracy~\cite{vargas2014improving}. We define $c(i) \sim $unif$(0,1)$. %Performance-wise it is similar to Bal. (Dynamic coverage), with better  gini performance.

%\subsubsection{Divide} This model applies the same strategy as the accuracy recommender, $c(i)= a(i)$, if  item $i$ is a long-tail item in the train set, otherwise $c(i)=0$. 

\iffalse
\subsubsection{Binary Complement  SVD} Here, we represent the user-item rating matrix using binary values, where 1 indicates the user rated the item and 0 indicates the item has not received any rating. We then apply SVD on the matrix. The resulting $\mathbf{\hat{R}}$ is estimated as in PureSVD. However to derive a model that  promote long-tail items,  we  need to flip the result $\hat{r}^{'} = 1 - \hat{r}$.
\fi

\item \textbf{Static (Stat)} focuses exclusively on promoting less popular items. We define $c(i)$ to be a monotone decreasing function of $f_i^{ \trainset}$,  the popularity  of $i$ in the train set $\trainset$. We use $c(i) =  \frac{1}{ \sqrt{f_i^{\trainset}+1 }}$ in our work.  %,i.e., $f(i) = |\usersofIteminTrainset|$. 
Note the gain of recommending an item is constant. 

%$\mathcal{A}$ a partial allocation where a subset of the users have been assigned top-$\size$ sets
%$c(\mathcal{P}_u,\mathcal{A}) = \frac{1}{\size} \sum_{i \in \mathcal{P}_u}\frac{1}{f(i,\mathcal{A}) + 1}\label{eq:coverage}$

\item \textbf{Dynamic (Dyn)}  allows us to better correct for the popularity bias in recommendations. In particular,  rather than the train set $\trainset$,  we  define $c(i)$ based on the set of recommendations  made so far. Let $\mathcal{P}  = \{ \mathcal{P}_u \}_{u=1}^ {|\mathcal{U}|} $ with $|\mathcal{P}_u| = \size$, denote the collection of top-$\size$ sets assigned to all users, and $\mathcal{A}  = \{ \mathcal{A}_u \}_{u=1}^ {|\mathcal{U}|} $  with $\mathcal{A}_u \subseteq \mathcal{P}_u$, denote  a partial collection where  a subset of users have been assigned  some items. We measure the long-tail appeal  of an item  using a monotonically decreasing function of the popularity of $i$ in $\mathcal{A}$, i.e.,  $f_i^{\mathcal{A}}$. We use 
$c(i) = \frac{1}{\sqrt{ f_i^{\mathcal{A}} + 1} }\label{eq:coverage}$  in our work.  The main intuition is that recommending an item  has a diminishing returns property:  the more the item is recommended, the less the gain of the item in coverage, i.e., $c(i)=1$ when $\mathcal{A} = \emptyset$, but decreases as the item is more frequently suggested.%, $c(i)$ decreases.  % Next, we discuss how to integrate this recommender into our framework. 

\end{itemize}
%This recommender (Bal.) encourages the balanced promotion of all items.  It  provides the upper bound  performance achievable in terms of gini and coverage.

\subsection{Optimization Algorithm for GANC }
\label{sec:optimizationAlgorithm}
The overall goal of the framework is to find an optimal top-$\size$ collection $\mathcal{P}  = \{ \mathcal{P}_u \}_{u=1}^ {|\mathcal{U}|} $  that  maximizes  the  aggregate of the user value functions:
\begin{align}
\small
\max_{\mathcal{P} } \ v(\mathcal{P}) = \sum_{u \in \mathcal{U}} v_{u}(\mathcal{P}_{u})
\label{eq:overallValueFunction}
\end{align}
%where $v_u(.)$ is the value function of user $u$, $\mathcal{P}_u$ is the top-$\size$ set for user $u$, and $\mathcal{P}$ is the collection of all top-$\size$ sets, i.e.,  $\mathcal{P}  = \{ \mathcal{P}_u \}_{u=1}^ {|\mathcal{U}|} $. 

The combination of Rand and Stat with the accuracy recommenders in Section~\ref{sec:accuracyRecommender}, result in value functions that can be optimized greedily and independently, for each user. Using Dyn, however,  creates a dependency between the optimization of user value functions, where the items suggested to one user depend on those suggested to previous users.  Therefore,  user value functions can no longer be optimized independently. 
\begin{algorithm}[t]
\KwIn{ $\size, S, \mathcal{U} , \mathcal{I}, \mathcal{R}$ }
\KwOut{$\mathcal{P} $}
$\mathbf{f} \gets  0 , \mathcal{P} \gets \emptyset, \boldsymbol{\theta} \gets $ Estimate from $\mathcal{R}$ (Section~\ref{sec:lt-pref})\;\label{algo:algSeqSample:line:initilize}
$\mathcal{S}  \gets$ Sample $S$ users from $\mathcal{U}$ acc. to KDE($\boldsymbol{\theta}$)\;\label{algo:algSeqSample:line:sample}
Sort $\mathcal{S}$ in increasing $\boldsymbol{\theta}$\;\label{algo:algSeqSample:line:sortUsers}
\tcp{Performed sequentially for users in sample}
\ForEach{ $u  \in \mathcal{S} $}{
		Update Dyn function parameter $\mathbf{f}$\; \label{algo:algSeqSample:line:updateCoverageFunction1}
		$\mathcal{P}_{u}$ = $\mathcal{P}_{u}$ $\cup$ $\argmax_{i \in \mathcal{I}}$ $v_{u}(\mathcal{P}_{u} \cup i)$ - $v_{u}(\mathcal{P}_{u})$\; \label{algo:algSeqSample:line:top-NAssign1}
		
		\lForEach{ $i \in \mathcal{P}_{u}$}{$\mathbf{f}_{i} = \mathbf{f}_{i} + 1$}\label{algo:algSeqSample:line:updateStep}	
		F($\theta_u$) $\gets \mathbf{f} $\;\label{algo:algSeqSample:line:storeshit}
		$\mathcal{P} = \mathcal{P} \cup \mathcal{P}_{u} $\;		
}
\tcp{Performed in parallel for users not in sample}
\ForEach{$ u \in  \mathcal{U} \setminus \mathcal{S}$  }{
	 $ \mathbf{\hat{f}} \gets F(\argmin(\theta_s - \theta_u))$ for $s \in \mathcal{S} $\; \label{algo:algSeqSample:line:getFrequency}
		Update Dyn function parameter $\mathbf{\hat{f}}$\;\label{algo:algSeqSample:line:updateCoverageFunction2}
	 $\mathcal{P}_{u}$ = $\mathcal{P}_{u}$ $\cup$ $\argmax_{i \in \mathcal{I}}$ $v_{u}(\mathcal{P}_{u} \cup i)$ - $v_{u}$	
	\label{algo:algSeqSample:line:top-NAssign2} 
	$\mathcal{P} = \mathcal{P} \cup \mathcal{P}_{u} $\;		
	\label{algo:algSeqSample:line:updateP} 
}
\Return $\mathcal{P}$
\caption{GANC with OSLG optimization}
\label{algo:algSeqSample}
\end{algorithm}

\vspace{4mm}
\noindent \textbf{Optimization algorithm for GANC with Dyn. }
\label{sec:optimizingFrameworkWithDynamic}
Because Dyn is monotonically decreasing in $f_i^{\mathcal{A}}$, when used in Eq.\ref{eq:vurelcov}, the overall objective in Eq.~\ref{eq:overallValueFunction} becomes submodular across users. Maximizing a submodular function is NP-hard~\cite{khuller1999budgeted}.  %Maximizing a submodular function is NP-hard~\cite{khuller1999budgeted}.
However, a key observation is that  the constraint of recommending $\size$ items to each user, corresponds to a partition matroid over the users.
 Finding  a top-$\size$ collection $\mathcal{P}$ that maximizes Eq.~\ref{eq:overallValueFunction} is therefore an instance of  maximizing a submodular function subject to a matroid constraint \iffullpaper
(see Appendix~\ref{sec:submodularmonotoneproof}).
\else
(see~\cite{ourFullVersion} for details).
\fi
A \textit{Locally Greedy} heuristic, due to Fisher et al.~\cite{fisher1978analysis}, can  be applied:  consider the users separately and in arbitrary order. At each step, select a user $u$ arbitrarily, and greedily construct $\mathcal{P}_u$  for that user. Proceed until all users have been assigned top-$\size$ sets. Locally Greedy produces a solution at least half the optimal value for maximizing a submodular monotone function subject to a matroid constraint~\cite{fisher1978analysis}.

\iffalse
\begin{theorem}
\label{thm:greedylocallygreedy}
The locally greedy heuristic is a $1/2$-approximation algorithm for maximizing Eq.~\ref{eq:overallValueFunction} when $v$ is submodular monotone increasing~\cite{fisher1978analysis}.
\end{theorem}
\fi

However, locally greedy is sequential and has a computational complexity of $O(\vert \mathcal{U} \vert . \vert \mathcal{I}\vert . \size)$ which is not scalable.  Instead, we introduce a heuristic we call \textit{Ordered Sampling-based Locally Greedy} (OSLG).  Essentially, we make two modifications based on the user long-tail preferences:  first, proportionate to the distribution of user long-tail preferences $\bm{\theta}$, we sample a subset of users, and run the sequential algorithm on this sample only.   Second, to allow the system to recommend  more established or popular products to users with lower long-tail preference, instead of arbitrary order,  we  modify  locally greedy to consider users in increasing long-tail preference.        
\iffalse
to improve the scalability of the algorithm, we only run the sequential part  on a representative sample  of the users $\mathcal{S} \subset \mathcal{U}$, and propagate the information for users not included in the sample. 
first,  locally greedy, operates on users  sequentially and in arbitrary order.   We  modify locally greedy to consider users in increasing risk degree. 
\fi

Algorithm~\ref{algo:algSeqSample} shows GANC with OSLG optimization:  We use $\mathbf{f}_i$ as a shorthand for $f_i^{\mathcal{A}}$, the  popularity of item $i$ in the current set of recommendations, used in Dyn. First, $\mathbf{f}$ is initialized and user preferences $\bm{\theta}$ are estimated (line~\ref{algo:algSeqSample:line:initilize}). Next, we  use  Kernel density estimation (KDE)~\cite{sheather1991reliable} to approximate the Probability density function (PDF) of $\bm{\theta}$, and use the PDF to draw a sample $\mathcal{S}$ of size $S$ from $\bm{\theta}$ and find the corresponding users in $\mathcal{U}$ (line~\ref{algo:algSeqSample:line:sample}). The sampled users are then sorted in increasing long-tail preference (line~\ref{algo:algSeqSample:line:sortUsers}), and the algorithm iterates over the users. In each iteration, it updates the Dyn function (line~\ref{algo:algSeqSample:line:updateCoverageFunction1}) and assigns a top-$\size$ set  to the current user by maximizing her value function (line~\ref{algo:algSeqSample:line:top-NAssign1}). The Dyn function parameter  $\mathbf{f}$ is then updated w.r.t. the recently assigned top-$\size$ set (line~\ref{algo:algSeqSample:line:updateStep}). Moreover, $\mathbf{f}$ is associated with the current long-tail preference estimate $\theta_u$, and stored (line~\ref{algo:algSeqSample:line:storeshit}). The algorithm then proceeds to the next user. Since Dyn  is monotonically decreasing in $\mathbf{f}_i$, frequently recommended items are weighted down by the value function of subsequent users. Consequently, as we reach users who prefer long-tail items and discovery, their value functions prefer relatively unpopular items that have not been recommended before. Thus, the induced user ordering, results in the promotion of long-tail items to the right group of users, such that we obtain better coverage while maintaining user satisfaction.

 For each user not included in the sample set, $u \not\in \mathcal{S}$, we find the most similar user $s \in \mathcal{S}$, where similarity  is defined as $|\theta_s-\theta_u|$ (line~\ref{algo:algSeqSample:line:getFrequency}), and use $F(\theta_s)$ to compute the coverage score (line~\ref{algo:algSeqSample:line:updateCoverageFunction2}), and assign a top-$\size$ set. Observe,  the value functions of $u \in \mathcal{U} \setminus \mathcal{S}$, are independent of each other, and  lines	\ref{algo:algSeqSample:line:getFrequency}-\ref{algo:algSeqSample:line:updateP}  can be performed in parallel. The computational complexity of the sequential part drops to $O(\vert \mathcal{S} \vert . \vert \mathcal{I}\vert . \size)$ at the  cost of  $O(\vert \mathcal{S} \vert . \vert \mathcal{I}\vert )$ extra memory.

%\vspace{4mm}
% For Dyn, we report  sample size e.g.,~GACN(Pop, Dyn500, $\theta^{G}$) uses 500 samples. 

\iffalse
\begin{align*}
\texttt{Pref. Model}_{\texttt{Acc. Rec.}}^{\texttt{Cov. Rec.}}
\end{align*}
\fi
\section{Empirical Evaluation}
\label{sec:Experiments}
\subsection{Experimental setup}
\label{sec:ExpSetup}
%\subsubsection{Datasets} %~\footnote{\url{https://github.com/sidooms/MovieTweetings/}}~\cite{dooms2013movietweetings}
%\vspace{4mm}
\noindent \textbf{Datasets and data split. }Table~\ref{tab:DatasetStatsticsSmall} describes our datasets. 
We use MovieLens 100K (ML-100K), 1 Million (Ml-1M), 10 Million (ML-10M) ratings~\cite{harper2016movielens}, Netflix, and 
MovieTweetings 200K (MT-200K)~\cite{dooms2013movietweetings}.  In the MovieLens datasets, every consumer has rated at least 20 movies ($\tau = 20$), with  $r_{ui} \in \{ 1, \dots, 5\} $ (ML-10M has half-star increments). % for ML-10m $r_{ui} \in \{ 1,1.5,2, \dots, 5\} $ 
MT-200K  contains voluntary movie ratings posted on twitter, with $r_{ui} \in \{ 0, \dots, 10\} $.  Following~\cite{hernandez2014probabilistic}, we preprocessed this dataset to map the ratings to  the interval $[1,5]$. Due to the extreme sparsity of this dataset and to ensure every user has some data to learn from, we   filtered the users to  keep those with at least 5 ratings ($\tau = 5$).

Our  selected datasets  have varying density levels. Additionally,  MT-200K and Netflix  include a large number of  difficult infrequent users, i.e., in MT-200K,  47.42\% (3.37\% in Netflix) of the users have rated fewer than 10 items, with the minimum being 4. 
We chose these datasets to study performance in settings  where  users  provide few feedback~\cite{kanagal2012supercharging,liu2017experimental}.

Next,  we randomly split each dataset into train and test sets by keeping a fixed ratio $\kappa$ of each user's ratings in the train set and moving the rest to the test set~\cite{lee2014local}.  
\iffullpaper 
 This way, when $\kappa =0.8$, an infrequent user with 5 ratings will have 4 train and 1 test rating,  while  a user with  100 ratings, will have 80 train ratings and the rest in test. 
 \fi
 For ML-1M and ML-10M, we set $\kappa=0.5$.  For MT-200K, we set $\kappa=0.8$.  For Netflix, we use their probe set as our test set, and remove the  corresponding ratings from train. We remove users in the probe set who do not appear in train set, and vice versa. 

\begin{table}[t]
\centering
\scriptsize
\begin{tabular}{llllllll}
  \toprule
  {\bf Dataset} & {\bf $|\mathcal{D}|$} & {\bf $|\mathcal{U}|$} & {\bf $|\mathcal{I}|$}  & d$\%$ & $\mathcal{L} \%$ & {\bf $\kappa$} & {\bf $\tau$ } \\ \midrule
  ML-100K  & 100K & 943& 1682  & 6.30 &  66.98   &0.5 & 20 \\
  
  ML-1M   & 1M & 6,040& 3,706 & 4.47 & 67.58  &  0.5 & 20  \\
  
  ML-10M & 10M & 69,878 & 10,677 & 1.34& 84.31  & 0.5 &  20 \\
  
  %MT-200k-m5  & 172,506 & 7,969 & 13,864 & 0.16\%& 86.84 \%   & 0.8 & 5 \\
  % MT-200k-m5-mapped
  MT-200k  & 172,506 & 7,969 & 13,864 & 0.16 & 86.84    & 0.8 & 5  \\ 
  
  %MT-200k-m20 & 96,788 & 1,564 &  11,729 & 0.53\% &  77.61\% & 0.5 & 30  \\
  
  %MT-200k-m20-mapped & 96,788 & 1,564 &  11,729 & 0.53\% &  77.61\% & 0.5 & 30 \\
  
  Netflix & 98,754,394 &  459,497 & 17,770  &1.21 & 88.27  & -& - \\
   \bottomrule
\end{tabular}
\caption{Datasets description.  $|\mathcal{D}|$ is number of ratings in dataset.  Density is  $ \text{d}\%=  |\mathcal{D}| / ( |\mathcal{U}| * |\mathcal{I}|) \times 100\%$.   Long-tail percentage is  $\mathcal{L}\% = (|\LT| / |\itemsinTrainset|) \times 100\%$. Train-test split ratio per user is $\kappa$,  $\tau$ is the minimum number of ratings per user. }
\label{tab:DatasetStatsticsSmall}
\end{table}

\begin{table}[t]
\centering
\small
\begin{tabular}{ll}
  \toprule
  %{\bf Measure} & Formula \\ \midrule
\multirow{5}{*}{\parbox{0.07\textwidth}{Local \\ Ranking \\ Accuracy \\ Metrics}}   
   %& $\predictionFormula$ \\ [0.2cm]
   &  $\precisionFormula$   \\[0.2cm]
   & $\recallFormula$  \\[0.2cm]
   & $\FmeasureFormula$ \\[0.2cm]
   %& \multirow{2} {*} {\parbox{0.3\textwidth} {$\DCGFormula$ \\ [0.1cm] $\nDCGFormula$}} \\[0.1cm]
   %& \\[0.1cm]
   \midrule  
   \multirow{4}{*}{\parbox{0.07\textwidth}{Longtail  \\ Promotion}} 
	%&$\LTPrecisionFormula$   \\[0.1cm]   
   & $\LTAccuracyFormula$ \\[0.2cm]
   & $\StratRecallFormula$ \\[0.2cm] % , $\StratWeightFormula$
	%&\\[0.1cm]   
   \midrule
  \multirow{3}{*}{\parbox{0.07\textwidth}{Coverage \\ Metrics}} 
  & $\coverageFormula$   \\[0.2cm]
  & $\giniFormula$ \\[0.2cm]
  %&$\LTCoverageFormula$ \\[0.1cm]  
  \iffalse
   \midrule 
   \multirow{3}{*}{\parbox{0.07\textwidth}{Weighted \\Harmonic \\ Mean}} 
   \\
    &$\WeightedHMFormula$ \\[0.1cm] 
    \\
   \fi 
  \bottomrule
  
\end{tabular}
\caption{Performance Metrics. %For accuracy,  we use local ranking metrics~\cite{agarwal_chen_2016}, where  each metric is computed per user and then averaged across all users.  
Notation is in Section~\ref{sec:Notation}. For gini, the vector 
$\mathbf{f}$ is  sorted in  non-decreasing order of recommendation frequency of items,~i.e., $f[j] \leq f[j+1]$.
}
\label{tab:perfMetrics}
\end{table}
\vspace{4mm}
%\label{sec:PerformanceMetrics}
%We focus on accuracy, coverage and novelty aspects of top-$\size$ recommendation. 

\noindent \textbf{Test ranking protocol and performance metrics. }For testing, we adopt the ``All unrated items test ranking protocol''~\cite{steck2013evaluation,vargas2014improving} where for each user, we generate the top-$\size$ set by  ranking all items that do not appear in the train set of that user \iffullpaper 
(details in Appendix~\ref{test-protocol-effect}).
\else 
(see~\cite{ourFullVersion} for details  and  experiments with different ranking protocol).
\fi %Note, this test protocol is a generalization of the one in~\cite{cremonesi2010performance}, and  
 
Table~\ref{tab:perfMetrics} summarizes the performance metrics. To measure how accurately an algorithm can rank items for each user,  we use local rank-based precision and recall~\cite{agarwal_chen_2016,vargas2014improving,steck2013evaluation}, where each metric is computed per user and then averaged across all users. Precision is the proportion of relevant test items in the top-$\size$ set, and recall is the proportion of relevant test items retrieved from among a user's relevant test items. As commonly done in the literature~\cite{agarwal_chen_2016,niemann2013new}, for each user $u$, we define her relevant test items as those that she rated highly,~i.e., $ \relevantItemsofUserinTestSet = \{i: i \in \itemsofUserinTestset ,  r_{ui} \geq 4\}$. Note,  because the collected datasets have many missing ratings, the hypothesis that only the observed test ratings are relevant,  underestimates the true precision and recall~\cite{steck2013evaluation}. But,  this  holds  for all algorithms, and the  measurements are known to reflect performance in real-world settings~\cite{steck2013evaluation}. F-measure is the harmonic mean of precision and recall.
\iffalse
For algorithms that re-rank rating-prediction models, we also report the prediction  metric (Prediction@$\size$) that is known to be correlated with precision~\cite{adomavicius2011maximizing}. It is the average predicted rating value of the top-$\size$ set, according to the  underlying rating-prediction model. It measures the deviation of any top-$\size$ from the default greedy strategy of a rating prediction model~\cite{adomavicius2011maximizing,ho2014likes}. %We normalized $\hat{r}_{ui}$ to $[0,1]$. 
\fi
%Furthermore, we use discounted cumulative gain (DCG)~\cite{voorhees2001overview}, which measures the accumulated  gain of items in the recommendation list while logarithmically discounting the gain of each item by its position. Note, $i$ ranges over positions $1$ to $\size$. We set $rel_{ui} = r_{ui}$ (the observed rating) when $i$ is in the test set ($i \in \itemsofUserinTestset $) and is recommendable  ($\{i: i \in \itemsinTrainset \setminus \itemsofUserinTrainset\}$), and zero otherwise. Normalized DCG (NDCG) is  obtained by normalizing by the ideal DCG measure.

%In our work, we consider the cold-start definition of novelty~\cite{Kaminskas:2016:DSN:3028254.2926720}. We therefore  focus exclusively on long-tail items and measure novelty using  Long-Tail Accuracy (LTAccuracy@$\size$)~\cite{ho2014likes},  the ratio of long-tail items in a top-$\size$ set.
We use Long-Tail Accuracy (LTAccuracy@$\size$)~\cite{ho2014likes} to measure the novelty of recommendation lists. It computes the proportion of the recommended items that are unlikely to be seen by the user.  Moreover, we use Stratified Recall (StratRecall@$\size$)~\cite{steck2013evaluation} which measures the ability of a model to compensate for the popularity bias of items w.r.t train set. Similar to~\cite{steck2013evaluation}, we set $\beta=0.5$. Note, LTAccuracy emphasizes a combination of novelty and coverage, while Stratified Recall emphasizes a combination of novelty and accuracy. 
%To evaluate long-tail item promotion, we use Long-Tail Accuracy (LTAccuracy@$\size$) and Long-Tail Coverage (LTCoverage@$\size$)~\cite{ho2014likes}. 

%Coverage  is a system-side metric  that indicates the overall number of distinct items  exposed to the entire users~\cite{vargas2014improving,ho2014likes}. 

For  coverage we use two metrics:  Coverage@$\size$  is the ratio of the total number of distinct  recommended  items to the total number of items~\cite{ho2014likes,vargas2014improving}. A maximum value of $1$ indicates  each item in $\mathcal{I}$ has been recommended at least once. 
%We also evaluate Long-Tail Coverage (LTCoverage@$\size$)~\cite{ho2014likes}.
%This measure is equivalent to normalized aggregate diversity@$\size$  proposed in ~\cite{adomavicius2012improving}. 
Gini~\cite{lorenz1905methods},  measures the inequality among values of a frequency distribution $\mathbf{f}$. It lies in $[0,1]$, with $0$ representing perfect equality, and larger  values representing  skewed distributions.  In Table~\ref{tab:perfMetrics},  $\mathbf{f}$ is the recommendation frequency of items,  and is sorted in non-decreasing order, i.e.,~$f[j] \leq f[j+1]$.% Let  $\mathbf{f}$ denote  the recommendation frequency of items. We sort the items in  non-decreasing order of  recommendation frequency, i.e., $\mathbf{f}_i \leq \mathbf{f}_{i+1}$.

\iffalse
To summarize performance of different algorithms w.r.t. multiple performance metrics,  we compute their weighted harmonic mean,  and rank the algorithms. Since different performance metrics can have different data ranges, we normalize  each metric to the range $[0,1]$.  In table~\ref{tab:perfMetrics},  $\mathbf{x}$ denotes the vector of performance measurements (e.g.$x_1$=precision, $x_2$=recall), and $\mathbf{w}$ their corresponding weights. Note, the harmonic mean of precision and recall is commonly known as F-measure. 
\fi

\vspace{4mm}
\noindent \textbf{Other algorithms and their configuration. } We compare against, or integrate the following methods  in our framework~\footnote{We report the default configurations specified in the original work for most algorithms.}.
 %For a more detailed analysis  refer to~\cite{ourFullVersion}.}.
 %experiments on other configurations  of some of the above frameworks, and report details for algorithm parameters in different datasets.,  and report the  parameters that led to the best  performance.
\begin{itemize}

\item \textbf{Rand} is  non-personalized and randomly suggests $\size$ unseen items from among all items.  It obtains high coverage and novelty,  but low accuracy~\cite{vargas2014improving}. 

\item \textbf{Pop}~\cite{cremonesi2010performance} is a non-personalized  algorithm. For  ranking tasks, it obtains high accuracy~\cite{cremonesi2010performance,vargas2014improving}, since it takes advantage of the popularity bias of the data. However, Pop makes trivial recommendations that lack novelty~\cite{cremonesi2010performance}.  

\item \textbf{RSVD}~\cite{koren2009matrix} is a latent-factor model for  rating prediction. We used \texttt{LIBMF}, with  L2-Norm as the loss function, and L2-regularization,  and Stochastic Gradient Descent (SGD) for optimization. We also tested the same model with non-negative constraints (RSVDN)~\cite{zhuang2013fast}, but did not find significant performance difference. We omit RSVDN from our results.  We performed 10-fold cross validation and tested:  number of latent factors  $g \in \{8,  20, 40, 50, 80, 100\}$, L2-regularization coefficients  $\lambda \in \{0.001, 0.005, 0.01, 0.05, 0.1\}$,  learning rate $\eta \in \{0.002,0.003,0.01, 0.03\}$. For each dataset, we used the parameters that led to best performance  
\iffullpaper
 (see Appendix~\ref{sec:configurationofR-SVD}).
\else
(see~\cite{ourFullVersion}).
\fi

\item \textbf{PSVD}~\cite{cremonesi2010performance} is a latent factor model,  known for  achieving high accuracy and novelty~\cite{cremonesi2010performance}. In PSVD,  missing values are imputed by zeros and conventional SVD is performed.  
\iffullpaper
 We used Python's \texttt{sparsesvd}  module and tested number of latent factors $g \in \{10, 20, 40, 60, 100, 150, 200, 300\}$. We report results for two configurations:  one with $10$ latent factors (PSVD10), and one with 100 latent factors (PSVD100). 
\else
We used Python's \texttt{sparsesvd}  module and tested two configurations:  one with $10$ latent factors (PSVD10), and one with 100 latent factors (PSVD100). 
\fi  

\item \textbf{CoFiRank}~\cite{weimer2007maximum} is a  ranking prediction model that can optimize directly the  Normalized Discounted Cumulative Gain (NDCG) ranking measure~\cite{voorhees2001overview}.   %We  tested dimensions $g \in \{10,100\}$,   $\lambda \in \{5,10\}$. However, 
We used the source code from~\cite{weimer2007maximum}, with parameters set according to~\cite{weimer2007maximum}:  100 dimensions and $\lambda =10$, and default values for other parameters. We experimented with both regression (squared) loss (CofiR100) and NDCG loss (CofiN100). 
Similar to~\cite{balakrishnan2012collaborative,volkovs2012collaborative}, we found CofiR100 to perform consistently better than CofiN100 in our experiments on ML-1M and ML-100K. We only report results for CofiR100.

\item \textbf{Ranking-Based Techniques (RBT)}~\cite{adomavicius2012improving} maximize coverage by re-ranking the output of a rating prediction model according to a re-ranking criterion. We implemented  two variants: one that re-ranks  a few of the items in the head according to their popularity (Pop  criterion), and another which re-ranks according to the  average rating (Avg criterion). As in~\cite{adomavicius2012improving}, we set $T_{max}=5$ in all datasets. The parameter $T_R$ controls the extent of re-ranking. We tested $T_{R}\in [4,4.2,4.5]$, and found $T_R=4.5$ to yield more accurate results. Furthermore, because our datasets contain a wider range of users compared to~\cite{adomavicius2012improving}, we set $T_H=1$ on all datasets, except ML-10M and Netflix, where we set $T_H=0$.  To refer to RBT variants, we  use  $\texttt{RBT}(\texttt{ARec}, \texttt{Re-ranking criterion})$.

\item \textbf{Resource~allocation}~\cite{ho2014likes} 
  is an method for re-ranking the output of a rating prediction model. It has two phases: \begin{enumerate*}
\item resources are  allocated to items according to the received ratings, and
\item the resources are distributed according to the relative preferences of the users, and top-$\size$ sets are generated by  assigning a a 5D score (for accuracy, balance, coverage, quality, and quantity of long-tail items) to every user-item pair.
\end{enumerate*}
We use the variants  proposed in~\cite{ho2014likes},  which are combinations of the scoring function (5D) with the rank by rankings (RR) and accuracy filtering (A) algorithms (Section 3.2.2 in~\cite{ho2014likes}). We use the template $\texttt{5D} (\texttt{ARec}, \texttt{A}, \texttt{RR})$ to show the different combinations, where A and RR are optional.  
We implemented and ran all four variants  with default values set according to~\cite{ho2014likes}: $k=3.|\mathcal{I}|$ and $q=1$. 
\iffullpaper
\else
We report results for 5D(ARec) and 5D(ARec, A, RR).%, with others in~\cite{ourFullVersion}.
\fi

\item \textbf{Personalized Ranking Adaptation (PRA)}~\cite{jugovac2017efficient} is  a generic re-ranking framework, that first estimates user tendency for various criteria like diversity and novelty, then  iteratively and greedily re-ranks  items in the head of the recommendations  to match the top-$\size$ set  with the user tendencies. We compare with the novelty-based variant of this framework, which relies  on item popularity statistics  to measure   user novelty tendencies. We use the  the mean-and-deviation based heuristic, that  is measured using the popularity of rated items, and was shown to provide comparable results with other heuristics in~\cite{jugovac2017efficient}.  For the configurable parameters, we followed~\cite{jugovac2017efficient}: Sample set size $S_u \in \min(|\mathcal{I}^{\mathcal{R}}_u|,10)$,  the exchangeable set size $|X_u| \in \{10,20\}$, and used ``optimal swap'' strategy with $maxSteps=20$. We use the  template $\texttt{PRA}(\texttt{ARec}, |X_u|)$ to refer to variants of PRA.
%then  and then  iteratively and greedily re-ranks items a small number of items  to match top-$\size$ set characteristics  with user preferences.%Recently,~\cite{jugovac2017efficient} proposed a generic Personalized Ranking Adaptation (PRA) framework. The idea is to  estimate user tendency for various objectives like diversity and item popularity using various heuristics, and to recommend top-$\size$ sets that match those tendencies.  We use \textbf{PRA} as a baseline.  The framework iteratively and greedily re-ranks items in the head of the recommendations to match them with the user preferences.
\end{itemize}

\subsection{Distribution of long-tail novelty preferences}
Figure~\ref{fig:Hist-LT-PreferenceModels} plots the histogram of various long-tail preference 
\iffullpaper
models. 
\else 
(results for other datasets are omitted due to space limitations, but have similar trends). 
\fi 
We observe $\theta^A_u$ is skewed to the right. This is due to the sparsity problem, where the majority of users rate a few items. $\theta^N_u$  is also skewed to the right across all datasets, due to both the popularity bias and sparsity problems~\cite{agarwal_chen_2016,Marlin07collaborativefiltering}. 
On the other hand, $\theta^G_u$ is normally distributed, with a larger mean and  a larger variance, on all datasets. In the experiments in Section~\ref{sec:perfOSLG}, we study the effect of these preference models on performance. 

\iffalse
One explanation for the bias of all preference estimates towards lower values, is the popularity bias of  training data. In some datasets, such as MovieLens, movies are presented to users to warm-start the recommender system. The presented movies, and the observed ratings, can be biased toward more popular movies~\cite{agarwal_chen_2016,Marlin07collaborativefiltering}. As a result, the preference estimates are biased towards lower values for datasets that exhibit this popularity bias. 

Note,  on the MovieTweetings dataset, where users voluntarily rate movies, and the effect of presentation bias is negligible,   $\theta^*$ is normally distributed  with $\mu_{\theta^*}=0.522$ and $\sigma_{\theta^*}=0.178 $, while $\theta^T$ and $\theta^N$ remain skewed toward lower values. 

Note, popularity bias and sparsity are inherent characteristics of the recommendation domain, although some datasets suffer more.  Totally eliminating this bias, may not be advantageous. 
\fi
%On Netflix ($\mu_{\theta^*}=0.323, \sigma_{\theta^*}=0.133$)
%On ML-10M ($\mu_{\theta^*}=0.312, \sigma_{\theta^*}=0.135$)
%On ML-1M ($\mu_{\theta^*}=0.349, \sigma_{\theta^*}=0.133$)

\iffullpaper
\begin{figure*}[t]
  %\vspace{-20pt}
\centering
        \subfloat[ML-100K]
        {\includegraphics[scale=0.20]{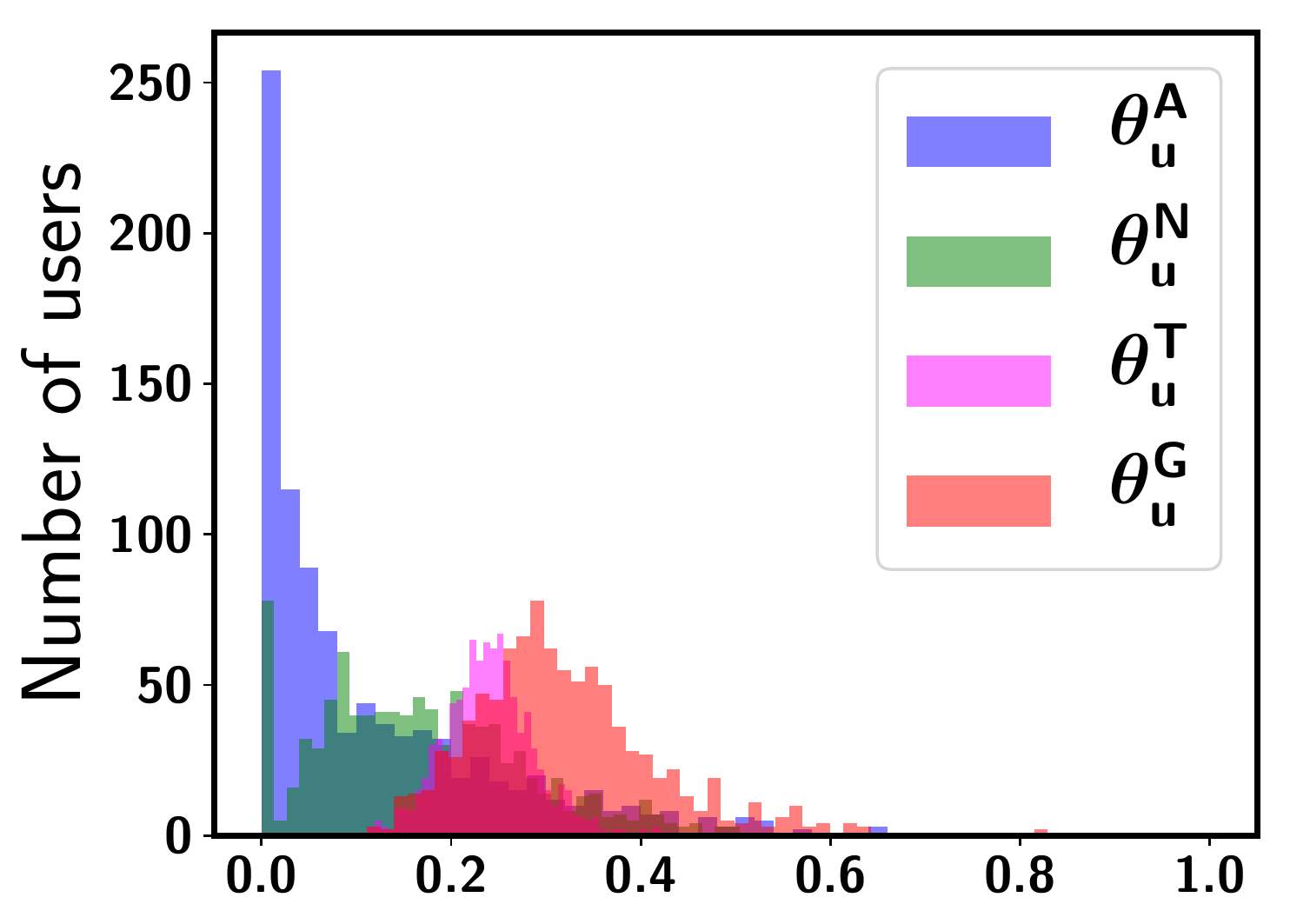}
		 \label{fig:ML-Users-vs-theta}
        } 
         \subfloat[ML-1M]
        {\includegraphics[scale=0.20]{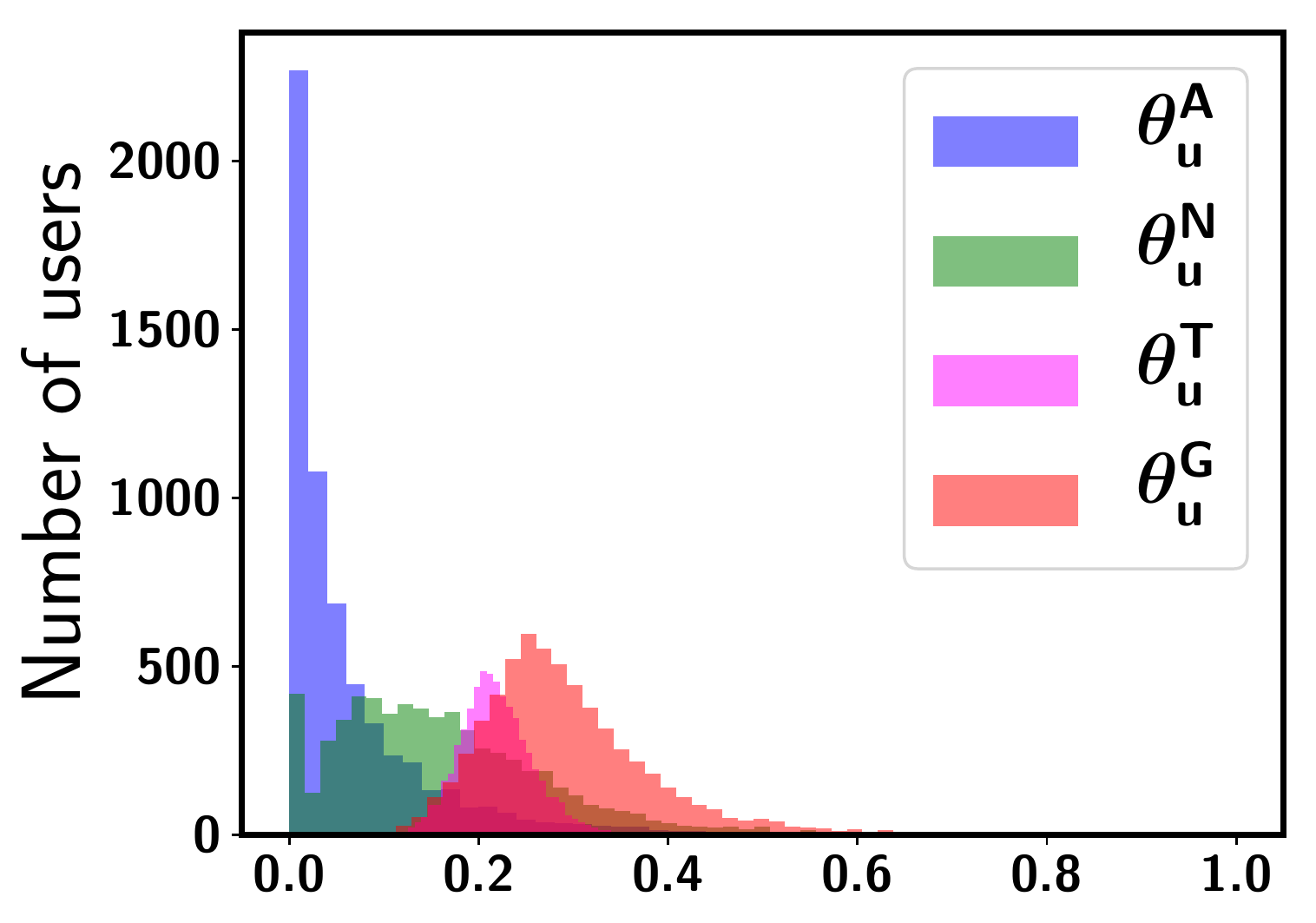}%twoHistogramsofthetaTNRthetaTFIDFthetaStarthetaNorNLT.pdf}
        
		 \label{fig:ML-Users-vs-theta}
        }         
        \subfloat[ML-10M]
	    {\includegraphics[scale=0.20]{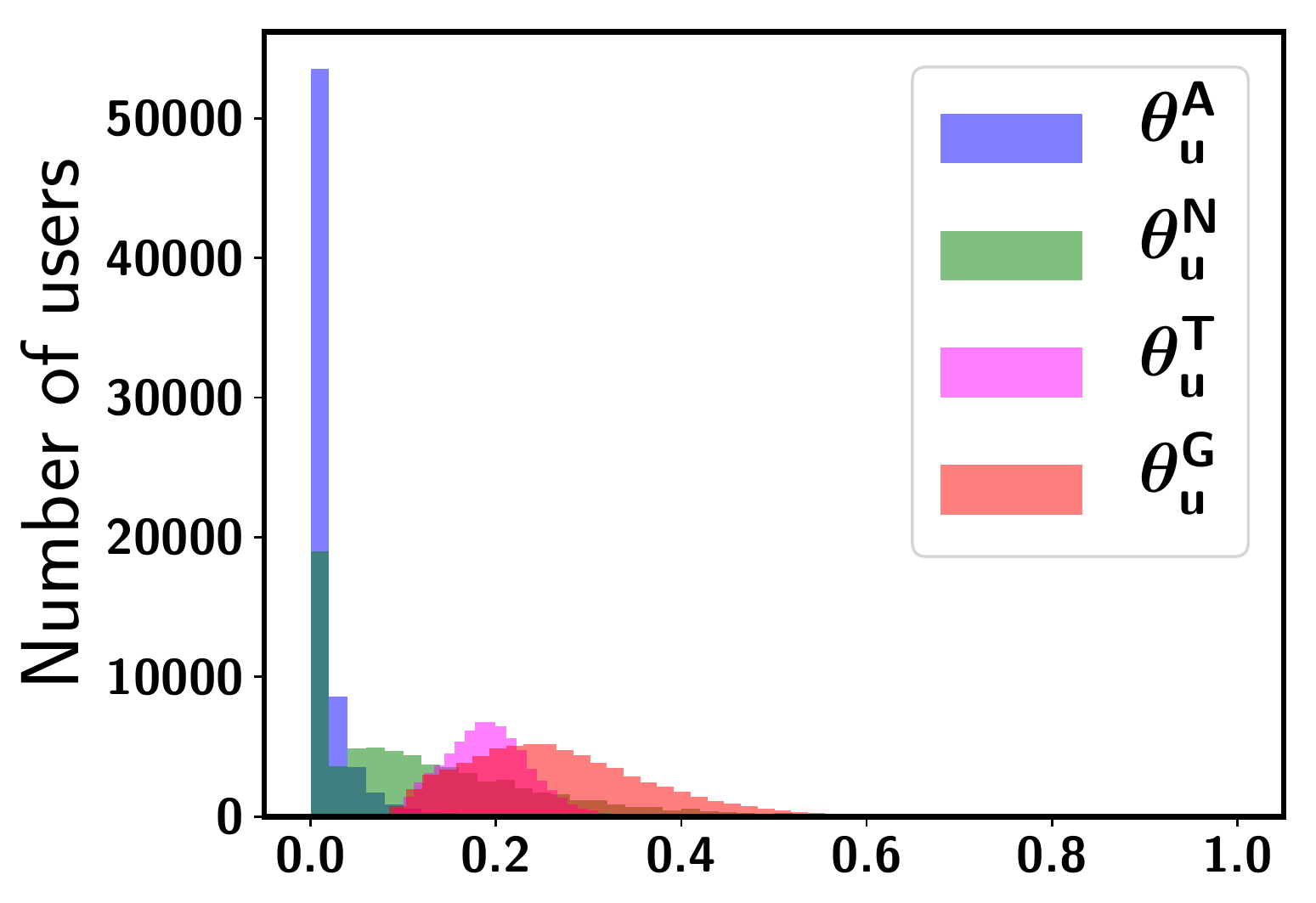}
		 \label{fig:ML-Users-vs-theta}
        }
        \subfloat[MT-200K]
        {\includegraphics[scale=0.20]{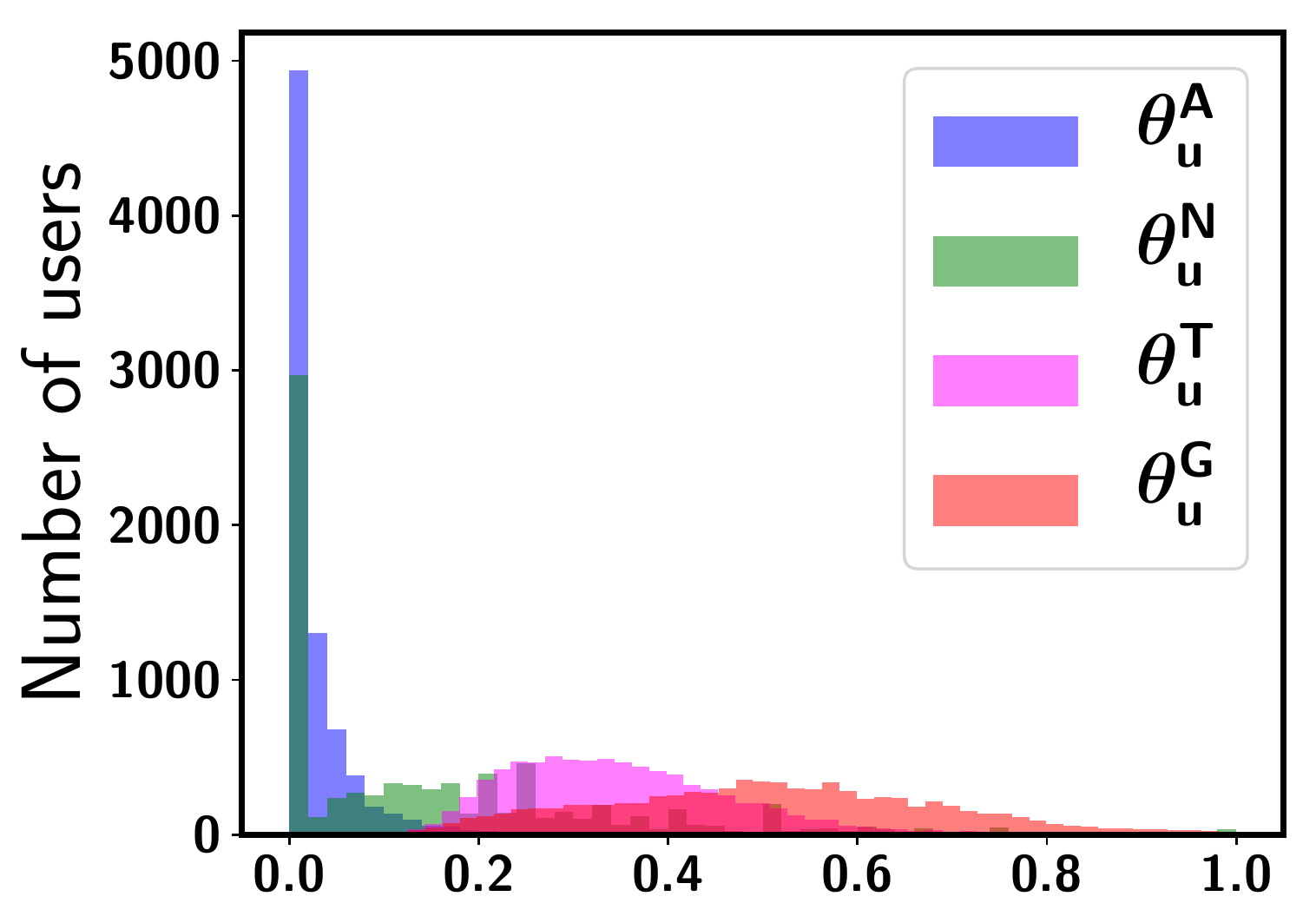}%twoHistogramsofthetaTNRthetaTFIDFthetaStarthetaNorNLT.pdf}
		 \label{fig:ML-Users-vs-theta}
        } 
        \subfloat[Netflix]
        {\includegraphics[scale=0.20]{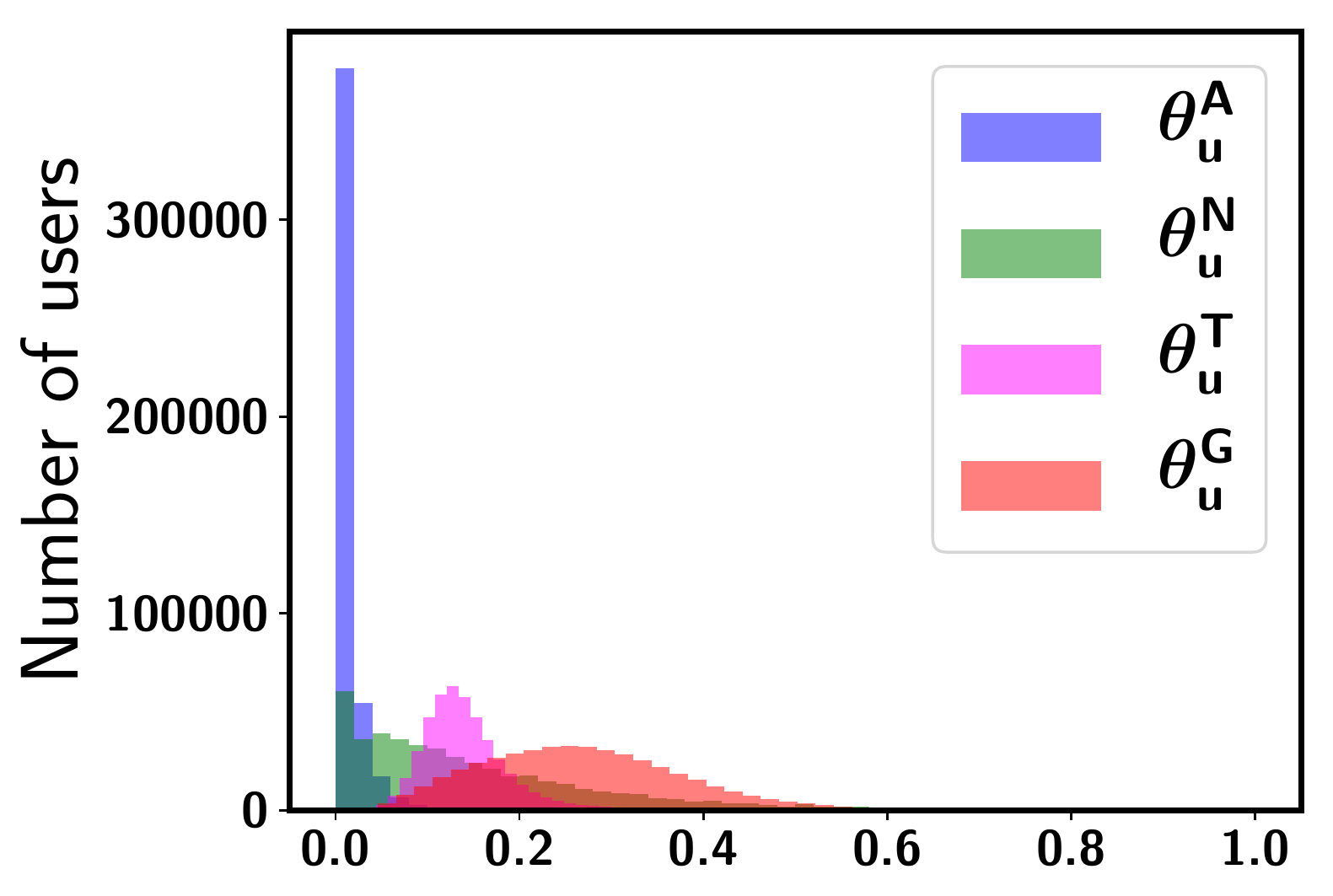}
		 \label{fig:ML-Users-vs-theta}
        } 

\caption{Histogram of long-tail novelty preference models. Observe $\theta^A_u$ is skewed toward smaller values because of sparsity, i.e.,~the majority of users rate a few items.  $\theta^N_u$ is also biased toward smaller values, due to a combination of popularity bias and sparsity. $\theta^T_u$ and $\theta^G_u$ are less biased and more normally distributed and  alleviate both problems.} %In addition to the disadvantages mentioned in Section~\ref{sec:simple-lt-pref},
\label{fig:Hist-LT-PreferenceModels}
\end{figure*}

\else

\begin{figure}[t]
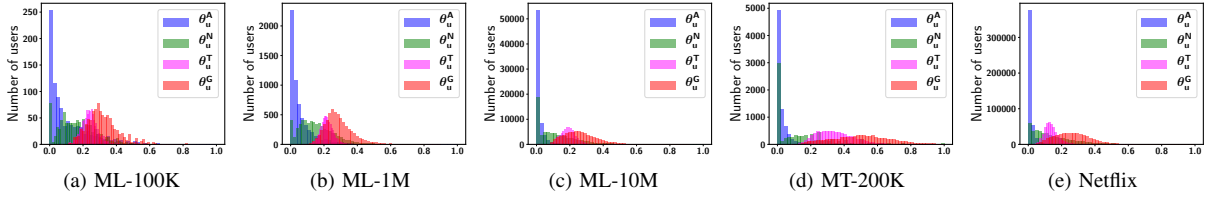

  %\vspace{-20pt}
\centering
         \subfloat[ML-1M]
        {\includegraphics[scale=0.28]{Figures-ml-1m/CombinedHistograms/twoHistogramsofthetaTNRthetaNorNLTthetaTFIDFthetaStar2.pdf}%twoHistogramsofthetaTNRthetaTFIDFthetaStarthetaNorNLT.pdf}
		 \label{fig:ML-Users-vs-theta}
        }          
        \subfloat[Netflix]
        {\includegraphics[scale=0.28]{FiguresSum/CombinedHistograms/twoHistogramsofthetaTNRthetaNorNLTthetaTFIDFthetaStar2.pdf}
		 \label{fig:ML-Users-vs-theta}
        } 

\caption{Histogram of long-tail novelty preference models. Observe $\theta^A_u$ is skewed toward smaller values because of sparsity, i.e.,~the majority of users rate a few items.  $\theta^N_u$ is also biased toward smaller values, due to a combination of popularity bias and sparsity. $\theta^T_u$ and $\theta^G_u$ are less biased and more normally distributed and  alleviate both problems.} %In addition to the disadvantages mentioned in Section~\ref{sec:simple-lt-pref},
\label{fig:Hist-LT-PreferenceModels}
\end{figure}

\fi

\iffullpaper

\begin{figure*}[t]
\centering
		\subfloat[PSVD100]
		{
		\includegraphics[width=0.22\textwidth]{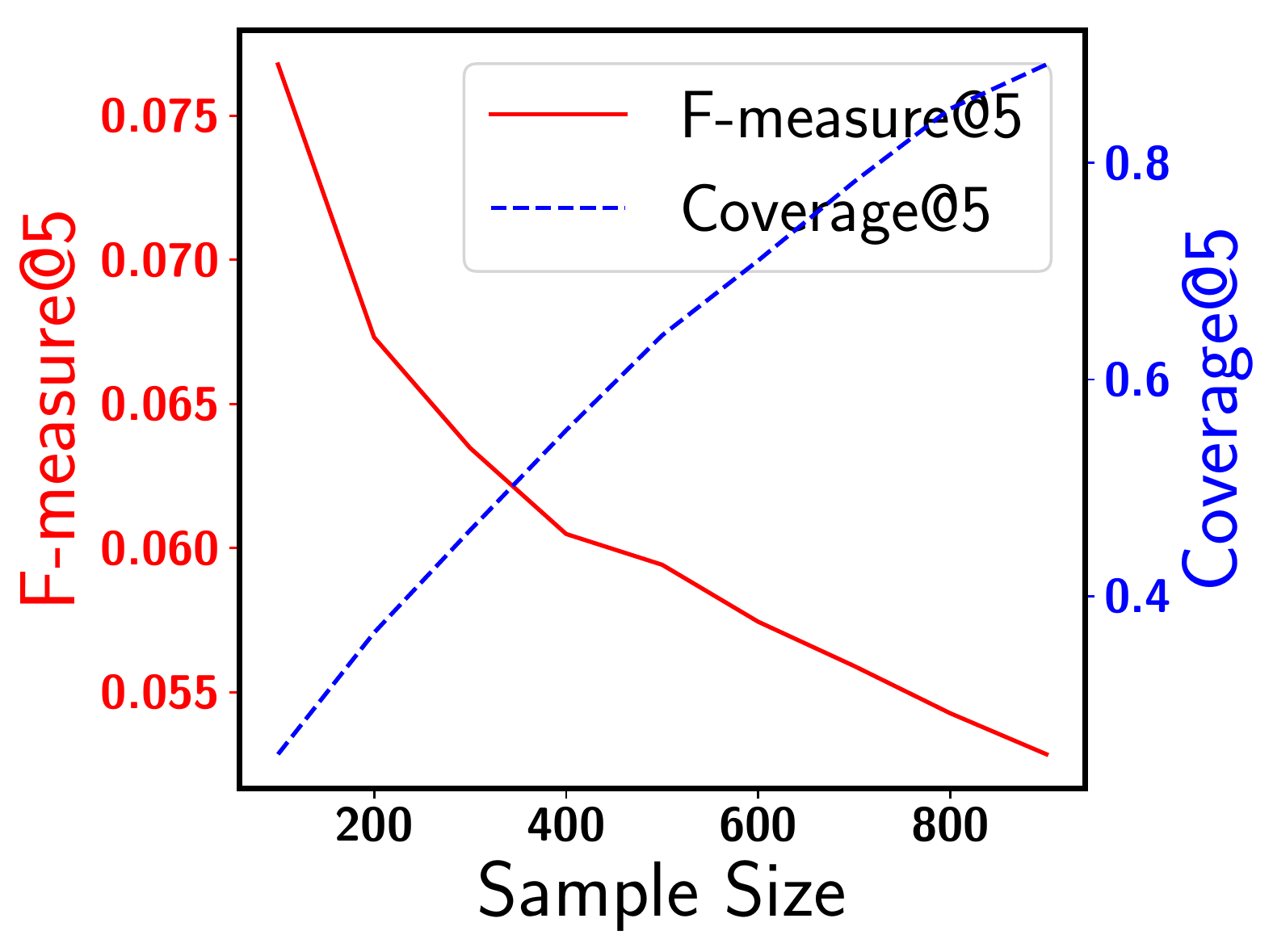} 
		}
		\subfloat[PSVD10]
		{
		\includegraphics[width=0.22\textwidth]{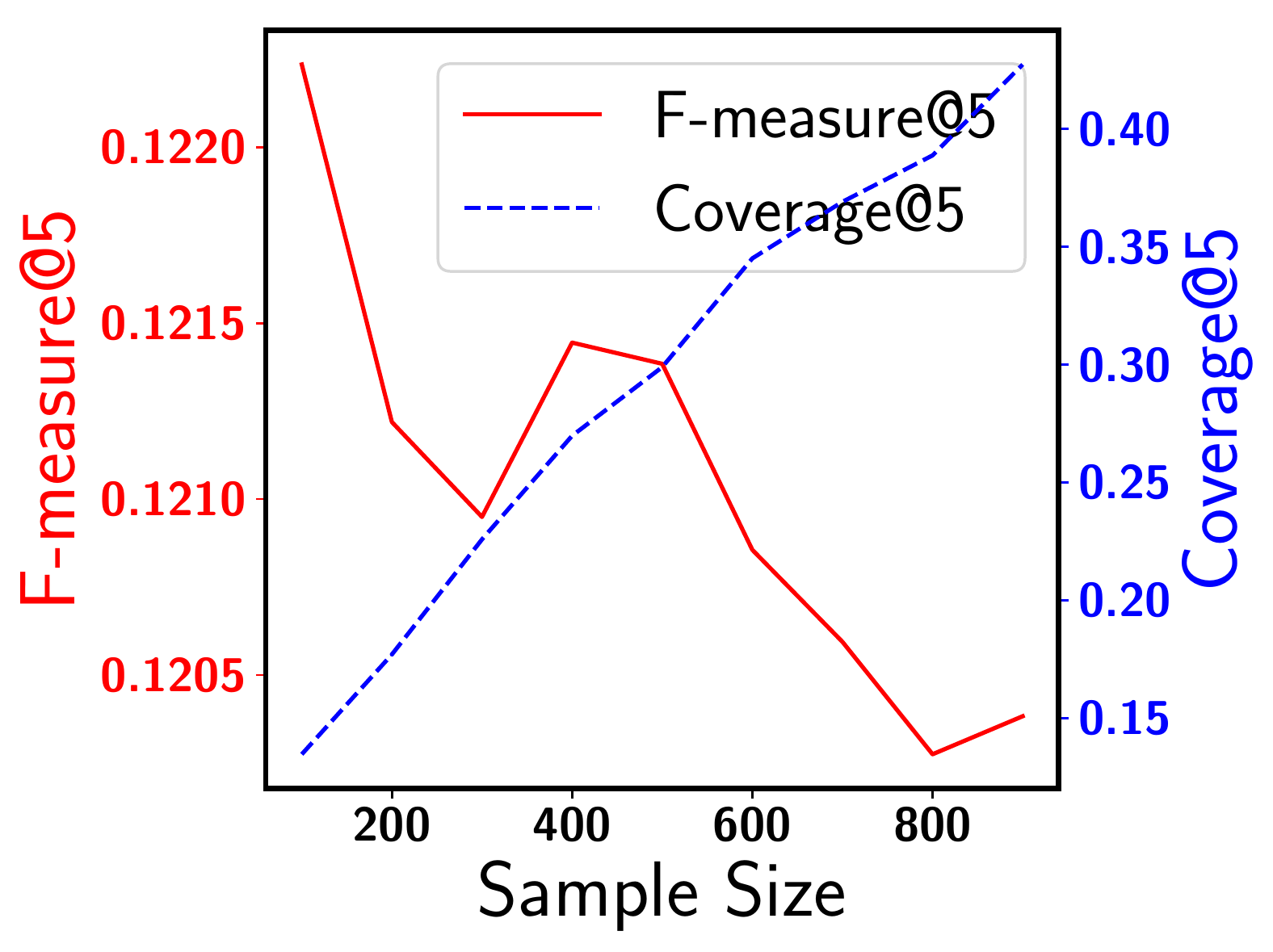} 
		}
		\subfloat[Pop]
		{
		\includegraphics[width=0.22\textwidth]{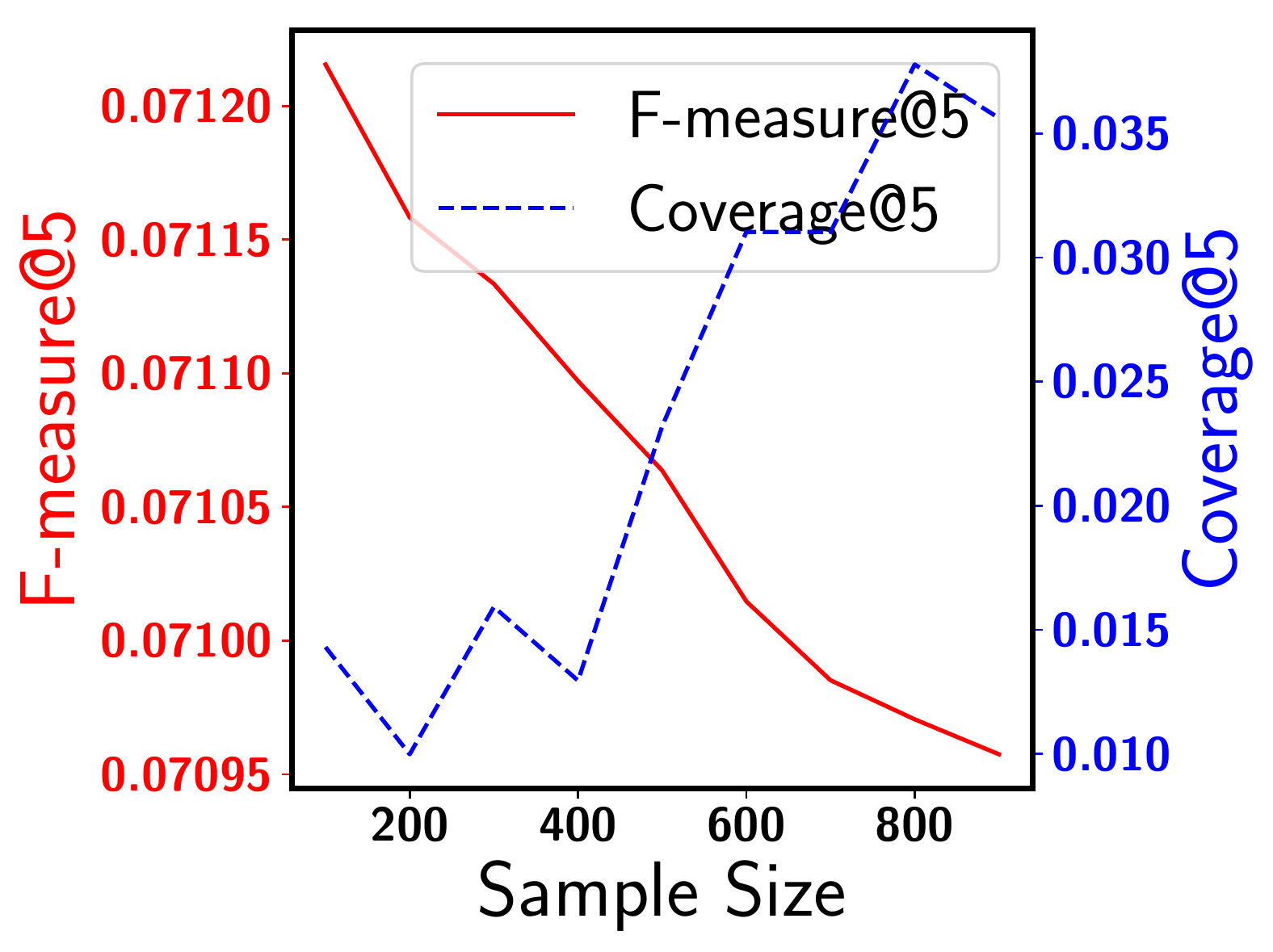} 
		}
        \subfloat[RSVD]
		{
		\includegraphics[width=0.22\textwidth]{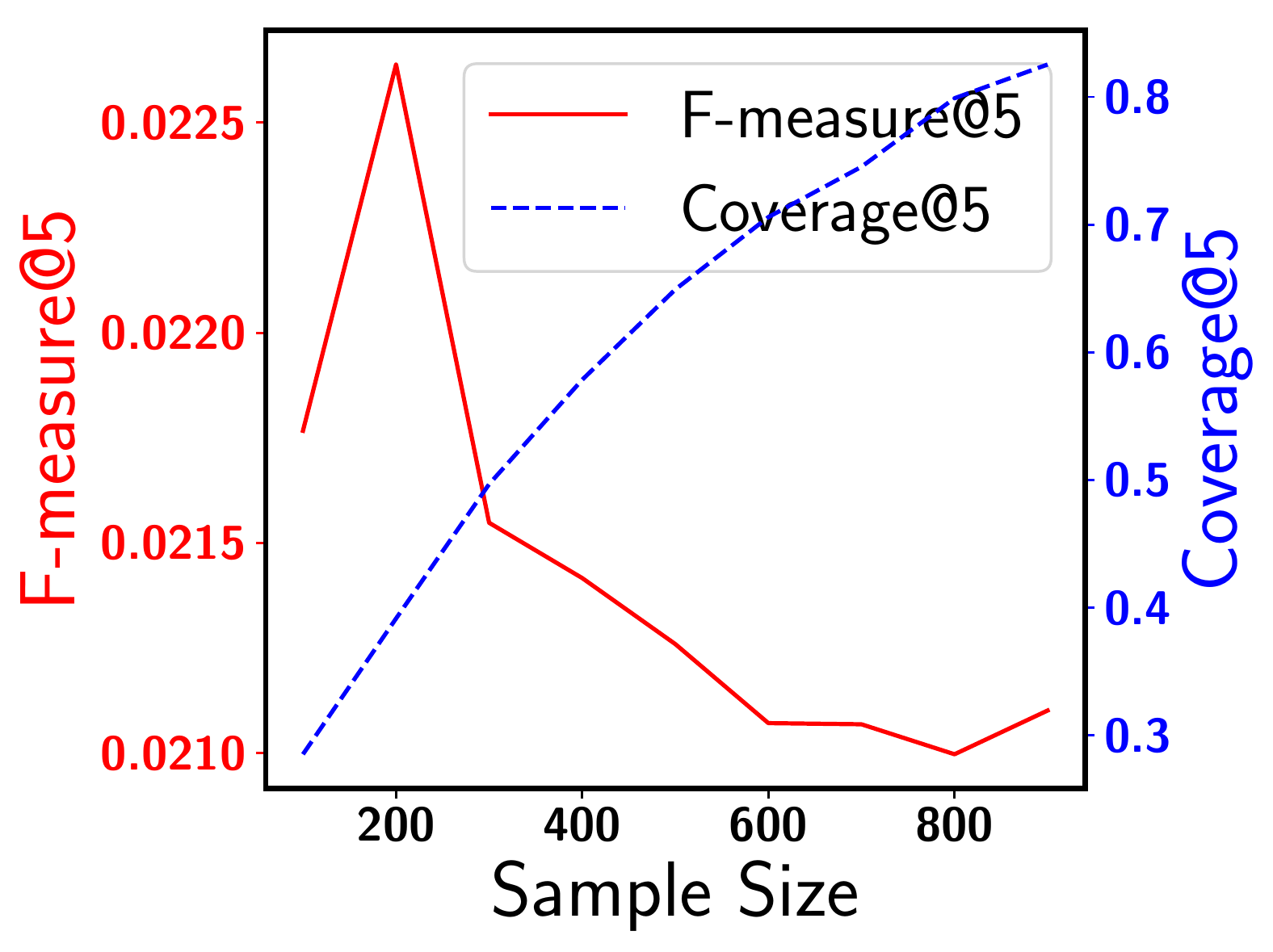} 
		}

\caption{Performance of GANC(ARec, $\bm{\theta}^{G}$, Dyn) with OSLG optimization, as sample size (S) is varied. The accuracy recommender ARec is indicated in each sub-figure. Dataset is ML-1M.}
\label{fig:SampleSize1}
\end{figure*}

\begin{figure*}[t]
\centering
		\subfloat[PSVD100]
		{
		\includegraphics[width=0.22\textwidth]{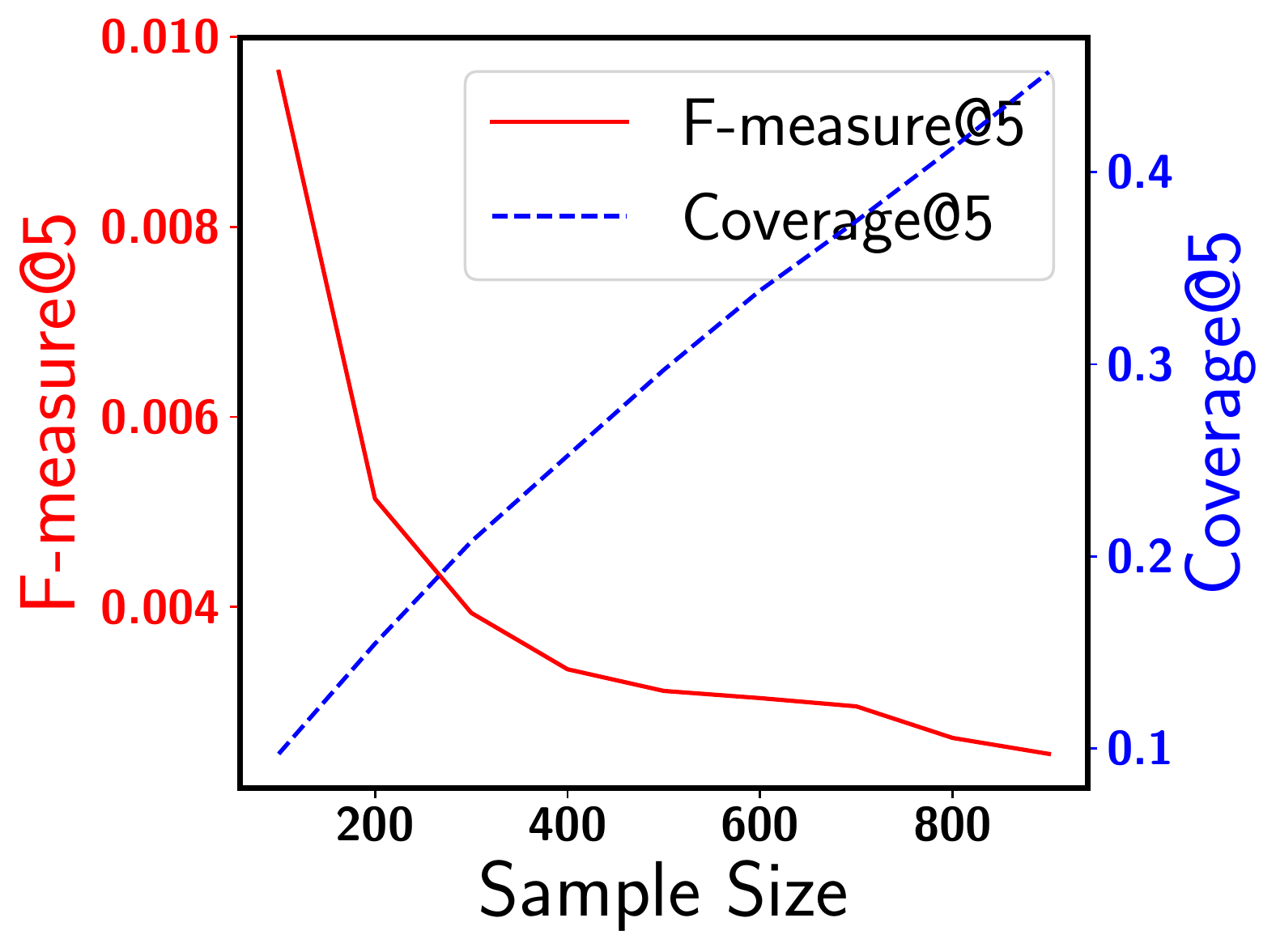} 
		}
		\subfloat[PSVD10]
		{
		\includegraphics[width=0.22\textwidth]{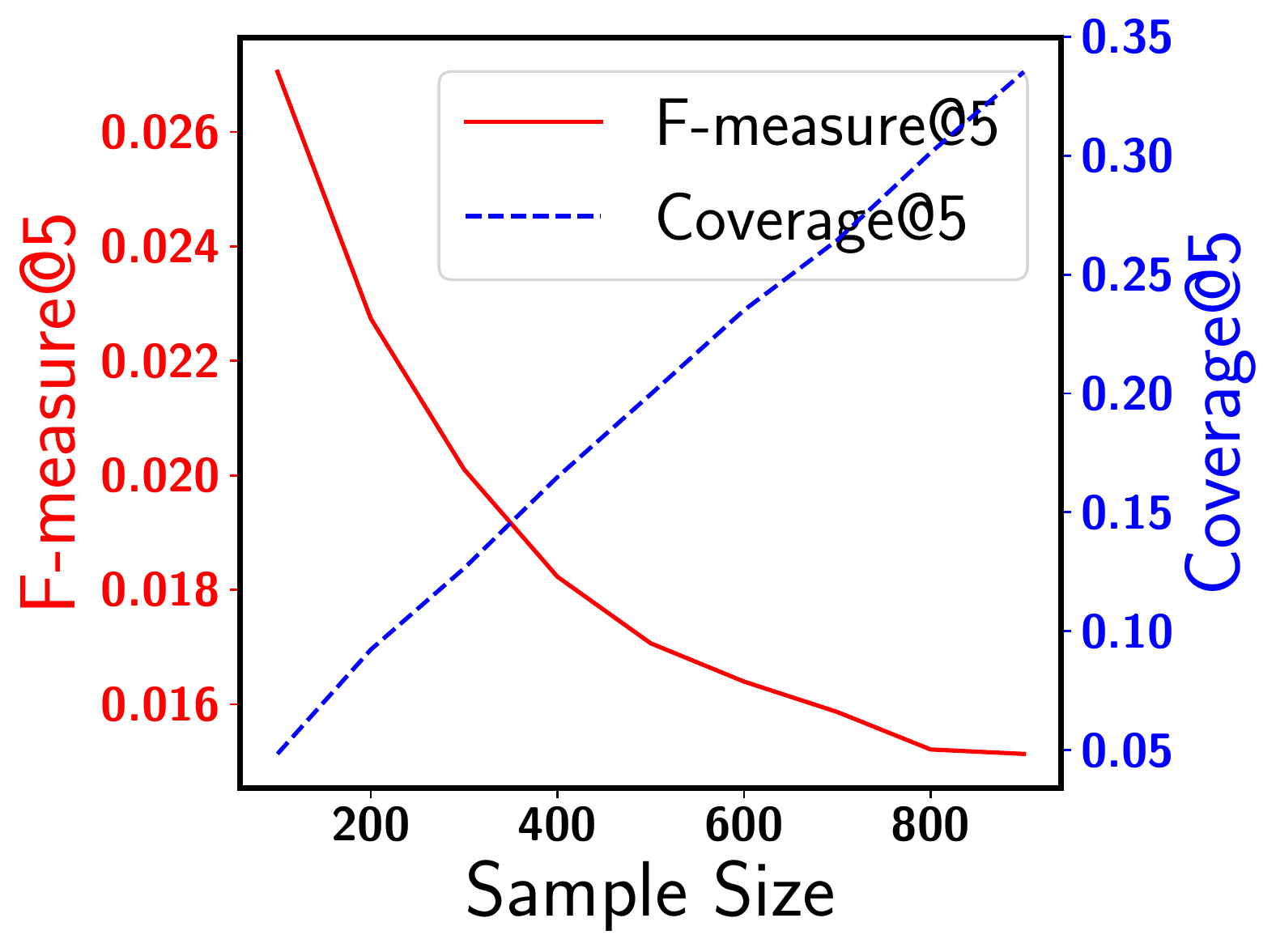} 
		}
		\subfloat[Pop]
		{
		\includegraphics[width=0.22\textwidth]{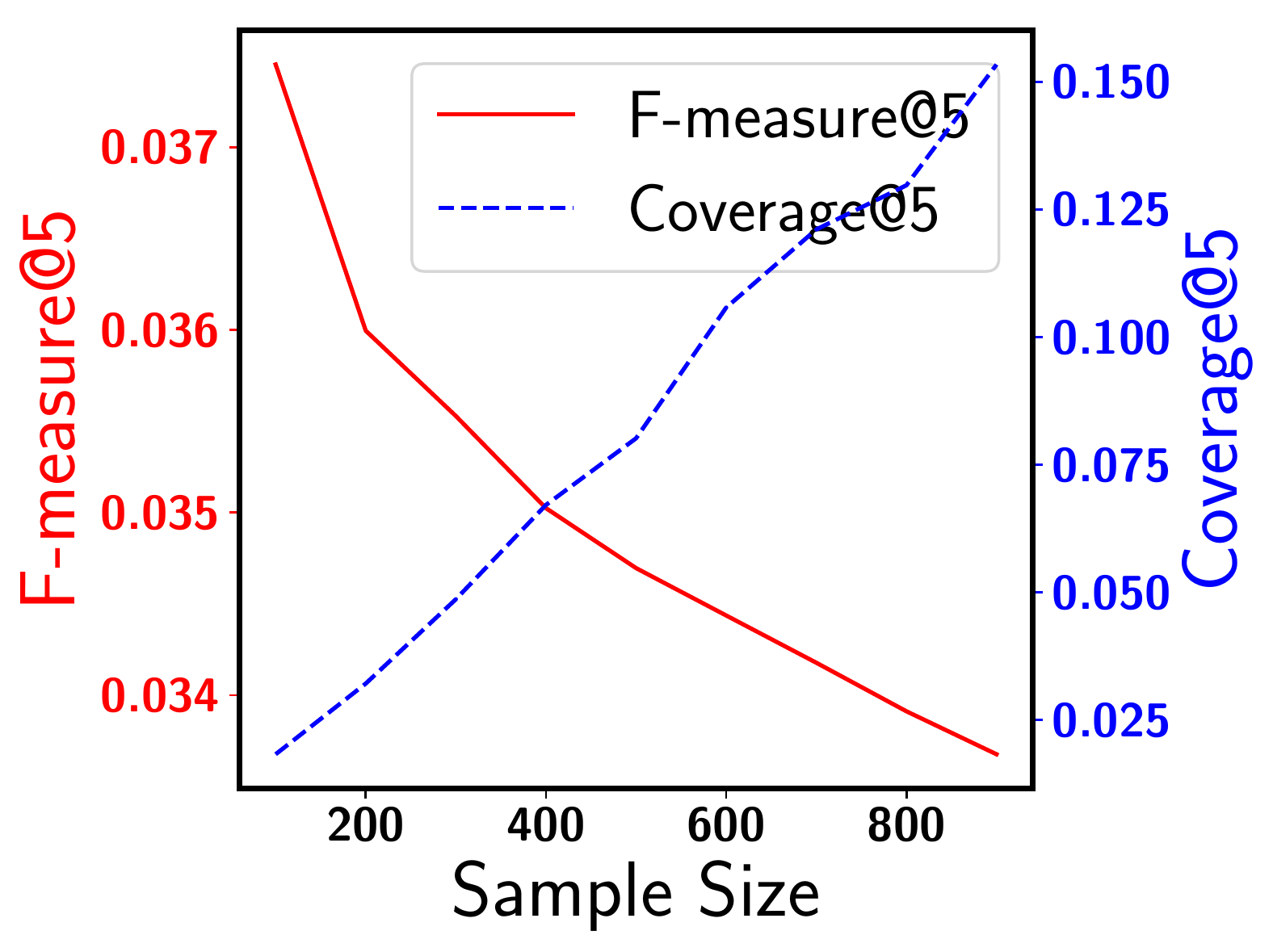} 
		}
        \subfloat[RSVD]
		{
		\includegraphics[width=0.22\textwidth]{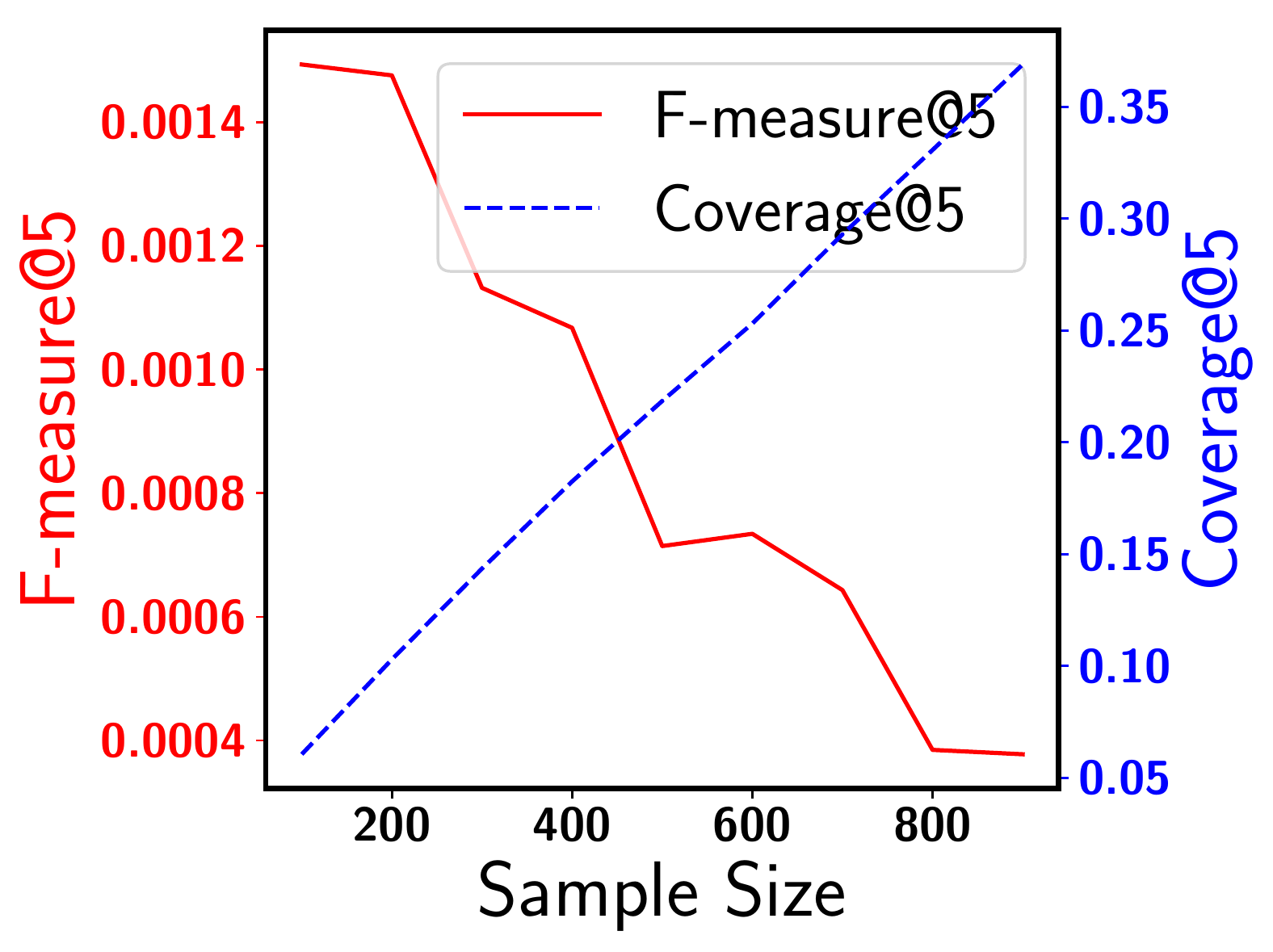} 
		}
		
\caption{Performance of GANC(ARec, $\bm{\theta}^{G}$, Dyn) with OSLG optimization, as sample size (S) is varied. The accuracy recommender ARec is indicated in each sub-figure. Dataset is MT-200K.}
\label{fig:SampleSize2}
\end{figure*}

\else
\begin{figure}[b]
\centering
		\subfloat[ARec is PSVD100]
		{
		\includegraphics[width=0.23\textwidth]{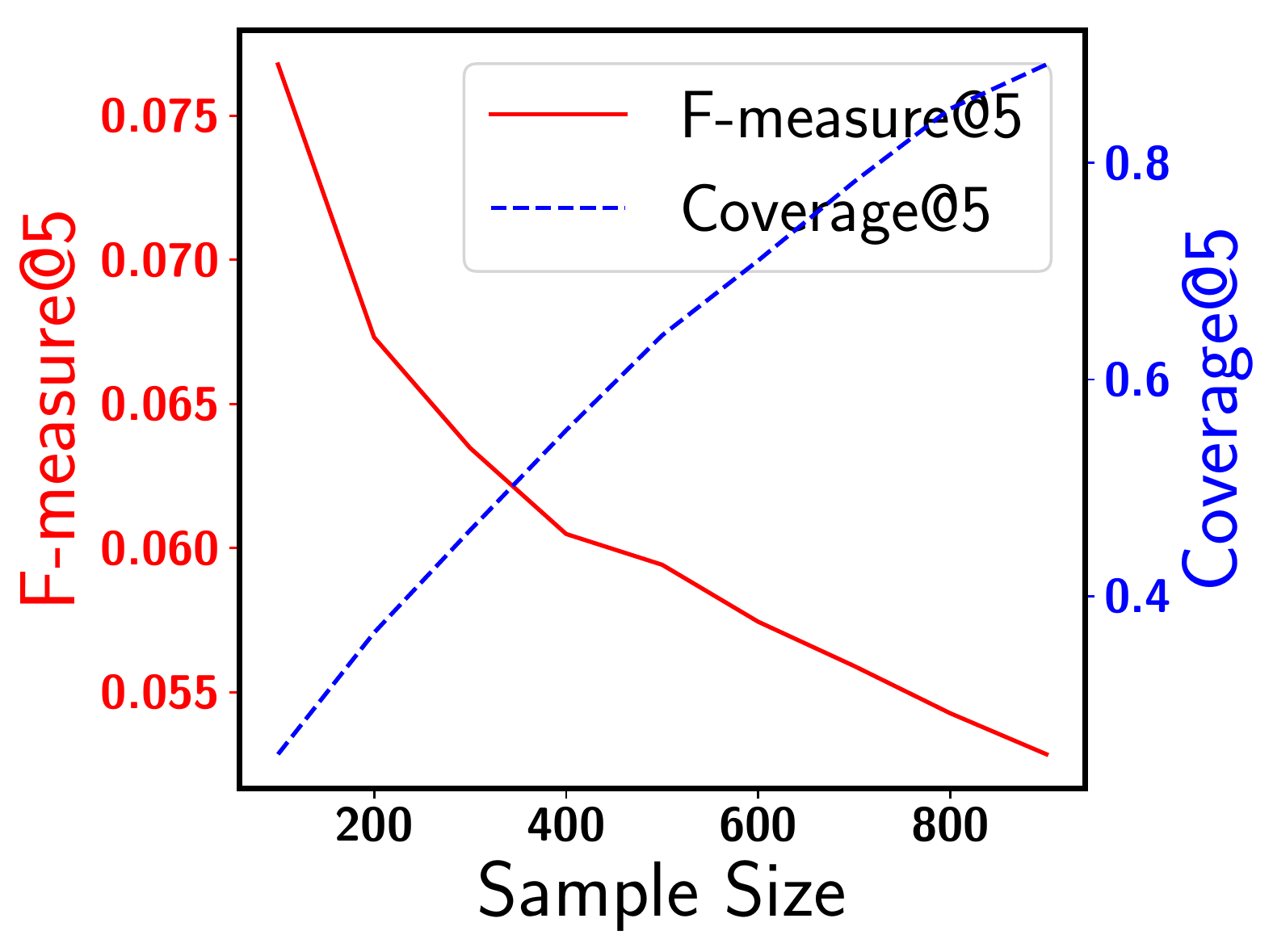} 
		}
		\subfloat[ARec is PSVD10]
		{
		\includegraphics[width=0.23\textwidth]{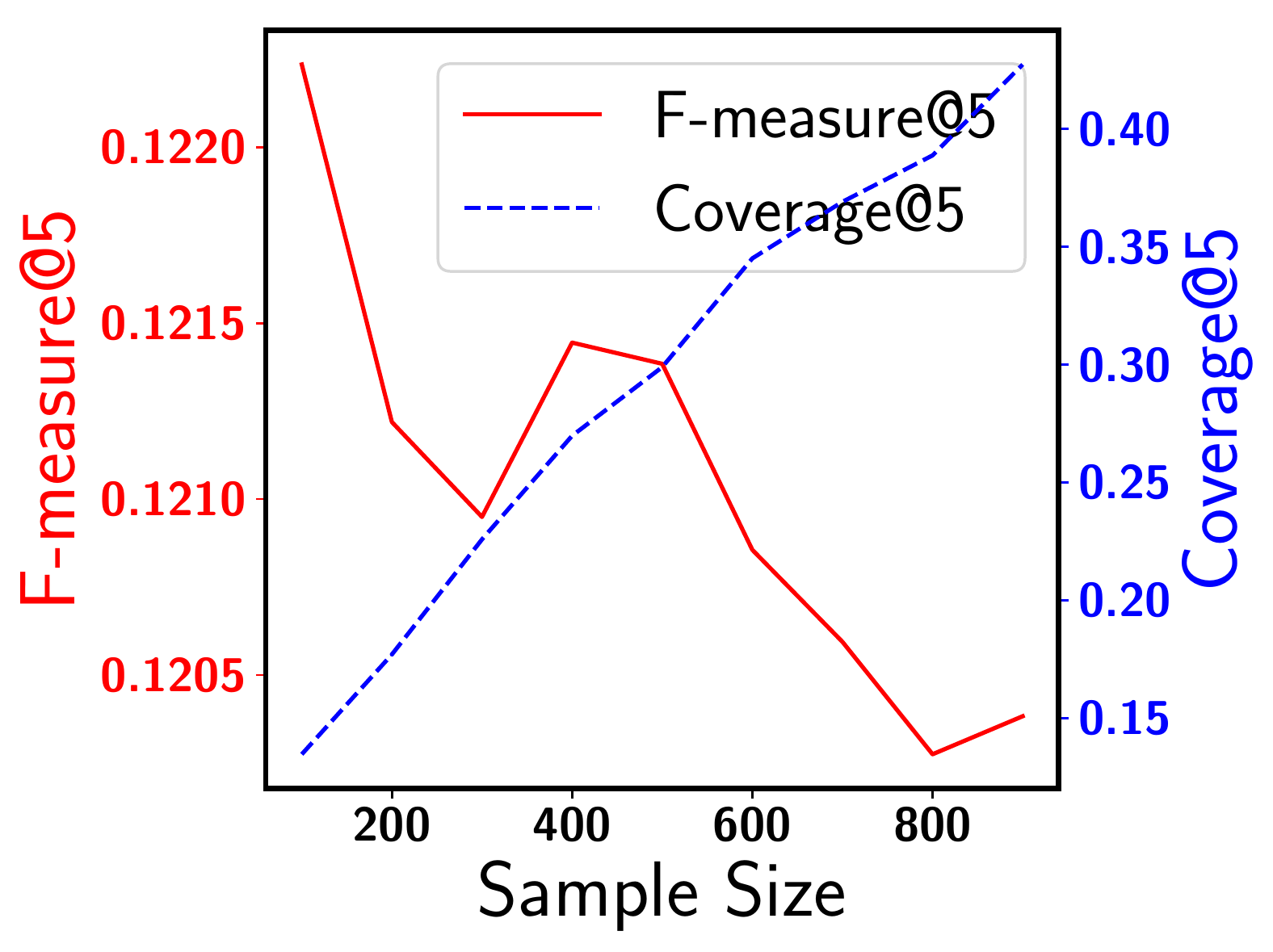} 
		}
\caption{Performance of GANC(ARec, $\bm{\theta}^{G}$, Dyn) with OSLG optimization, as sample size (S) is varied. The accuracy recommender ARec is indicated in each sub-figure. Dataset is ML-1M.}
\label{fig:SampleSize}
\end{figure}
\fi
\subsection{Performance of GANC with Dyn coverage recommender}
\label{sec:perfOSLG}
When Dyn is integrated in GANC, we use OSLG optimization. We run variants of GANC that involve   randomness (e.g., sampling-based variants) 10 times and report the average. 

\vspace{4mm}
\noindent \textbf{Effect of sample size. }The  sample size $S$ is a  system-wide hyper-parameter in the OSLG algorithm,  and should be determined w.r.t. preference for accuracy or coverage, and the accuracy recommender.  For tuning $S$, we run experiments on ML-1M, and  assess its effect   on F-measure  and coverage. 
\iffullpaper
As shown in Figures~\ref{fig:SampleSize1} and~\ref{fig:SampleSize2}, increasing $S$, increases coverage and decreases the F-measure of  most  accuracy recommenders.   Regarding RSVD, the scores output by this model lead to the initial decrease and later increase of  F-measure.  Since  we want  to maintain  accuracy, we fix $S=500$ in the rest of our experiments,  although a larger $S$ can be used. 
\else
Figure~\ref{fig:SampleSize} shows the effect of increasing $S$ on coverage and F-measure for two of our base accuracy recommenders (other results are omitted but the trends are similar~\cite{ourFullVersion}). Since  we want  to maintain  accuracy, we fix $S=500$ in the rest of our experiments,  although a larger $S$ can be used. 
\fi

\vspace{4mm}
\noindent \textbf{Effect of the user long-tail novelty preference model, the accuracy recommender, and their interplay. } 
We evaluate GANC with Dyn coverage, i.e., GANC(ARec, $\bm{\theta}$, Dyn),  while varying the accuracy recommender ARec, and the long-tail novelty preference model $\bm{\theta}$. We examine the following long-tail preference models: 
\begin{itemize}
\item \textbf{Random} $\bm{\theta}^{R}$ randomly initializes $\theta^{R}_u$ (10 times per user).%, run the algorithm and average over the  10 runs. 
\item \textbf{Constant} $\bm{\theta}^{C}$ assigns the same  constant value  $C$ to all users. We report results for $C=0.5$. %We tested $C \in {0.2,0.5,0.7}$, and report the best result. %on each dataset. \fi
\item \textbf{Normalized Long-tail} $\bm{\theta}^{N}$ (Eq.~\ref{eq:nlt-risk}) measures the proportion of long-tail items the user has rated in the past. 
\item \textbf{Tfidf-based} $\bm{\theta}^{T}$ (Eq.~\ref{eq:tfidf-risk}) incorporates user interest and popularity of items. 
\item \textbf{Generalized} $\bm{\theta}^{G}$  (Eq.~\ref{eq:overallRiskUpdate}) incorporates user interest, popularity of items, and the preferences of other users. 
\end{itemize}

Due to sampling ($S=500$), we run each variant 10 times (with random shuffling for  $\bm{\theta}^C$) and average over the runs.  We run the experiments on ML-1M.  %We set $S=|\mathcal{U}|$, so as to remove the effect of sampling, and run the experiments on ML-1M and MT-200K.

\begin{figure*}[p]
\centering 
        \subfloat[Accuracy recommender (ARec) is RSVD]
        {
		\includegraphics[width=0.88\textwidth]{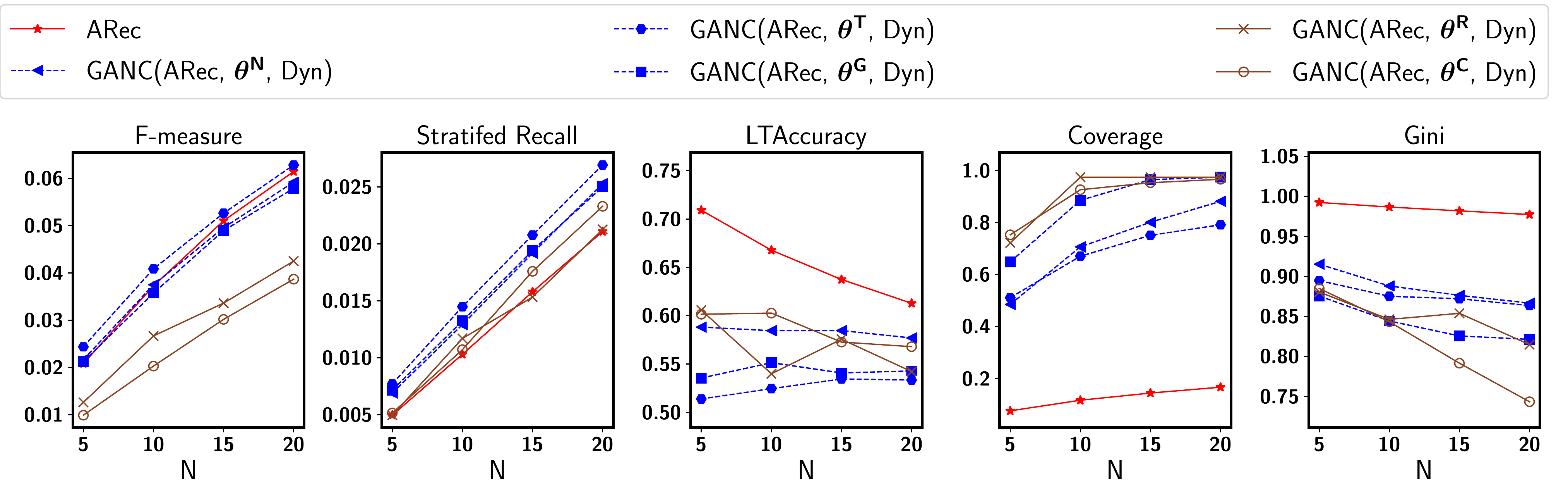}
		\label{fig:ThetaConfigs}
        }
        
        \subfloat[Accuracy recommender (ARec) is PSVD100]
        {
		\includegraphics[width=0.88\textwidth]{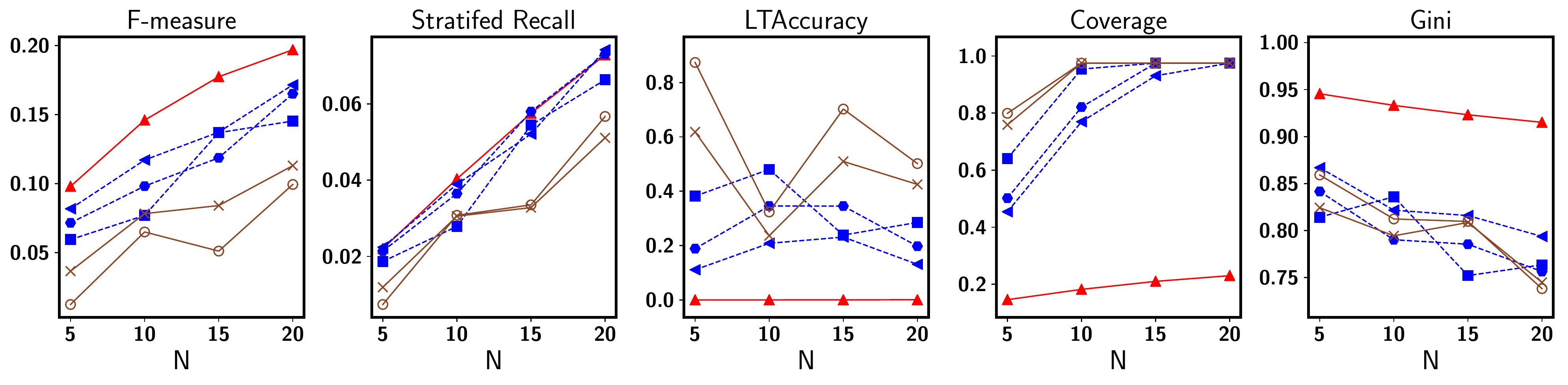}
		\label{fig:ThetaConfigs}
        }
        
        \subfloat[Accuracy recommender (ARec) is PSVD10]
        {
		\includegraphics[width=0.88\textwidth]{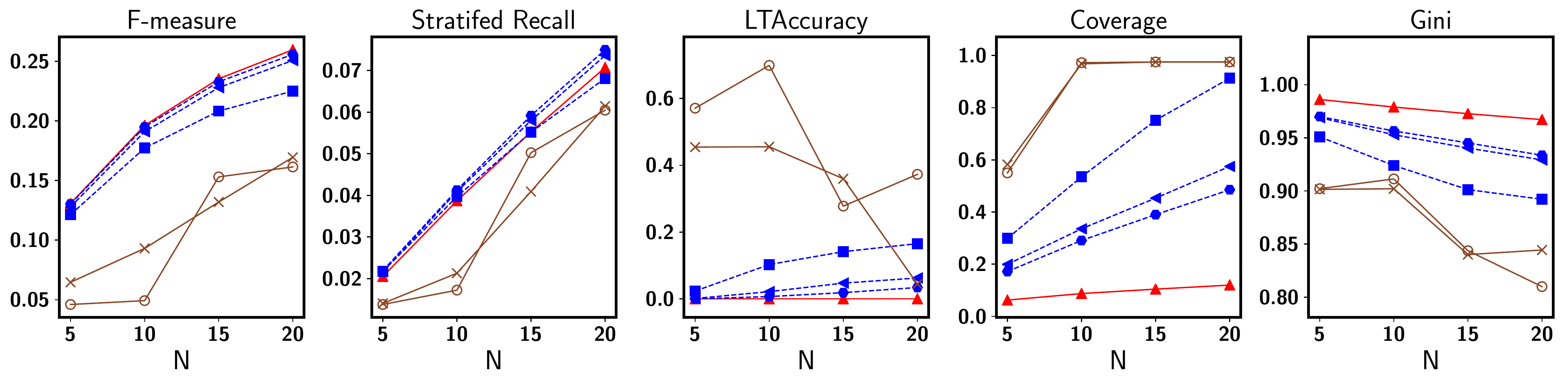}
		\label{fig:ThetaConfigs}
        }
        
        \subfloat[Accuracy recommender (ARec) is Pop]
        {
		\includegraphics[width=0.88\textwidth]{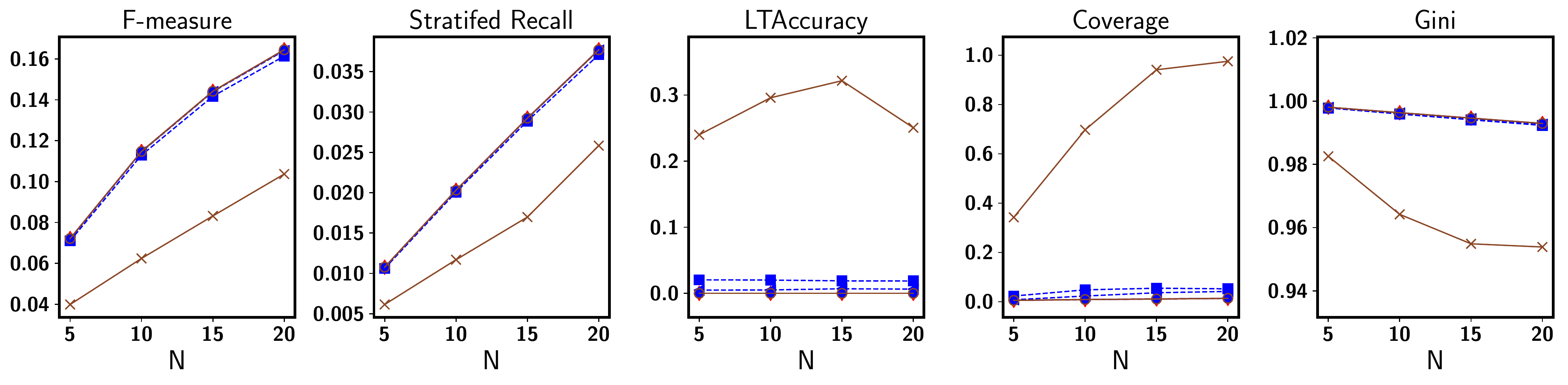}
		\label{fig:ThetaConfigs}
        }
        
\caption{Performance of GANC(ARec, $\bm{\theta}$, Dyn), with fixed sample size $S =|500|$,  different accuracy recommenders (ARec), and different long-tail preference models ($\bm{\theta}$). Dataset is ML-1M. The trends are approximately the same as we vary  the accuracy recommender  in Figures~\ref{fig:thetavsthetaStar}.a,~\ref{fig:thetavsthetaStar}.b,~\ref{fig:thetavsthetaStar}.c, and~\ref{fig:thetavsthetaStar}.d. In each row, ARec typically achieves the highest F-measure, but performs poorly w.r.t. coverage and gini.  Variants of our framework that use $\bm{\theta}^{G}$, $\bm{\theta}^{T}$, $\bm{\theta}^{N}$, obtain higher f-measure levels compared to those that use    $\bm{\theta}^{R}$ and $\bm{\theta}^{C}$. They also improve stratified recall, independent of the accuracy recommender. Stratified Recall emphasizes novelty and accuracy}.%, while LTAccuracy emphasizes novelty and coverage.}
\label{fig:thetavsthetaStar}
\end{figure*}

Figure~\ref{fig:thetavsthetaStar} shows performance of GANC(ARec, $\bm{\theta}$, Dyn) as $\bm{\theta}$ and the accuracy recommender (ARec) are varied. Across all rows, as expected, the accuracy  recommender on its own, typically obtains the best F-measure.  Moreover, $\bm{\theta}^{R}$ and $\bm{\theta}^{C}$ often have the lowest F-measure. Different variants of GANC with $\bm{\theta}^{N},\bm{\theta}^{T}$, and $\bm{\theta}^{G}$ are generally in the middle, which shows that these  preference estimates are better than  $\bm{\theta}^{R}$ and $\bm{\theta}^{C}$ in terms of accuracy.  For Stratified Recall, similar trends as accuracy are observed in the ranking of the algorithms.   The trends are approximately the same as we vary  the accuracy recommender  in Figures~\ref{fig:thetavsthetaStar}.b,~\ref{fig:thetavsthetaStar}.c, and~\ref{fig:thetavsthetaStar}.d.

%Re-read
%In Figure~\ref{fig:thetavsthetaStar}.a,  we observe $\theta^{R}$ and $\theta^{C}$  are able to  increase the novelty and coverage of recommendations.  On the other hand, $\theta^{T}$ and $\theta^{N}$, are not able to sufficiently increase coverage and F-measure since they are biased toward lower values (Figure~\ref{fig:Hist-LT-PreferenceModels}), and are more restrained in promoting long-tail items. Compared to $\theta^{T}$ and $\theta^{N}$,  $\theta^{*}$  obtains the highest coverage and novelty levels. 
%By using better estimates, we can increase novelty and coverage while  maintaining reasonable levels of  precision and recall.

\iffalse 
\input{otherResults}
\fi

\section{Comparison with other recommendation models}
We conduct two rounds of experiments: first, we evaluate re-ranking frameworks that post-process rating-prediction models, then we  study general top-$\size$ recommendation algorithms. For GANC, we run variants that involve randomness (e.g., sampling-based variants)  10 times, and report the average.

\subsection{Comparison with re-ranking methods for rating-prediction}
\label{sec:re-rankingratingprediction}
Standard re-ranking frameworks typically use a rating prediction model as their  accuracy recommender. In this section, we use RSVD as the underlying rating prediction model, and analyze performance across datasets with varying density levels.   We  compared the standard greedy ranking strategy of RSVD with 
 5D(RSVD), 5D(RSVD, A, RR),
 %5D(RSVD), 5D(RSVD, A), 5D(RSVD, RR), 5D(RSVD, A, RR),
 RBT(RSVD, Pop), RBT(RSVD, Avg),  PRA(RSVD,10),  PRA(RSVD,20), GANC(RSVD, $\bm{\theta}^T$, Dyn), and  GANC(RSVD, $\bm{\theta}^G$, Dyn). We report results for $\size=5$, since users rarely look past the items at the top of  recommendation sets.  Table~\ref{tab:re-rankingREGSVD} shows top-$5$ recommendation performance.  
% We have omitted ML-100K and Netflix due to space limitations, although their results are similar to ML-1M, and MT-200K, respectively.

\begin{table*}[pt]
\centering
\footnotesize
%\begin{tabular}{lllllllll}
%  \toprule
%&Alg.&HM-Rank & HM-Score&PR@5&F@5&L@5&C@5&G@5 \\

%\begin{tabular}{lllllllllll}
%  \toprule
%&Alg.&HM-Rank & HM-Score&PR@5&F@5&P@5&R@5&L@5&C@5&G@5 \\

\begin{tabular}{lllllllc}
  \toprule
&Alg.&F@5&S@5&L@5&C@5&G@5& Score\\
 \toprule 
\iffullpaper
\multirow{9}{*}{\rotatebox[origin=c]{90}{~ML-100K}} 
%&Alg.&F@5&S@5&L@5&C@5&G@5&Score \\
 & RSVD & 0.0279 (2) & 0.0098 (4) & 0.6649 (4) & 0.0707 (8) & 0.9886 (9) & 5.4 (4)\\
& 5D(RSVD)& 0.0013 (9)& 0.0014 (9)& \textbf{0.9260} (1)& 0.2586 (3)& 0.9000 (3)& 5.0 (2)\\
 & 5D(RSVD, A, RR) & 0.0071 (8) & 0.0035 (8) & 0.7669 (2) & 0.1171 (7) & 0.9783 (8) & 6.6 (6)\\
 & RBT(RSVD, Pop) & 0.0136 (7) & 0.0043 (7) & 0.7362 (3) & 0.1284 (6) & 0.9759 (7) & 6.0 (5)\\
 & RBT(RSVD, Avg) & 0.0201 (6) & 0.0075 (6) & 0.6271 (5) & 0.1938 (4) & 0.9583 (5) & 5.2 (3)\\
 & PRA(RSVD, 10) & 0.0252 (5) & 0.0103 (3) & 0.5642 (6) & 0.1171 (7) & 0.9674 (6) & 5.4 (4)\\
 & PRA(RSVD, 20) & 0.0255 (4) & 0.0095 (5) & 0.5544 (7) & 0.1379 (5) & 0.9581 (4) & 5.0 (2)\\
& GANC(RSVD, $\bm{\theta}^{T}$, Dyn)& \textbf{0.0310} (1)& \textbf{0.0127} (1)& 0.5064 (9)& 0.6260 (2)& 0.7669 (2)& 3.0 (1)\\
& GANC(RSVD, $\bm{\theta}^{G}$, Dyn)& 0.0260 (3)& 0.0122 (2)& 0.5501 (8)& \textbf{0.8716} (1)& \textbf{0.6242} (1)& 3.0 (1)\\

\midrule
\fi
\multirow{9}{*}{\rotatebox[origin=c]{90}{~ML-1M}}
%&Alg.&F@5&S@5&L@5&C@5&G@5&Score \\
 & RSVD & 0.0208 (3) & 0.0050 (6) & 0.7091 (3) & 0.0758 (9) & 0.9923 (9) & 6.0 (6)\\
& 5D(RSVD)& 0.0008 (9)& 0.0006 (9)& \textbf{0.9579} (1)& 0.1927 (4)& 0.9468 (3)& 5.2 (4)\\
 & 5D(RSVD, A, RR) & 0.0167 (6) & 0.0052 (5) & 0.6649 (5) & 0.1360 (6) & 0.9752 (6) & 5.6 (5)\\
 & RBT(RSVD, Pop) & 0.0091 (8) & 0.0022 (8) & 0.8019 (2) & 0.1125 (8) & 0.9872 (8) & 6.8 (7)\\
 & RBT(RSVD, Avg) & 0.0155 (7) & 0.0044 (7) & 0.6816 (4) & 0.2261 (3) & 0.9704 (4) & 5.0 (3)\\
 & PRA(RSVD, 10) & 0.0207 (4) & 0.0053 (4) & 0.6268 (6) & 0.1171 (7) & 0.9800 (7) & 5.6 (5)\\
 & PRA(RSVD, 20) & 0.0205 (5) & 0.0055 (3) & 0.5976 (7) & 0.1436 (5) & 0.9714 (5) & 5.0 (3)\\
& GANC(RSVD, $\bm{\theta}^{T}$, Dyn)& \textbf{0.0244} (1)& \textbf{0.0077} (1)& 0.5139 (9)& 0.5113 (2)& 0.8947 (2)& 3.0 (2)\\
& GANC(RSVD, $\bm{\theta}^{G}$, Dyn)& 0.0213 (2)& 0.0072 (2)& 0.5355 (8)& \textbf{0.6492} (1)& \textbf{0.8754} (1)& 2.8 (1)\\

\midrule
\multirow{9}{*}{\rotatebox[origin=c]{90}{~ML-10M}} 
%&Alg.&F@5&S@5&L@5&C@5&G@5&Score \\
& RSVD& \textbf{0.0147} (1)& \textbf{0.0021} (1)& 0.6775 (5)& 0.0066 (9)& 0.9992 (9)& 5.0 (4)\\
& 5D(RSVD)& 0.0000 (9)& 0.0000 (7)& \textbf{1.0000} (1)& 0.1248 (3)& \textbf{0.9609} (1)& 4.2 (2)\\
 & 5D(RSVD, A, RR) & 0.0024 (8) & 0.0007 (6) & 0.9421 (2) & 0.0489 (5) & 0.9968 (5) & 5.2 (5)\\
 & RBT(RSVD, Pop) & 0.0086 (6) & 0.0012 (5) & 0.8062 (3) & 0.0210 (6) & 0.9973 (7) & 5.4 (6)\\
 & RBT(RSVD, Avg) & 0.0087 (5) & 0.0013 (4) & 0.8039 (4) & 0.0614 (4) & 0.9945 (4) & 4.2 (2)\\
 & PRA(RSVD, 10) & 0.0116 (2) & 0.0020 (2) & 0.5888 (7) & 0.0085 (8) & 0.9978 (8) & 5.4 (6)\\
 & PRA(RSVD, 20) & 0.0110 (3) & 0.0020 (2) & 0.5992 (6) & 0.0115 (7) & 0.9972 (6) & 4.8 (3)\\
 & GANC(RSVD, $\bm{\theta}^{T}$, Dyn) & 0.0091 (4) & 0.0019 (3) & 0.5861 (8) & 0.2158 (2) & 0.9920 (3) & 4.0 (1)\\
& GANC(RSVD, $\bm{\theta}^{G}$, Dyn)& 0.0057 (7)& 0.0012 (5)& 0.5704 (9)& \textbf{0.2477} (1)& 0.9910 (2)& 4.8 (3)\\

\midrule
 \multirow{9}{*}{\rotatebox[origin=c]{90}{~MT-200K}}
%&Alg.&F@5&S@5&L@5&C@5&G@5&Score \\
 & RSVD & 0.0002 (5) & 0.0004 (4) & 0.9991 (2) & 0.0029 (9) & 0.9995 (9) & 5.8 (9)\\
& 5D(RSVD)& 0.0000 (6)& 0.0000 (5)& \textbf{0.9996} (1)& 0.0597 (3)& 0.9789 (3)& 3.6 (3)\\
 & 5D(RSVD, A, RR) & 0.0002 (5) & 0.0005 (3) & 0.9433 (5) & 0.0206 (5) & 0.9970 (6) & 4.8 (6)\\
 & RBT(RSVD, Pop) & 0.0002 (5) & 0.0005 (3) & 0.9988 (3) & 0.0154 (6) & 0.9968 (5) & 4.4 (5)\\
& RBT(RSVD, Avg)& 0.0005 (2)& \textbf{0.0008} (1)& 0.9701 (4)& 0.0273 (4)& 0.9957 (4)& 3.0 (2)\\
& PRA(RSVD, 10)& 0.0003 (4)& \textbf{0.0008} (1)& 0.9202 (6)& 0.0058 (8)& 0.9985 (8)& 5.4 (8)\\
& PRA(RSVD, 20)& 0.0004 (3)& \textbf{0.0008} (1)& 0.8038 (8)& 0.0082 (7)& 0.9974 (7)& 5.2 (7)\\
 & GANC(RSVD, $\bm{\theta}^{T}$, Dyn) & 0.0004 (3) & 0.0005 (3) & 0.7720 (9) & 0.2143 (2) & 0.9775 (2) & 3.8 (4)\\
& GANC(RSVD, $\bm{\theta}^{G}$, Dyn)& \textbf{0.0007} (1)& 0.0006 (2)& 0.8106 (7)& \textbf{0.2185} (1)& \textbf{0.9755} (1)& 2.4 (1)\\

\iffullpaper
\midrule
\multirow{9}{*}{\rotatebox[origin=c]{90}{~Netflix}} 
%&Alg.&F@5&S@5&L@5&C@5&G@5&Score \\
& RSVD& \textbf{0.0023} (1)& 0.0019 (2)& 0.6772 (6)& 0.0062 (8)& 0.9997 (8)& 5.0 (3)\\
& 5D(RSVD)& 0.0000 (8)& 0.0001 (7)& \textbf{0.9968} (1)& \textbf{0.3523} (1)& \textbf{0.9463} (1)& 3.6 (1)\\
 & 5D(RSVD, A, RR) & 0.0012 (5) & 0.0011 (5) & 0.7862 (4) & 0.1854 (2) & 0.9928 (2) & 3.6 (1)\\
 & RBT(RSVD, Pop) & 0.0010 (7) & 0.0010 (6) & 0.8199 (2) & 0.0044 (9) & 0.9991 (7) & 6.2 (6)\\
 & RBT(RSVD, Avg) & 0.0011 (6) & 0.0010 (6) & 0.8054 (3) & 0.0290 (5) & 0.9978 (5) & 5.0 (3)\\
 & PRA(RSVD, 10) & 0.0020 (3) & 0.0017 (3) & 0.6697 (8) & 0.0115 (7) & 0.9991 (7) & 5.6 (5)\\
 & PRA(RSVD, 20) & 0.0018 (4) & 0.0017 (3) & 0.6722 (7) & 0.0158 (6) & 0.9987 (6) & 5.2 (4)\\
& GANC(RSVD, $\bm{\theta}^{T}$, Dyn)& 0.0021 (2)& \textbf{0.0020} (1)& 0.5792 (9)& 0.0979 (4)& 0.9975 (4)& 4.0 (2)\\
 & GANC(RSVD, $\bm{\theta}^{G}$, Dyn) & 0.0012 (5) & 0.0016 (4) & 0.6938 (5) & 0.1522 (3) & 0.9962 (3) & 4.0 (2)\\

\fi
\bottomrule
\end{tabular}  
\caption{Top-5 recommendation performance for re-ranking a rating prediction model, RSVD. \iffullpaper  \else We have omitted ML-100K and Netflix, since they obtain similar results as ML-1M and MT-200K, respectively. \fi    
 The metrics are (F)measure@5, (S)tratified Recall@5, (L)TAccuracy@5, (C)overage@5, and (G)ini@5. Bolded entries show the best value for each metric, with relative rank of each algorithm on each metric inside parenthesis. For all models,  improving trade-offs is better on dense datasets. Regarding our two variants of GANC  (with fixed sample size $S=500$), they outperform others in all metrics except LTAccuracy, in dense settings (ML-1M).  On all datasets, our models obtain the lowest average rank across all metrics (last column). Overall, the results suggest a different accuracy recommender should be used in sparse settings.   }
\label{tab:re-rankingREGSVD}
\end{table*}

 Regarding RSVD, the model obtains high LTAccuracy, but under-performs all other models in coverage and gini. Essentially, RSVD picks a small subset of the items, including long-tail items,  and recommends them to all the users. After all, RSVD model is trained by minimizing Root Mean Square Error (RMSE) accuracy measure, which is  defined w.r.t. available data and not the complete data~\cite{steck2010training}. Therefore, when the model is used to choose a few items ($\size$)  from among \emph{all}  available items, as is required in top-$\size$ recommendation problems, it does not obtain good overall performance. 
 
In dense settings (ML-1M), GANC outperforms other models in all metrics except LTAccuracy. In sparse settings, except on ML-10M,  GANC has at least one variant in the top 2 methods w.r.t.  \iffullpaper F-measure. \else F-measure (see~\cite{ourFullVersion}). \fi  In both sparse and dense settings, except on ML-10M, GANC has at least one variant in the top 2 methods w.r.t.  stratified recall. Other methods, e.g.,~5D, focus exclusively on LTAccuracy and  reduce F-measure and stratified recall. 

In summary, the performance of  RSVD depends on the dataset density. On the sparse datasets, it  does not provide accurate suggestions w.r.t. F-measure, and subsequent re-ranking models make less accurate suggestions.  Although another reason for the smaller F-measure values on  datasets like  ML-10M (and Netflix), is the larger item space size. Top-5 recommendation, corresponds to a very small proportion of the overall item space. Re-ranking a rating prediction model like RSVD, is mostly effective for  dense settings.  However, our framework is generic,  and enables us to plug-in a different accuracy recommender. We show the results for this in the next section.   Moreover, all re-ranking techniques increase coverage, but reduce accuracy. 5D(RSVD) obtains the highest novelty among all models, but reduces accuracy significantly. On most datasets, GANC significantly increases coverage and decreases gini, while maintaining reasonable levels of accuracy.  
%(3.37\% in Netflix) 

\subsection{Comparison with top-N item recommendation models}
\label{sec:top-nrecommendation}

\begin{figure*}[t]	
\centering
	\subfloat[]
	{
		\includegraphics[width=1\textwidth]{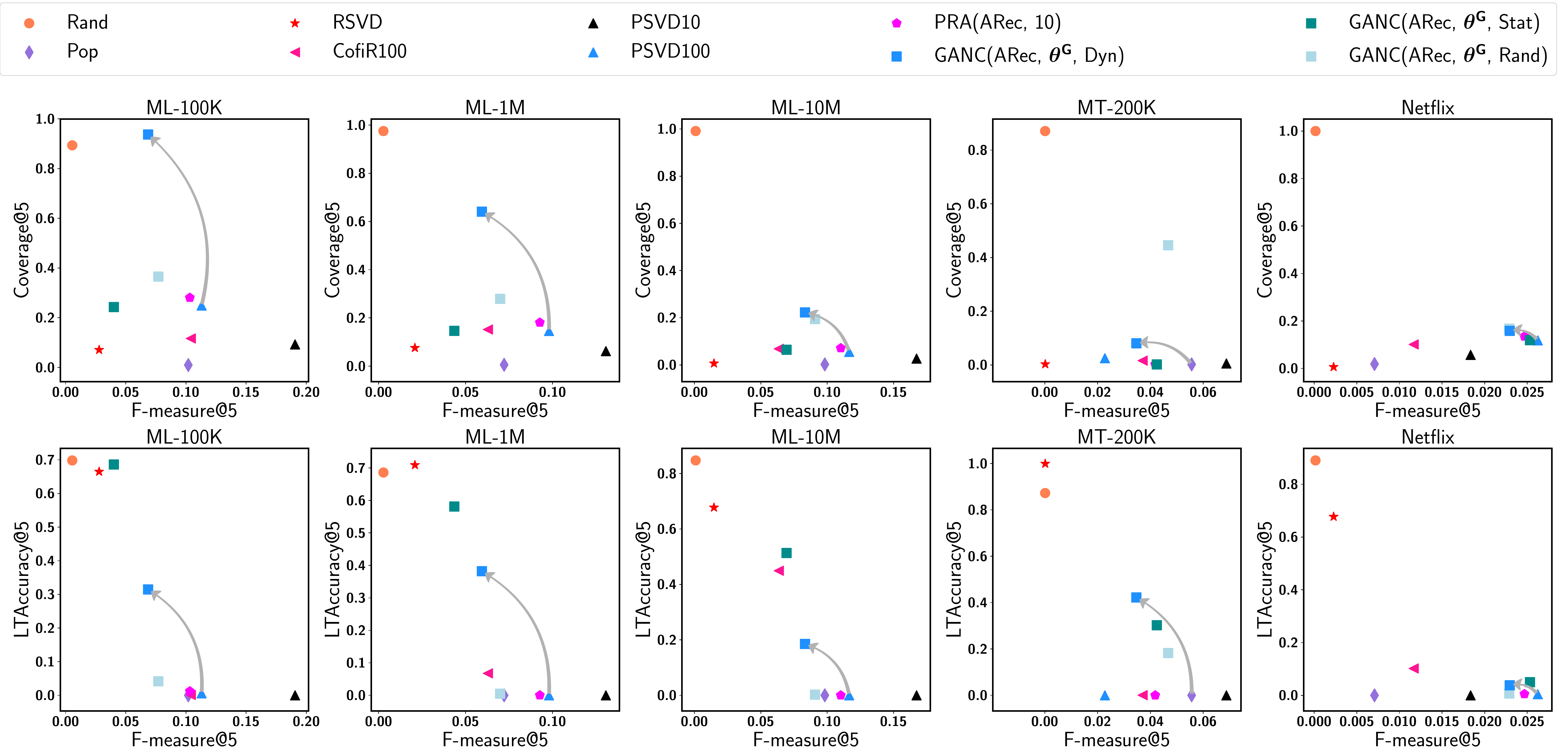} 
	}

\caption{Accuracy vs Coverage vs Novelty. The head of the arrow shows our main model GANC(ARec, $\bm{\theta}^G$, Dyn) with sample size  $S=500$, while the bottom shows the underlying accuracy recommender (ARec).   On MT-200K ARec is Pop. On other datasets it is PSVD-100.  Note, RSVD is consistently dominated by all other models in F-measure and coverage.  }
\label{fig:pcn}
\end{figure*}	

As shown previously, in sparse settings, re-ranking frameworks that rely on rating prediction models do not generate accurate solutions. In this section we plug-in a different accuracy recommender w.r.t. dataset density. 
On MT-200K,  we  plug-in  Pop. On all other datasets we plug-in  PSVD100 as the accuracy recommender.
 For GANC, we use three variants which differ in their coverage recommender: GANC(ARec, $\bm{\theta}^G$, Dyn), GANC(ARec, $\bm{\theta}^G$, Stat) and GANC(ARec, $\bm{\theta}^G$, Rand). We also compare to the generic re-ranking framework, PRA(ARec, 10). For both GANC and PRA, we always plug-in the same accuracy recommender. Furthermore, we compare with standard top-$\size$ recommendation algorithms:  Rand, Pop, RSVD, CofiR100, PSVD10, PSVD100.

Figure~\ref{fig:pcn} compares  accuracy, coverage and novelty trade-offs. On all datasets,  Rand achieves the best coverage and gini, but  the lowest  accuracy.  Similar to~\cite{cremonesi2010performance,vargas2014improving}, we find the non-personalized method Pop,  which takes advantage of the  popularity bias of the data, is a strong contender in accuracy, but under-performs in coverage and novelty. 

Regarding GANC,  Figure~\ref{fig:pcn} shows the best improvements in coverage are obtained when we use either Rand or Dyn coverage recommenders. Stat is generally not a strong coverage recommender,~e.g, GANC(ARec, $\bm{\theta}^G$, Stat) obtains high novelty (LTAccuracy)  on ML-10M, but does not lead to significant improvements in coverage.  This is because Stat has constant gain for long-tail promotion and focuses exclusively on recommending (a small subset of the) long-tail items.  Our main model is GANC(ARec, $\bm{\theta}^G$, Dyn). An interesting observation is that on MT-200K, both GANC(ARec, $\bm{\theta}^G$, Dyn) and GANC(ARec, $\bm{\theta}^G$, Rand), that use the non-personalized accuracy recommender Pop as ARec, are competitive with algorithms like  PSVD100 or CofiR100.

%END
%%%%%%%%%%%%%%%%%%%%%%%%%%%%%%%%%%%%%%5
%%%%%%%%%%%%%%%%%%%%%%%%%%%%%%%%%%%%%%%%
%%%%%%%%%%%%%%%%%%%%%%%%%%%%%%%%%%%%%%%%%%
%%%%%%%%%%%%%%%%%%%%%%%%%%%%%%%%%%%%%%%%%%%

\section{Related Work}
\label{sec:related-work}
We provide an overview of related work here. A detailed description of  baselines and methods integrated in our framework is provided in Section~\ref{sec:Experiments}.

%Methods that are integrated in our framework, and baselines,  are bolded in this section. 
%\subsection{Popularity bias and Sparsity of CF  Data}
%\label{Popularity-bias}
%\vspace{3mm}

%\subsection{Accuracy-focused CF Methods} 
%\label{sec:acc-focused-CF-methods}
%Standard CF methods target predictive accuracy and are grouped into rating prediction or ranking prediction CF~\cite{lee2012comparative}. %herlocker2002empirical,

%\vspace{4mm}
%\noindent \textbf{Accuracy-focused CF Methods} are grouped into rating prediction or ranking prediction CF~\cite{lee2012comparative}. We next describe two  groups. 

\vspace{4mm}
\noindent \textbf{Accuracy-focused rating prediction CF models} aim to accurately predict unobserved user ratings~\cite{koren2008factorization,lee2012comparative}.  % target predictive accuracy formulate  recommendation as a rating prediction problem. 
Accuracy is measured using  error metrics such as RMSE or Mean Absolute Error that are defined w.r.t. the observed user feedback.
%RAP: I took out the short versions RMSE and MAE because we never used them anywhere else. I also took out the discussion of training and testing phase since I don't think people will care and it saved a live.
%In the training phase,  the value of unobserved user ratings are predicted so that they accurately reflect the user's observed ratings. In the testing phase, the value of withheld user ratings are predicted and accuracy is measured. %The scoring component predicts the value of unobserved user ratings so that they accurately reflect the user's observed ratings.  The  ranking component then greedily selects the top $\size$ items based  on their predicted value. 
%These  methods can be categorized into memory-based, model-based, or hybrid combinations~\cite{koren2008factorization}. Memory-based CF include the neighbourhood models, user-based CF~\cite{herlocker1999algorithmic}, and item-based CF~\cite{sarwar2001item} that make recommendations based on similarities between users, or items. Model-based methods train a parametric model on the user-item matrix and base their recommendations on the model. Matrix Factorization-based (MF) models~\cite{koren2009matrix, koren2008factorization} %, paterek2007improving 
%, a.k.a.~Singular Value Decomposition (SVD) models, belong to this  group. 
Generally, these methods are grouped into memory-based or model-based~\cite{koren2008factorization}. Earlier memory-based or neighbourhood models used the rating data directly to compute similarities between users~\cite{herlocker1999algorithmic} or items~\cite{sarwar2001item}. However,  these models are not scalable on large-scale datasets like Netflix. Model-based methods  instead train a parametric model on the user-item matrix.  Matrix Factorization models~\cite{koren2008factorization,koren2009matrix}, a.k.a.~Singular Value Decomposition (SVD) models, belong to this group and are  known for both  scalablility and accuracy~\cite{cremonesi2010performance}.  %herlocker1999algorithmic, paterek2007improving
% We consider four models from this category
 \iffullpaper 
These models factorize the user-item matrix into a product of two matrices; one  maps the users ($\mathcal{U}$), and the other maps items ($\mathcal{I}$), into a latent factor space of dimension $g \ll \min(\vert \mathcal{U} \vert, \vert \mathcal{I} \vert )$. Here,  each user $u$, and each item $i$, are  represented by a factor vector, $\mathbf{p}_u \in \mathbb{R}^{g}$, and $\mathbf{q}_i \in \mathbb{R}^{g}$, respectively. Ratings are estimated using $\hat{r}_{ui} = \mathbf{q}^{T}_i \mathbf{p}_u$. 
%However, due to the large number of missing values in the user-item matrix,  conventional SVD cannot be applied to learn the factors.  Instead, the  regularized squared error on the observed ratings is minimized.
Due to the large number of missing values in the user-item matrix,  the  regularized squared error on the observed ratings is minimized. The resulting objective function is  non-convex, and iterative methods such as Stochastic Gradient Descent (SGD) or Alternating Least Squares (ALS)~\cite{koren2009matrix, mnih2007probabilistic,rendle:tist2012} %srebro2004maximum,paterek2007improving
 can be used to find a local minimum.
\else
Using the user-item interaction data, these models learn latent factors for representing the users and items. Unobserved user ratings are estimated using the latent factors~\cite{koren2009matrix, mnih2007probabilistic,rendle:tist2012,zhuang2013fast}.   
\fi 
%From this category, we use a Regularized SVD  (RSVD) model and the same model with non-negative constraints (RSVDN)~\cite{koren2009matrix,zhuang2013fast}, in Section~\ref{sec:Experiments}.
From this group, we use  RSVD and RSVDN~\cite{koren2009matrix,zhuang2013fast}, in Section~\ref{sec:Experiments}.

\vspace{2mm}
\noindent \textbf{Accuracy-focused ranking prediction CF models} focus on accurately compiling ranked lists.  
The intuition is that since predicted rating values are not shown to the user, the focus should be on accurately selecting and ranking items.
Accordingly, accuracy is measured using  ranking metrics,  like recall and precision,  that can be   measured either on the observed  user feedback, or on \emph{all} items~\cite{cremonesi2010performance,steck2013evaluation}. 
%For ranking tasks, Most Popular (Pop) obtains high precision and recall~\cite{cremonesi2010performance,vargas2014improving}. PureSVD (PSVD)~\cite{cremonesi2010performance} and CofiRank (Cofi)~\cite{weimer2007maximum} are well-known  latent factor models for ranking,  with others in~\cite{balakrishnan2012collaborative,lee2014local}. We use Pop, PSVD, and Cofi in our experiments.
For ranking tasks, Pop obtains high precision and recall~\cite{cremonesi2010performance,vargas2014improving}. PSVD~\cite{cremonesi2010performance} and CoFiRank~\cite{weimer2007maximum} are well-known  latent factor models for ranking,  with others in~\cite{balakrishnan2012collaborative,lee2014local}. We use Pop, PSVD, and CofiRank in  Section~\ref{sec:Experiments}.

\vspace{2mm}
\noindent \textbf{Multi-objective methods}  devise new models that optimize several objectives, like  coverage and novelty, in addition to accuracy~\cite{vargas2014improving,steck2011item, yin2012challenging,niemann2013new,shi2013trading}.  In~\cite{niemann2013new},  items are assumed to be similar if they significantly co-occur with the same items. This leads to better representations for  long-tail items, and increases their chances of being recommended. A new performance measure that combines accuracy and popularity bias is proposed in~\cite{steck2011item}. This measure can be gradually tuned towards recommending long-tail items.  More recently, the idea of recommending users to items as a means of improving sales diversity, and implicitly, recommendation novelty, while retaining precision, has been explored in~\cite{vargas2014improving}.  In comparison to both~\cite{steck2011item,niemann2013new}, we focus on targeted promotion of long-tail items. While~\cite{vargas2014improving} focuses on neighbourhood models for top-$\size$ recommendation, our framework is generic and independent of the recommender models. 
Graph-based approaches for long-tail item promotion are studied in~\cite{yin2012challenging,shi2013trading}. They  construct a bipartite user-item graph and use a random walk to trade-off between popular and long-tail items. Rather than devise new  multi-objective models, we post-process existing models.
% A graph-based approach that  simultaneously trades-off among accuracy, sales diversity, and long-tail criteria, is proposed~\cite{shi2013trading}.  Here, each edge is associated with a cost and a transition probability. The latter encodes the likelihood of a user rating/purchasing an item whereas the former trades off relevance of recommendations and promotion of long-tail items. A similar approach  is employed in~\cite{yin2012challenging} where edge transition probabilities are discounted by item popularity. In both, a random walk leads to a trade-off between popular and long-tail items.  Other examples include clustering  of items according to their popularity~\cite{park2013adaptive}, learning classifiers for the items and ranking potential users of rare items~\cite{huang2005item}, and combining CF with decision tree models in people-to-people recommendations~\cite{krzywicki2012using}. 

\vspace{2mm}
\noindent \textbf{Re-ranking methods}  post-process the output of a standard model to account for additional objectives like coverage and diversity rather than devising a new model. These algorithms are very efficient. 
%Re-ranking to  simultaneously combine accuracy, diversity, novelty, and serendipity is  studied in~\cite{zhang2012auralist}.Re-ranking to maximize diversity within individual top-$\size$ is explored in~\cite{qin2013promoting,zhang2008avoiding,ziegler2005improving}.%ashkan2015optimal,
\cite{ziegler2005improving} explores re-ranking to maximize diversity within individual top-$\size$. It shows that users preferred diversified lists despite their lower accuracy. However, diversifying individual top-$\size$ sets does not necessarily increase  coverage~\cite{jambor2010optimizing,niemann2013new}. %(aggregate diversity (or coverage))
Re-ranking techniques that directly maximize coverage  and promote long-tail items are explored in~\cite{adomavicius2011maximizing,adomavicius2012improving,ho2014likes,jugovac2017efficient}. In contrast to~\cite{adomavicius2011maximizing,adomavicius2012improving,ho2014likes} that re-rank rating prediction models, our framework is generic and is independent of the base recommendation  model. Furthermore,  our long-tail  personalization is independently learned from interaction data.  PRA~\cite{jugovac2017efficient} is also generic framework for re-ranking, although we differ in our long-tail novelty prefernce modelling. We use RBT~\cite{adomavicius2012improving}, Resource allocation~\cite{ho2014likes}, PRA~\cite{jugovac2017efficient} as baselines since we share similar objectives.

\vspace{2mm}
\noindent \textbf{Modelling user novelty preference} is studied in~\cite{jambor2010optimizing,jugovac2017efficient,kapoor2015like,oh2011novel,Zhao:2016:MNR:2911451.2911488}.  As explained in~\cite{kapoor2015like}, an item can be novel in three ways: \begin{enumerate*}
\item it is new to the system and is unlikely to be seen by most users (cold-start), \item it existed in the system but is new to a single user, \item  it existed in the system, was previously known by the user, but is a forgotten item.
\end{enumerate*}
\cite{kapoor2015like,Zhao:2016:MNR:2911451.2911488} focus on definitions 2 and 3 of novelty, which are useful in settings where seen items can be recommended again, e.g.,~music recommendation. 
For defining users' novelty preferences, tag and temporal data are used in~\cite{kapoor2015like,Zhao:2016:MNR:2911451.2911488}.
\iffullpaper
A logistic regression model is used to predict user novelty preferences in~\cite{kapoor2015like}, while~\cite{Zhao:2016:MNR:2911451.2911488} learn  a curiosity model for each user based on her access history and item tags.
\fi
\iffalse
The curiosity model is a function of item novelties. The item  novelty for that user is defined based on the user's item access frequency, recency of access, and item tags. ~\cite{Zhao:2016:MNR:2911451.2911488} proposed a framework to accurately recommend items that the user is curious about. Gross movie earnings are used for defining users' tendencies for novelty in~\cite{oh2011novel}.  In particular, to measure the tendency of users toward popular items,~\cite{oh2011novel} define a notion of personal popularity tendency (PPT) per user using the average gross earnings of each movie and based on the past ratings of the user. The computation of PPT, however, is based on the  assumption that  the total earnings of a movie reflects its popularity. The idea is to match the PPT of each user with the PPT of recommended items, thereby increasing the novelty of recommendation lists in a targeted manner. 
\fi
In~\cite{oh2011novel}, gross movie earnings are assumed to reflect  movie popularity, and are used to define a personal popularity tendency (PPT) for each user. The idea is to match the PPT of each user with the PPT of recommended items~\cite{oh2011novel}. 
The major difference between our work and~\cite{kapoor2015like,oh2011novel,Zhao:2016:MNR:2911451.2911488} is that we focus on the  cold-start definition of novelty (definition 1), we do not use contextual or temporal information to define users' tendencies for novelty, and we consider domains where  each item is accessed at most once (seen items cannot be recommended).

In~\cite{jambor2010optimizing},  users are  characterized  based on their tendency towards disputed items, defined as items with  high average rating and high variance. These items  are claimed to  be in the long-tail. We differ in terms of long-tail novelty definition, and consequently our preference estimates.  \iffullpaper Furthermore,~\cite{jambor2010optimizing}  independently solve a constrained convex optimization problem for each user, with the user's disputed item tendency as a constraint. \fi \cite{zhang2013personalize,wang2009portfolio} use a user risk indicator to decide  between a personalized  and a non-personalized model; both focus on accuracy. In contrast, we combine accuracy and coverage models. Furthermore, while their risk indicators are optimized via cross validation, we learn the users' long-tail preferences.

PRA~\cite{jugovac2017efficient}   models user  preference for  long-tail novelty using on  item popularity statistics. In contrast,  we consider  additional information, like if the user found the item interesting,  and the long-tail preferences of other users  of the item.

\vspace{2mm}
\noindent \textbf{CF interaction data properties and test ranking protocol } %\footnote{RAP: this is a very weird start to the related work. I would expect you to talk about related algorithms, instead you're talking about bias in the dataset. However, as mentioned, you also bias the dataset by removing those with fewer than 5 ratings. Further, in reading this, it's not clear what the point is of this sub-section. I think that you're trying to tell us that your data preparation is better than others, but that doesn't come all the way across. I'd also explicitly state that you discuss algorithms that you compare against in the experimental section}
are two important aspects to consider in recommendation setting. CF interaction data suffers from  popularity bias~\cite{agarwal_chen_2016}.  In the movie rating domain, for instance, users are more likely to rate movies they know and like~\cite{cremonesi2010performance,steck2011item,steck2013evaluation,agarwal_chen_2016,steck2010training}. As a result, the partially observed interaction data is  not a  random subset of the (unavailable) complete interaction data. %(it is Missing Not At Random (MNAR)).  
Furthermore, many real-world CF interaction datasets~\cite{dooms2013movietweetings,kanagal2012supercharging,liu2017experimental,zolaktaf2015learn} are sparse, and the majority of  items and users  have few observations available. Due to  the popularity bias and sparsity of datasets,  many accuracy-focused  CF models are also biased toward recommending popular items. 
 % For example, almost $87\%$ of the movies in MovieTweetings~\cite{dooms2013movietweetings} are long-tail, and approximately $47.42\% $ of the users  have rated fewer than 10 items. 
% almost $88\%$ of the items in the Netflix rating dataset are long-tail, and approximately, $3.37\%$ of users have rated fewer than 10 items.  MovieTweetings~\cite{dooms2013movietweetings}is a movie rating dataset collected from Twitter.
  
 Moreover, some accuracy evaluation protocols are also  biased and tend to reward  popularity biased algorithms~\cite{cremonesi2010performance,vargas2014improving,steck2013evaluation}.  In~\cite{steck2013evaluation}, the main evaluation protocols are assessed in  detail, and the ``All unrated items test ranking protocol'' is described to be closer to the accuracy the user experience in real-world recommendation setting, where performance is measured using the complete data rather than available data~\cite{steck2013evaluation}. Following~\cite{steck2013evaluation,vargas2014improving}, and w.r.t. the additional experiments we conducted 
\iffullpaper
in Appendix~\ref{test-protocol-effect},
\else 
in~\cite{ourFullVersion},
\fi
 we chose the ``All unrated  items test ranking protocol'' for experiments in  Section~\ref{sec:Experiments}.

\section{Conclusion}
\label{sec:conclusion}
This paper presents a generic top-$\size$ recommendation framework for  trading-off accuracy, novelty, and coverage. To achieve this, we profile the users according to their preference for long-tail novelty. We examine various measures, and formulate an optimization problem to learn these user preferences from interaction data.  We then integrate the user preference estimates in our generic framework, GANC.  Extensive experiments on several datasets confirm that there are trade-offs between accuracy, coverage, and novelty. Almost all re-ranking models increase coverage and novelty at the cost of accuracy. However, existing re-ranking models typically rely on rating prediction models, and are hence more effective in dense settings. Using a generic approach, we can easily incorporate a suitable base accuracy recommender to devise an effective solution for both sparse and dense settings.  %Our results  also indicate there is no single method that outperforms other methods in all metrics. However our techniques obtain a significant improvement in coverage, while  . 
Although we integrated the  long-tail novelty preference estimates into a re-ranking framework, their use-case is not limited to these frameworks. In  the future, we intend to explore the temporal and topical dynamics of long-tail novelty preference, particularly in settings where contextual information is  available.  

\section*{Acknowldegments}
We thank  Glenn Bevilacqua,  Neda Zolaktaf,  Sharan Vaswani, Mark Schmidt, Laks V. S. Lakshmanan, and Tamara Munzner  for their help and valuable discussions. We also thank  the anonymous reviewers for their constructive feedback. This work has been funded by NSERC Canada,  and supported in part by the Institute for Computing, Information and Cognitive Systems at UBC.

%For zainab, backup of files
%\input{TopRec}
%\input{submit-wsdm}
%\input{IC-Alg}
%\input{MarginalIncrement}
%\input{FocusingDegree}
%\input{figuresBackup}
%\input{backup}
% NOTE keywords are not used for conference papers so do not populate them
% \begin{keywords}
% keyword-1, keyword-2, keyword-3
% \end{keywords}
%

%\section*{Acknowledgment}
%We thank the anonymous reviewers.

\bibliographystyle{IEEEtran}
\small
\bibliography{ref-rec} 
\iffullpaper 
\appendix

\subsection{Regularized SVD configuration}
\label{sec:configurationofR-SVD}
\begin{table}[ht]
\centering
\scriptsize
\begin{tabular}{lllllllll}
  \toprule
  &\multicolumn{4}{c}{R-SVD} & \multicolumn{4}{c}{R-SVDN} \\
 \cmidrule(r){2-5} \cmidrule(r){6-9} 
  {\bf Dataset}  &  $\eta$ & $\lambda$ & $g$ & RMSE  &  $\eta$ & $\lambda$ & $g$ & RMSE  \\ \midrule
  ML-100K &0.03 & 0.05  & 100 &  0.935   & 0.03 &0.05 & 100& 0.935\\
  
  ML-1M  & 0.03 & 0.05 & 100 &  0.868   & 0.03 & 0.05 & 100 & 0.875   \\
  
  ML-10M  &0.003 & 0.005 &  20 & 0.872   & 0.003 & 0.005& 20 &0.872    \\
  
  %MT-200k-m5 & 0.01 &  0.1 & 40 & 1.4786   &0.01 &0.1& 40 & 1.4785  \\
  
  MT-200k  & 0.01 &  0.01 & 40 & 0.761   &0.01 &0.01& 40 & 0.761 \\ 
  
  %MT-200k-m20 &  0.01 & 0.01 & 8 & 1.521  & 0.01 & 0.01 & 8 &1.5209\\
  
  %MT-200k-m20-mapped &  0.01 & 0.01 & 8 & 0.7822  & 0.01 & 0.01 & 8 & 0.7822 \\
  
  Netflix & 0.002 & 0.05 &  100 & 0.979  & 0.002 & 0.05 & 100 & 0.979 \\
   \bottomrule
\end{tabular}
\caption{R-SVD and R-SVDN  parameters on different datasets. $g$ is the number of latent factors, $\eta$ is the learning rate, $\lambda$ is the L2-reqularization coefficient.}
\label{tab:DatasetStatsticsFull}
\end{table}

Table~\ref{tab:DatasetStatsticsFull} provides details for the setup of Regularized SVD (R-SVD) and the same model with non-negative constraints (R-SVDN). We use \texttt{LIBMF}~\cite{zhuang2013fast} with  L2-Norm as the loss function, and L2-regularization,  and SGD for optimization. We performed 10-fold cross validation and tested:  number of latent factors  $g \in \{8,  20, 40, 50, 80, 100\}$, L2-regularization coefficients  $\lambda \in \{0.001, 0.005, 0.01, 0.05, 0.1\}$,  learning rate $\eta \in \{0.002,0.003,0.01, 0.03\}$. For each dataset, we used the parameters that led to best performance.

\iffalse
\subsection{Other baselines configuration}
 provides the details for algorithm parameters in different datasets. We conducted  experiments over the parameter space of the various algorithms and report the  parameters that led to the best  performance in  Table~\ref{tab:DatasetStatsticsFull}.
 For CofiRank, we experimented with both regression (squared) loss (Cofi$^R$) and NDCG loss (Cofi$^N$).  We use the parameter values  provided in the paper~\cite{weimer2007maximum},  100 dimensions and $\lambda =10$, and default values for other parameters (source code). Similar to \cite{balakrishnan2012collaborative,volkovs2012collaborative}, we found Cofi$^R$ to perform consistently better than Cofi$^N$ in our experiments on ML-1M and MT-200K. We only report results for Cofi$^R$.   For the Resource allocation re-ranking~\cite{ho2014likes} method, we implemented and ran all four variants with default values set according to~\cite{ho2014likes} and MF parameters set according to Table~\ref{tab:DatasetStatsticsFull}. For PureSVD we use Python's \texttt{sparsesvd} module and tested $g^{'} \in \{10, 20, 40, 60, 100, 150, 200, 300\}$.
\fi
\subsection{Analysis of Fully Sequential  Dynamic Coverage}
\label{sec:submodularmonotoneproof}
%We assume the same item can be recommended to many users. Let $\mathcal{U}$ denote the users,  $\mathcal{I}$ the items.
In this section, we show the problem of finding  a top-$\size$ collection $\mathcal{P}  = \{ \mathcal{P}_u \}_{u=1}^ {|\mathcal{U}|} $ that maximizes Eq.~\ref{eq:overallValueFunction}, is an instance of  maximizing a submodular function subject to a matroid constraint.  

\vspace{4mm}
\noindent \textbf{Matroids.}
A set system is a pair ($\mathcal{I}, \mathcal{F}$), where $\mathcal{I}$ denotes a ground set of elements and $\mathcal{F} = \{\mathcal{A}: \mathcal{A} \subseteq \mathcal{I}\}$, is a collection of subsets of $\mathcal{I}$. % that are feasible sets. 
A set system  is an independence system if it satisfies 
\begin{enumerate*}
\item $\emptyset \in \mathcal{F}$,
\item  $\mathcal{A} \subseteq \mathcal{B} \in \mathcal{F} \text{ then } \mathcal{A} \in  \mathcal{F}$. 
\end{enumerate*}
A \textit{matroid} is an independence system that also satisfies the property $\mathcal{A} ,\mathcal{B} \in \mathcal{F} \text{ and } |\mathcal{B}| > |\mathcal{A}| \text{ then } \exists i \in \mathcal{B} \setminus \mathcal{A} \text{ with } \mathcal{A} \cup \{i\} \in \mathcal{F}$. A \textit{uniform matroid}, is a special class of matroids that satisfies   $\mathcal{F} = \{\mathcal{A}: \mathcal{A} \subseteq \mathcal{I}, |\mathcal{A}| \leq \size \}$, that is all basis are maximal. A \textit{partition matroid} satisfies  $\mathcal{F} = \{\mathcal{A}: \mathcal{A} = \cup_{i=1}^{\size} \mathcal{A}_i, \mathcal{A}_i \subseteq \mathcal{I}_i, |\mathcal{A}_i| \leq  l_i, \cup \mathcal{I}_i = \mathcal{I}\}$. 
\begin{lemma}
\label{lemma1}
The constraint of recommending $\size$ items to each user, corresponds to a partition matroid over the users. 
\end{lemma}
\begin{proof} 
Define a new ground set $\mathcal{N} = \{(u,i) :  u \in \mathcal{U}, i \in \mathcal{I} \}$. Define $\mathcal{N}_u = \{(u,i): i \in \mathcal{I} \} , u \in \mathcal{U}$ and let $l_u=\size, \forall u \in \mathcal{U}$. Let $\mathcal{M} =(\mathcal{U}, \mathcal{F})$ where $\mathcal{F} = \{ \mathcal{P}^{'}: \mathcal{P}^{'}=\cup_{u \in \mathcal{U}} \mathcal{P}^{'}_u, \mathcal{P}^{'}_u \subseteq \mathcal{N}_u, |\mathcal{P}^{'}_u| \leq l_u, \cup \mathcal{N}_u = \mathcal{N}\}$. $\mathcal{P}^{'}$ form independent sets of a partition matroid.
\end{proof}

\vspace{4mm}
\noindent \textbf{Submodularity and Monotonicity.}
Let $\mathcal{I}$  denote a ground set  of items.  Given a  set function $f:2^{\mathcal{I}} \rightarrow \mathbb{R}$,  $\delta(i|\mathcal{A}) :=  f(\mathcal{A} \cup\{i\}) - f(\mathcal{A})$ is the marginal gain of $f$ at $\mathcal{A}$ with regard to item $i$.  Furthermore, $f$ is  submodular if and only if  $\delta(i|\mathcal{A}) \geq \delta(i|\mathcal{B}), \forall \mathcal{A}\subseteq \mathcal{B} \subseteq \mathcal{I}, \forall i \in \mathcal{I}\setminus \mathcal{B}$. It is modular if $f(\mathcal{A} \cup {i})$ = $f(\mathcal{A}) + f(i)$, $\forall \mathcal{A} \subset \mathcal{I}, i \in \mathcal{I} \setminus \mathcal{A}$. In addition,  $f$ is  monotone increasing if $ f(\mathcal{A}) \le f(\mathcal{B}), \forall \mathcal{A}\subseteq \mathcal{B}\subseteq \mathcal{I}$. Equivalently, a function is monotone increasing if and only if $\forall \mathcal{A} \subseteq \mathcal{I}$ and $i \in \mathcal{I}$, $\delta(i|\mathcal{A}) \geq 0$~\cite{krause2012submodular}.
Submodular functions have the following concave composition property:

\begin{theorem}
Using dynamic coverage, the objective function $v(.)$ in Eq.\ref{eq:overallValueFunction} is submodular monotone increasing w.r.t. sets of user-item pairs. 
\end{theorem}

\begin{proof}Consider the ground set $\mathcal{N}$,  defined in Lemma~\ref{lemma1}. Based on any set $\mathcal{P}^{'} \subseteq \mathcal{N}$, define $\mathcal{P}^{'}_{u} = \{ i | (u,i) \in \mathcal{P}^{'}\}$. We can rewrite the objective function with dynamic coverage as 
%The objective function can be written as
\begin{align} 
v(\mathcal{P}^{'}) &= \sum_u v_u(\mathcal{P}^{'}_{u})\nonumber \\  
&=  \sum_u  (1-\theta_u) a(\mathcal{P}^{'}_{u})  + \theta_{u} c(\mathcal{P}^{'}_{u}) \nonumber \\
&= \sum_u  (1-\theta_u) \sum_{i \in \mathcal{P}^{'}_{u}} a(i)  + \theta_{u} \sum_{i \in\mathcal{P}^{'}_{u}} c(i) \label{formula-l1} \\
&= \sum_u  (1-\theta_u) \sum_{i \in \mathcal{P}^{'}_{u}} \hat{r}_{ui}  + \theta_{u} \sum_{i \in \mathcal{P}^{'}_{u}} \frac{1}{\sqrt{1+f_i^{\mathcal{P}^{'}}}}
\end{align}
where $f_{i}^{\mathcal{P}^{'}}$ is the number of times item $i$ is recommended  in $\mathcal{P}^{'}$. For submodularity consider any $\mathcal{A} \subseteq \mathcal{B} \subset \mathcal{N}$, and a pair $(u,i) \in \mathcal{N} \setminus \mathcal{B}$.  We have 
\begin{align*}
f_i^{\mathcal{A}} &  \leq f^{\mathcal{B}}_i \\
%& \frac{1}{1+f_i^{\mathcal{A}}} \geq \frac{1}{1+f^{\mathcal{B}}_i}  \\
\frac{1}{\sqrt{1+f_i^{\mathcal{A}}}} & \geq \frac{1}{\sqrt{1+f^{\mathcal{B}}_i}} \\
(1-\theta_u) \hat{r}_{ui} + \theta_u \frac{1}{\sqrt{1+f_i^{\mathcal{A}}}} & \geq (1-\theta_u) \hat{r}_{ui} + \theta_u \frac{1}{\sqrt{1+f^{\mathcal{B}}_i}} \\
\delta(i|\mathcal{A}) & \geq \delta(i|\mathcal{B})
\end{align*}
%So the coverage function $c(.)$ is submodular.  In addition, the accuracy function $a(.)$ is a modular function. 
Therefore, due to the submodularity of the coverage function, the overall value function  $v(.)$ is submodular. %, which is a non-negative linear combination of a modular accuracy function and a submodular coverage function.
  
For monotonicity, both $a(.)$ and $c(.)$ map a set of items $\mathcal{P}_u$ to the $[0,1]$ range, and  are additive in terms of the number of items (line~\ref{formula-l1}). So, they are both monotonically increasing, i.e., adding a new element $i \in \mathcal{I} \setminus \mathcal{P}_u$ to the set $\mathcal{P}_{u}$ can only increase their value. Since  $\theta_{u} $ is also  in $[0,1]$, $v_{u}(.)$ is monotonically increasing. $v(.)$ is therefore submodular monotone increasing since it is a sum of submodular monotone increasing functions.\end{proof}

\newpage
\subsection{Effect of test ranking protocol on performance metrics}
\label{test-protocol-effect}

\begin{figure}[t]
\centering

\subfloat[All unrated items ranking protocol]{\includegraphics[width=0.4\textwidth]{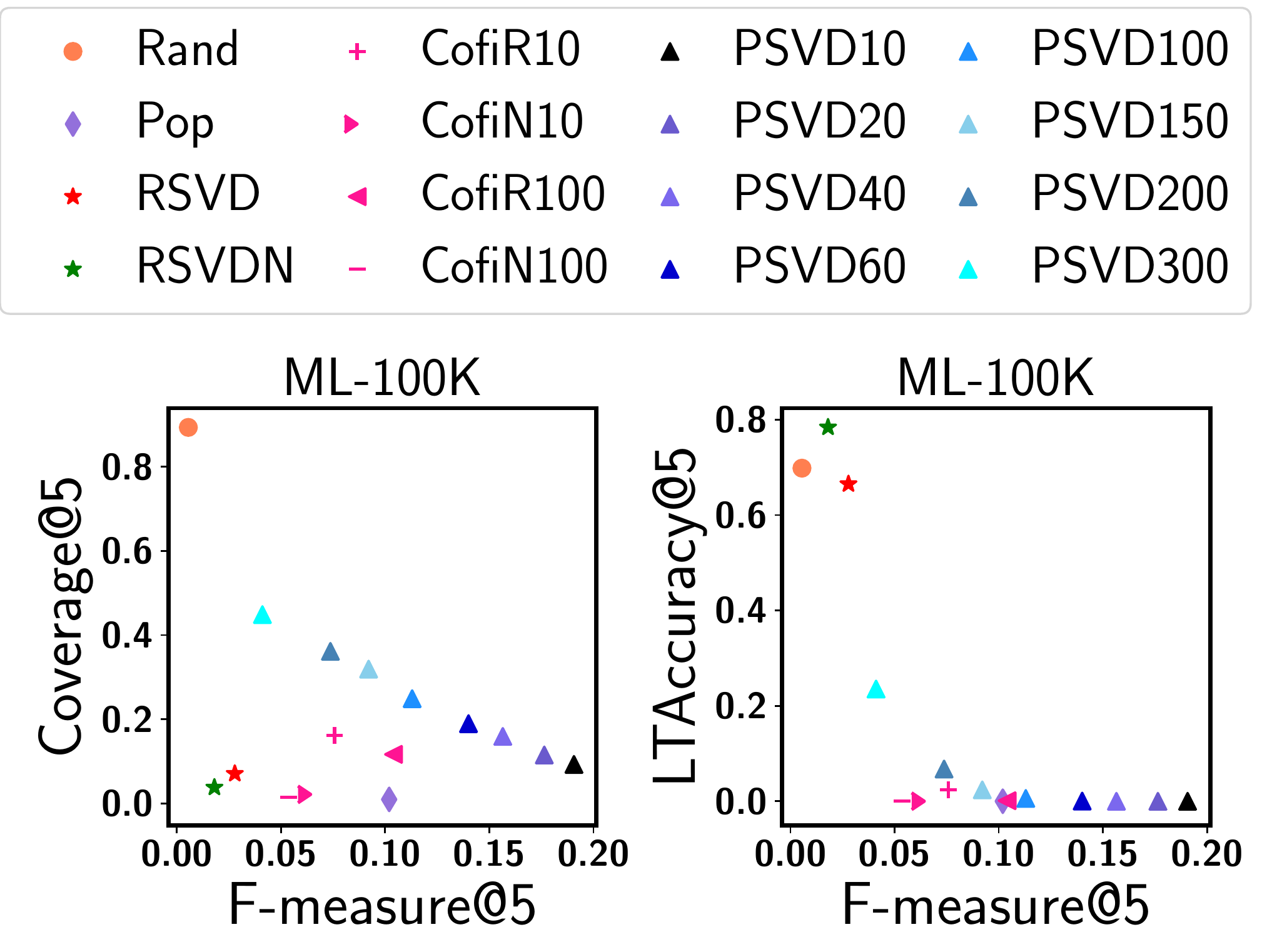}}

\subfloat[Rated test-items ranking protocol]{\includegraphics[width=0.4\textwidth]{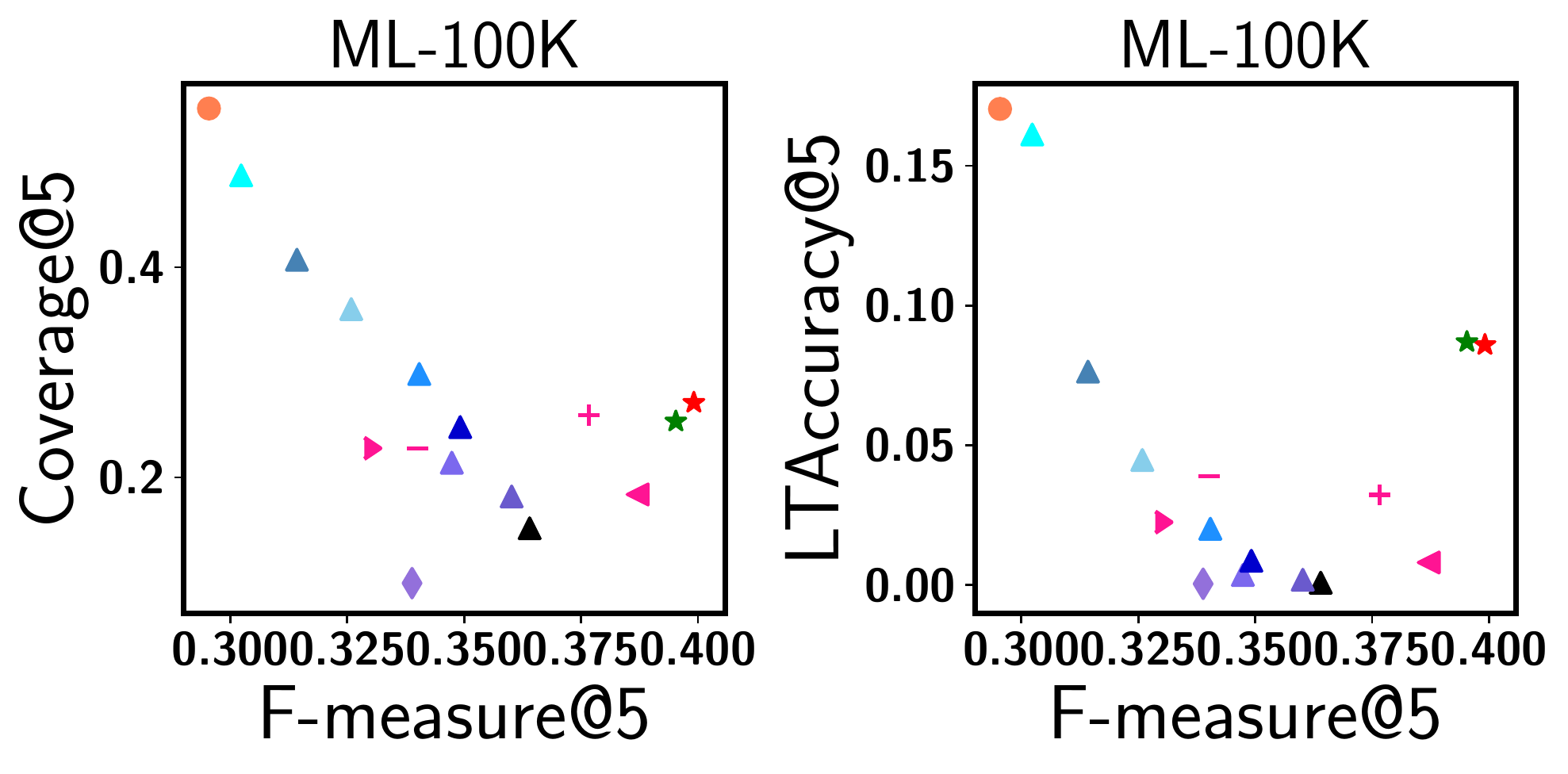}}

\subfloat[All unrated items ranking protocol]{\includegraphics[width=0.4\textwidth]{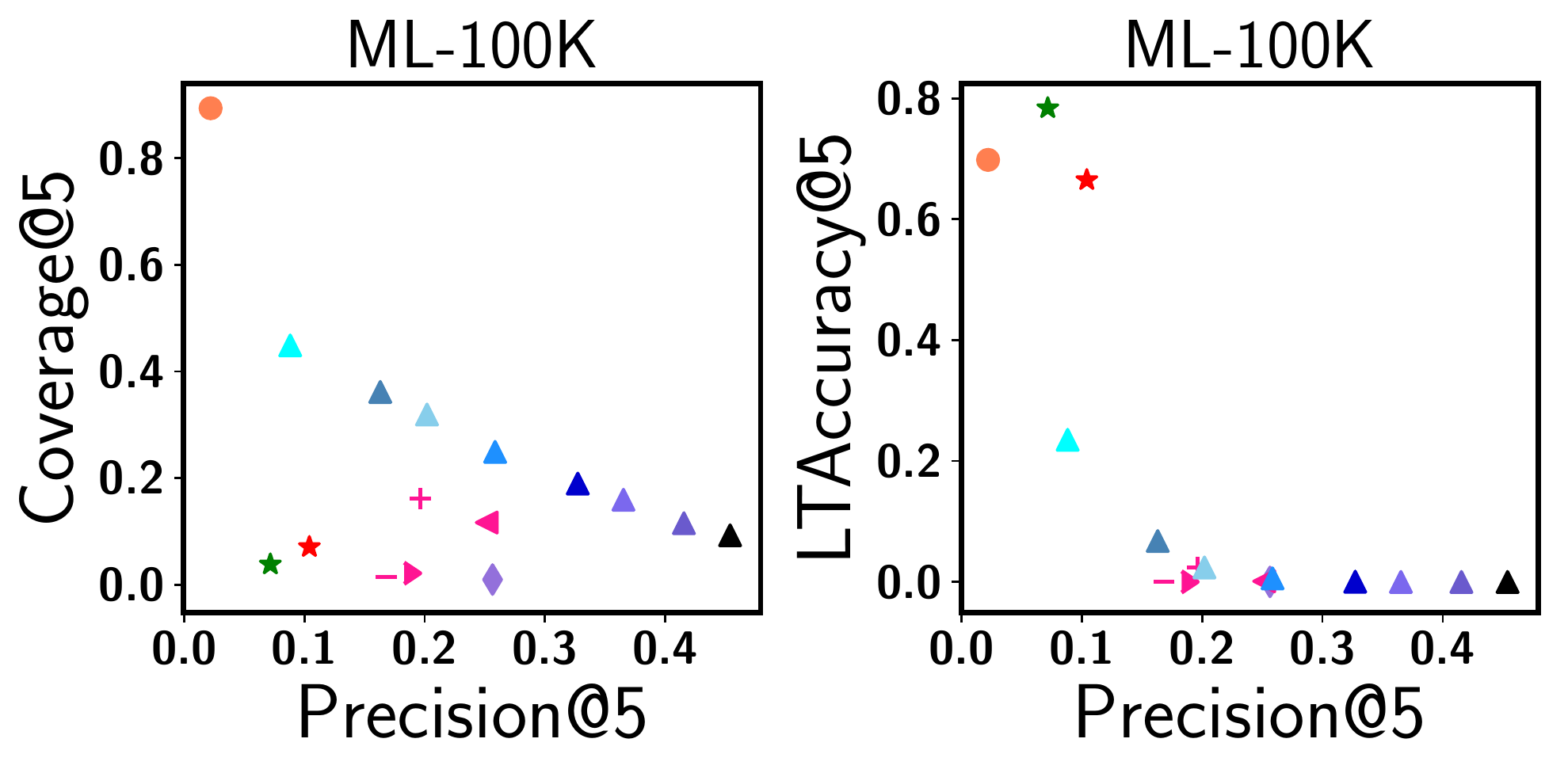}}

\subfloat[Rated test-items ranking protocol]{\includegraphics[width=0.4\textwidth]{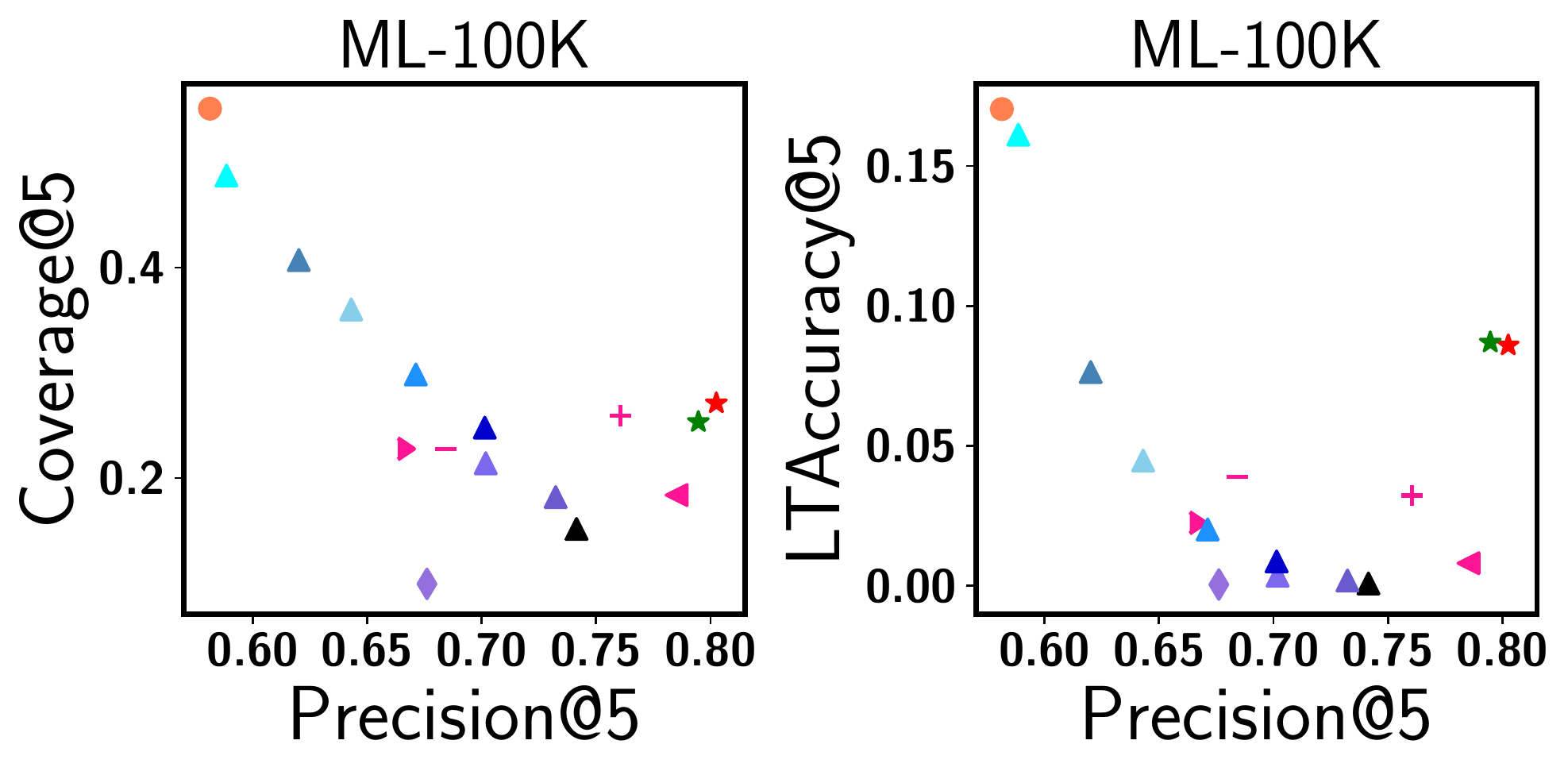}}

\caption{Comparing the trade-offs  for top-5 recommendation, when using different ranking protocols. Dataset is ML-100K.}

\label{fig:CP-LC-tradeoffs-ml-100k}
\end{figure}

In our empirical study, we studied off-line recommendation performance of top-$\size$ recommendation algorithms from three perspectives: accuracy, novelty, and coverage (Section~\ref{sec:ExpSetup}). The  choice of \emph{test ranking protocol}~\cite{steck2013evaluation} is also an important aspect in off-line evaluations. The  test ranking protocol describes which items in the test set are ranked for each user~\cite{steck2013evaluation}. We use the definitions in~\cite{steck2013evaluation}: 
\begin{itemize}
\item \textbf{Rated test-items ranking protocol}: for each user, we only rank  the observed items in the test set of  that user. %with we refer to a ranking of items that  are observed in the test set of a user.
\item \textbf{All unrated items ranking protocol}: for each user, we rank all items that do not appear in the train set of that user.
%With \emph{all unrated items} ranking protocol we refer to a ranking  of all unrated test items, regardless of whether the user has provided feedback on them or not. 
\end{itemize}

\begin{figure}[t]
\centering

\subfloat[All unrated items ranking protocol]{\includegraphics[width=0.4\textwidth]{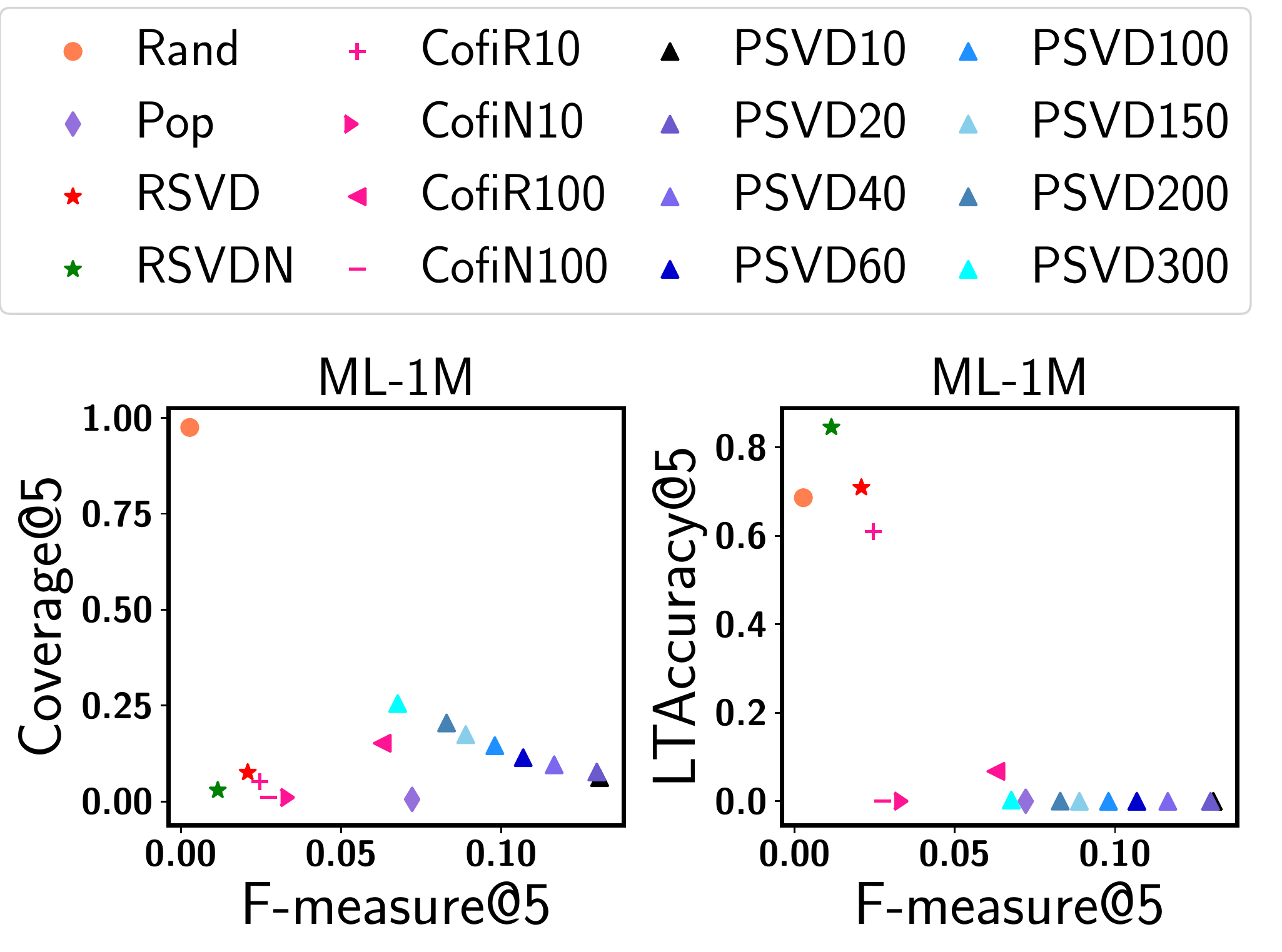}}

\subfloat[Rated test-items ranking protocol]{\includegraphics[width=0.4\textwidth]{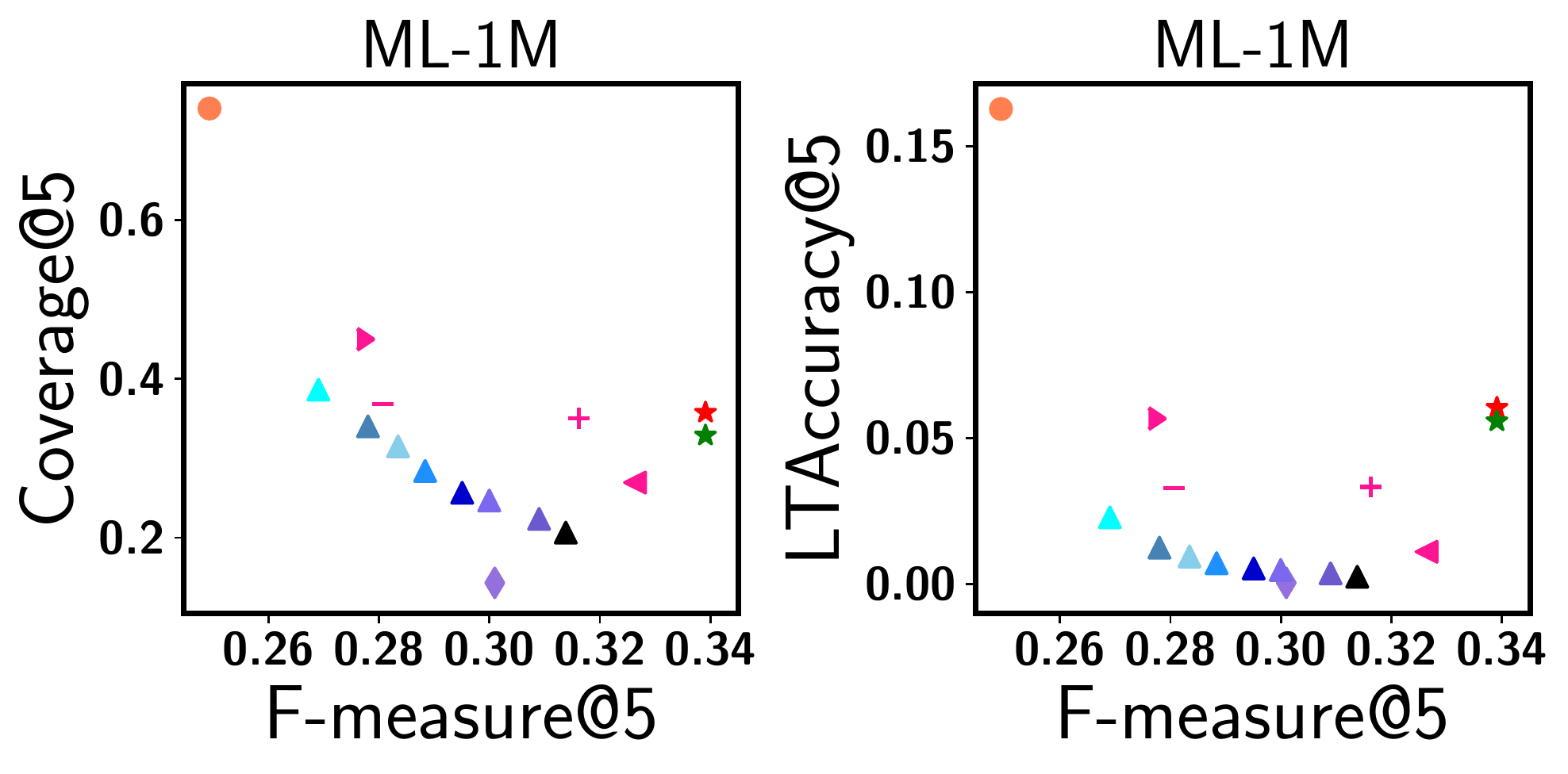}}

\subfloat[All unrated items ranking protocol]{\includegraphics[width=0.4\textwidth]{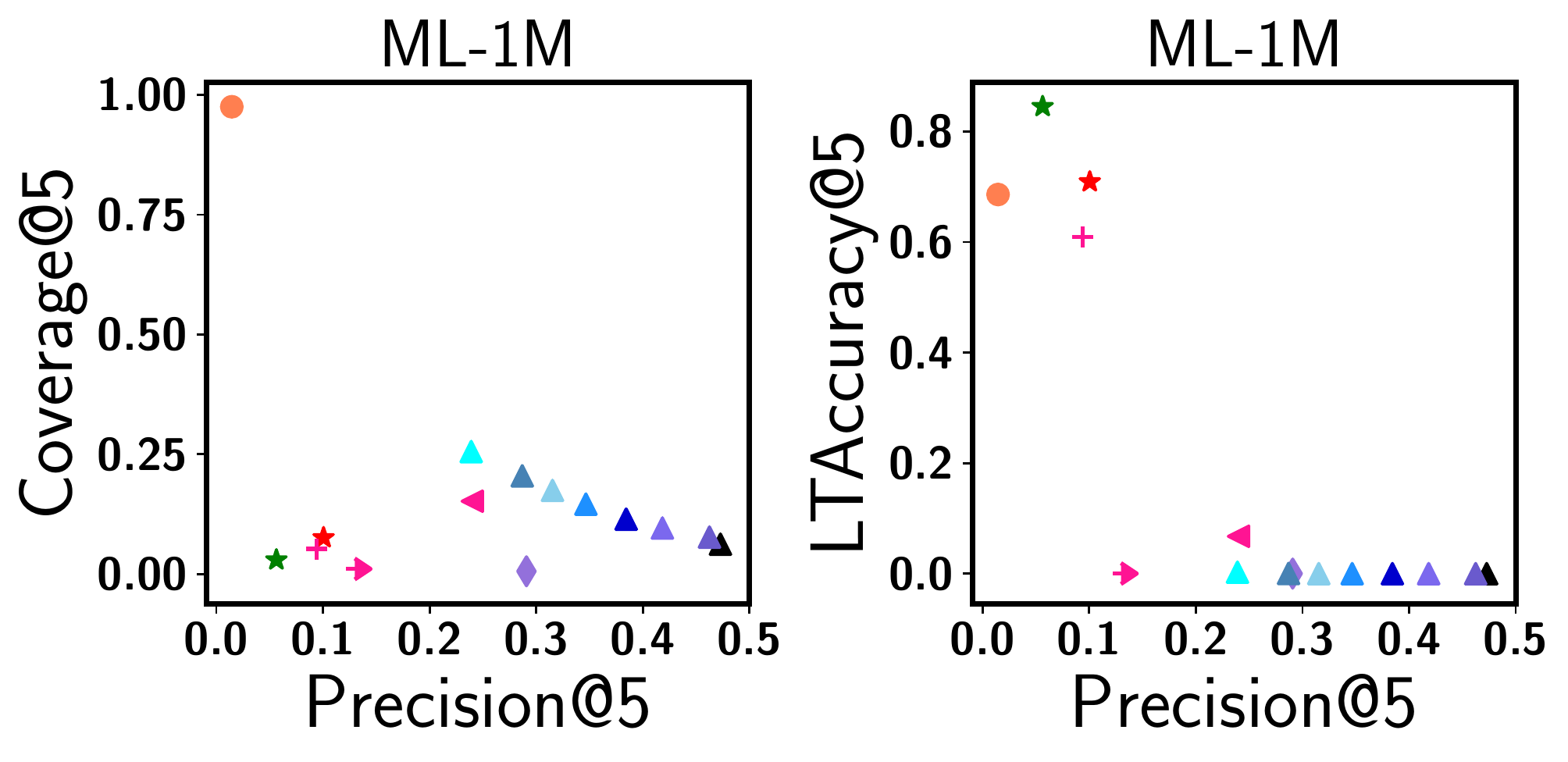}}

\subfloat[Rated test-items ranking protocol]{\includegraphics[width=0.4\textwidth]{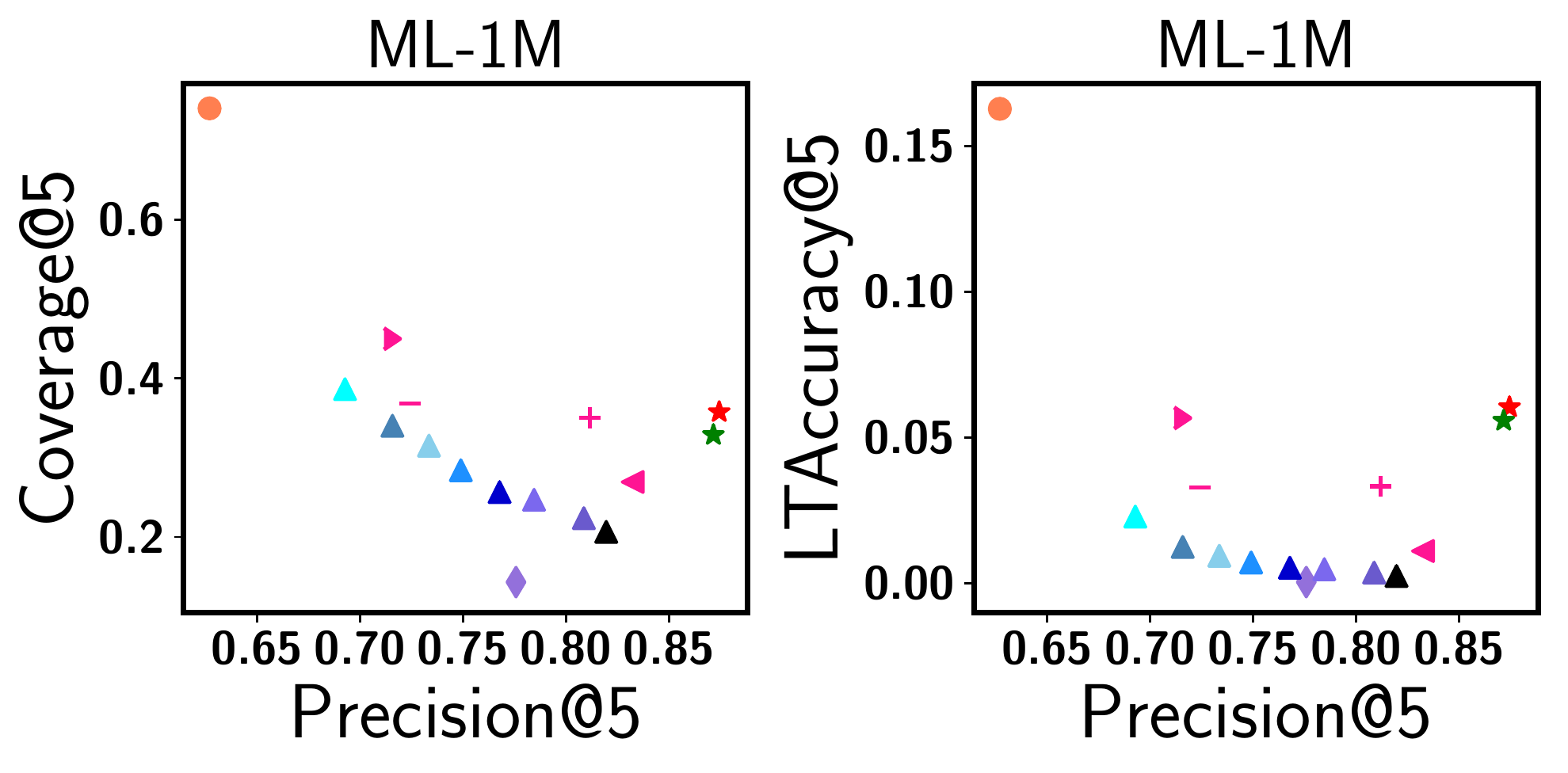}}

\caption{Comparing the trade-offs  for top-5 recommendation, when using different ranking protocols. Dataset is ML-1M.}

\label{fig:CP-LC-tradeoffs-ml-1m}
\end{figure}

%In other words, their measurements only consider  the subset of  items that have received feedback. Naturally
%We begin this section by introducing different test ranking protocols. We then  use them to  evaluate off-line performance of standard accuracy focused CF models,  from accuracy, coverage, and novelty perspectives.
%However, since by definition,  error metrics are measured only on the observed user feedback (subset of the items), they are not a precise indicator of accuracy as experienced by the user~\cite{cremonesi2010performance}.

In~\cite{steck2013evaluation}, it was shown  that  the choice of test ranking protocol can affect accuracy measurements considerably. Specifically,  accuracy is  measured either  by error metrics (e.g., RMSE and MAE) or ranking metrics (e.g., recall and precision) in  recommendation settings ~\cite{steck2013evaluation}. Error metrics are defined w.r.t.  the observed user feedback only.  Rating prediction models that optimize these metrics in the train phase, e.g., R-SVD, typically  adopt  the rated test-items ranking protocol in the test phase. However,  error metrics are not  precise indicators of  accuracy,  since they only consider the subset of items that have feedback. But  in real-world  applications, the system must find and rank  a  few items ($\size$)  from among \emph{all} items~\cite{cremonesi2010performance,steck2013evaluation}. 

Ranking metrics can be  measured on the observed  user feedback, or on \emph{all} items.  When  measured on the observed  user feedback, these metrics can  be strongly biased because of the  popularity bias of datasets~\cite{agarwal_chen_2016,cremonesi2010performance,steck2011item,vargas2014improving,steck2013evaluation}.  Therefore,  in top-$\size$ recommendation settings, these metrics are measured  using  the all-items ranking protocol,  to better reflect  accuracy as experienced by users  in real-world applications~\cite{steck2013evaluation,vargas2014improving}.

We extend the empirical study of~\cite{steck2013evaluation} by evaluating  the effect of the test ranking protocol on  accuracy, coverage, and novelty.  We use standard  accuracy-focused CF models (introduced in Section~\ref{sec:related-work}).  We set $\size=5$, and ran experiments on  ML-100K and Ml-1M  datasets, shown in  Figures~\ref{fig:CP-LC-tradeoffs-ml-100k}  and~\ref{fig:CP-LC-tradeoffs-ml-1m}, respectively.    %Details of experimental set up is provided in Section~\ref{sec:Experiments}. 

\balance

We analyze Figure~\ref{fig:CP-LC-tradeoffs-ml-1m}, which shows the results on the ML-1M dataset (Figure~\ref{fig:CP-LC-tradeoffs-ml-100k} has similar trends).  The first observation is that  all  algorithms obtain higher F-measure scores using the rated test-item ranking protocol. In particular, for the all unrated items ranking protocol, Figure~\ref{fig:CP-LC-tradeoffs-ml-1m}.a,  F-measure lies in $[0,0.2]$ (with the corresponding precision in $[0,0.5]$ in Figure~\ref{fig:CP-LC-tradeoffs-ml-1m}.c ). 
For the rated test-items ranking protocol  F-measure lies in $[0.2,0.4]$ (with precision in $[0.6,0.9]$) .  As a specific example, consider Rand which  randomly  suggests items according to  the ranking protocol.   As expected, random suggestion from among all items has low F-measure and precision. However, random suggestion from among the test items of each user, leads to an average F-measure of almost $0.25$ (precision approximately $0.6$). This demonstrates the bias  of the rated test-items ranking protocol. 
Similar to~\cite{cremonesi2010performance}, we observe Pop  is a strong contender in accuracy metrics, using both test protocols~\cite{cremonesi2010performance}. Recent work in~\cite{liu2017experimental} also confirmed that Pop outperformed more sophisticated algorithms for tourists datasets,   that are sparse  and where the users have few and  irregular interests (only visit popular locations). 
In addition, although R-SVD and R-SVDN are less accurate using the all unrated items ranking protocol, they obtain the highest F-measure scores using the rated test-items ranking protocol. This is because these models are optimized w.r.t. the  observed user feedback. Therefore, the rated test-items ranking protocol is to their advantage because it also considers  only the observed user feedback.

Coverage is also affected by the ranking protocol. The rated test-items ranking protocol  results in better coverage for all algorithms except Random.  This is because random suggestion from among a user's test items, is a more constrained algorithm compared to random from among all train items. 
Regarding the effect of the ranking protocol on  LTAccuracy,  it lies in $[0,1]$ for the all unrated test items ranking protocol,  but drops to $[0,0.2]$ for the rated test-items ranking protocol.  %By definition,  long-tail items are those that receive fewer ratings,  therefore, they receive fewer ratings in the  test set, and since the observed test-items are ranked per user, LTAccuracy  scores are generally lower. 

Regarding the  trade-off between  metrics, irrespective of the ranking protocol, we observe that  Pop makes accurate yet trivial recommendations that lack novelty,   as indicated by the low LTaccuracy and coverage in Figure~\ref{fig:CP-LC-tradeoffs-ml-1m}. For the PureSVD models (P-SVD), on ML-100K and ML-1M, increasing the number of factors reduces  F-measure, and results in an increase in coverage and LTAccuracy.  Among all baselines, CoFiRank with regression loss (CofiR10, CofiR100) has reasonable coverage, precision, and long-tail accuracy. 

Overall, our experiments in this section confirm the findings of prior work~\cite{cremonesi2010performance,steck2013evaluation,agarwal_chen_2016}:  due to the popularity bias of recommendation datasets, rank-based precision  is strongly biased when it is measured using the rated test-items ranking  protocol.  This is demonstrated by the results of Pop, which  achieves an F-measure of $0.3$  on ML-100K  and  ML-1M, with even higher precision scores on those datasets.    It outperforms personalized models like CoFiRank.  However, using the all-items ranking protocol, the performance of these models aligns better with expectation.   Following prior research on top-$\size$ recommendation ~\cite{steck2013evaluation,vargas2014improving}, we conducted  the experiments in Section~\ref{sec:Experiments}  using the all-items ranking protocol.

\fi
\end{document}